\documentclass[iop]{emulateapj}

\usepackage{lineno}

\usepackage{natbib,aas_macros,amsmath}
\usepackage{multirow,color}
\usepackage{natbib}

\newcommand{\carcsec}{$\!\!\arcsec$}
\newcommand{\m}[1]{\mathrm{#1}}

\newcommand{\redc}[1]{\textcolor{black}{#1}}
\newcommand{\redcc}[1]{\textcolor{black}{#1}}
\newcommand{\redccc}[1]{\textcolor{black}{#1}}

\begin{document}
\shortauthors{Harikane et al.}
\slugcomment{Accepted for Publication in ApJS}

\shorttitle{
Census of 4 Million Star-forming Galaxies at $z=2-7$
}

\title{
GOLDRUSH. IV. Luminosity Functions and Clustering Revealed with $\sim$4,000,000 Galaxies at $z\sim2-7$: Galaxy-AGN Transition, Star Formation Efficiency, and Implication for Evolution at $z>10$ 
}

\email{hari@icrr.u-tokyo.ac.jp}
\author{
Yuichi Harikane\altaffilmark{1,2},
Yoshiaki Ono\altaffilmark{1},
Masami Ouchi\altaffilmark{3,4,1,5},
Chengze Liu\altaffilmark{6},
Marcin Sawicki\altaffilmark{7,8},
Takatoshi Shibuya\altaffilmark{9},\\
Peter S. Behroozi\altaffilmark{10},
Wanqiu He\altaffilmark{3},
Kazuhiro Shimasaku\altaffilmark{11,12},
Stephane Arnouts\altaffilmark{13},
Jean Coupon\altaffilmark{14},
Seiji Fujimoto\altaffilmark{15,16},\\
Stephen Gwyn\altaffilmark{17},
Jiasheng Huang\altaffilmark{18,19},
Akio K. Inoue\altaffilmark{20,21},
Nobunari Kashikawa\altaffilmark{11,12},
Yutaka Komiyama\altaffilmark{3,4},\\
Yoshiki Matsuoka\altaffilmark{22}, 
and
Chris J. Willott\altaffilmark{17}
}

\affil{$^1$
Institute for Cosmic Ray Research, The University of Tokyo, 5-1-5 Kashiwanoha, Kashiwa, Chiba 277-8582, Japan
}
\affil{$^2$
Department of Physics and Astronomy, University College London, Gower Street, London WC1E 6BT, UK
}
\affil{$^3$
National Astronomical Observatory of Japan, 2-21-1 Osawa, Mitaka, Tokyo 181-8588, Japan
}
\affil{$^4$
Graduate University for Advanced Studies (SOKENDAI),
2-21-1 Osawa, Mitaka, Tokyo 181-8588, Japan
}
\affil{$^5$
Kavli Institute for the Physics and Mathematics of the Universe (Kavli IPMU, WPI), The University of Tokyo, 5-1-5 Kashiwanoha,Kashiwa, Chiba, 277-8583, Japan
}
\affil{$^6$
Department of Astronomy, School of Physics and Astronomy, and Shanghai Key Laboratory for Particle Physics and Cosmology, Shanghai Jiao Tong University, Shanghai 200240, P. R. China
}
\affil{$^7$
Institute for Computational Astrophysics and Department of Astronomy and Physics, Saint Mary’s University, 923 Robie Street, Halifax, Nova Scotia, B3H 3C3, Canada
}
\affil{$^8$
Canada Research Chair
}
\affil{$^9$
Kitami Institute of Technology, 165, Koen-cho, Kitami, Hokkaido 090-8507, Japan
}
\affil{$^{10}$
Department of Astronomy and Steward Observatory, University of Arizona, Tucson, AZ 85721, USA
}
\affil{$^{11}$
Department of Astronomy, School of Science, The University of Tokyo, 7-3-1 Hongo, Bunkyo-ku, Tokyo 113-0033, Japan
}
\affil{$^{12}$
Research Center for the Early Universe, The University of Tokyo, 7-3-1 Hongo, Bunkyo-ku, Tokyo 113-0033, Japan
}
\affil{$^{13}$
Aix Marseille University, CNRS, CNES, Laboratoire d’Astrophysique de Marseille, Marseille, France
}
\affil{$^{14}$
Astronomy Department, University of Geneva, Chemin d’Ecogia 16, CH-1290 Versoix, Switzerland
}
\affil{$^{15}$
Cosmic Dawn Center (DAWN), Copenhagen Denmark
}
\affil{$^{16}$
Niels Bohr Institute, University of Copenhagen, Jagtvej 128, Copenhagen, Denmark
}
\affil{$^{17}$
NRC Herzberg Astronomy and Astrophysics, 5071 West Saanich Road, Victoria, BC, V9E 2E7, Canada
}
\affil{$^{18}$
CASSACA, National Astronomical Observatories of China
}
\affil{$^{19}$
Harvard-Smithsonian Centre for Astrophysics, USA
}
\affil{$^{20}$
Department of Physics, School of Advanced Science and Engineering, Faculty of Science and Engineering, Waseda University, 3-4-1, Okubo, Shinjuku, Tokyo 169-8555
}
\affil{$^{21}$
Waseda Research Institute for Science and Engineering, Faculty of Science and Engineering, Waseda University, 3-4-1, Okubo, Shinjuku, Tokyo 169-8555, Japan
}
\affil{$^{22}$
Research Center for Space and Cosmic Evolution, Ehime University, Bunkyo-cho, Matsuyama, Ehime 790-8577, Japan
}

\begin{abstract}
We present new measurements of rest-UV luminosity functions and angular correlation functions from 4,100,221 galaxies at $z\sim2-7$ identified in the Subaru/Hyper Suprime-Cam survey and CFHT Large-Area $U$-band Survey.
The obtained luminosity functions at $z\sim4-7$ cover a very wide UV luminosity range of $\sim0.002-2000L_\m{UV}^*$ combined with previous studies, \redc{confirming} that the dropout luminosity function is a superposition of the AGN luminosity function dominant at $M_\m{UV}\lesssim-24\,\mathrm{mag}$ and the galaxy luminosity function dominant at $M_\m{UV}\gtrsim-22\,\mathrm{mag}$, consistent with galaxy fractions based on 1037 spectroscopically-identified sources.
Galaxy luminosity functions estimated from the spectroscopic galaxy fractions show the bright end excess beyond the Schechter function at $\gtrsim2\sigma$ levels, possibly made by inefficient mass quenching, low dust obscuration, and/or hidden AGN activity.
By analyzing the correlation functions at $z\sim2-6$ with halo occupation distribution models, we find a weak redshift evolution (within $0.3\,\mathrm{dex}$) of the ratio of the star formation rate (SFR) to the dark matter accretion rate, $SFR/\dot{M}_\mathrm{h}$, indicating the almost constant star formation efficiency at $z\sim2-6$, as suggested by our earlier work at $z\sim4-7$.
Meanwhile, the ratio gradually increases with decreasing redshift at $z<5$ within $0.3\,\mathrm{dex}$, which quantitatively reproduces the cosmic SFR density evolution, suggesting that the redshift evolution is primarily driven by the increase of the halo number density due to the structure formation, and the decrease of the accretion rate due to the cosmic expansion.
Extrapolating this calculation to higher redshifts assuming the constant efficiency suggests a rapid decrease of the SFR density at $z>10$ with $\propto10^{-0.5(1+z)}$, which will be directly tested with {\it{JWST}}.
\end{abstract}


\section{Introduction}\label{ss_intro}

Studying statistical properties of galaxies is important to understand the overall picture of galaxy formation and evolution.
To quantify galaxy build-up in the early universe, many studies have investigated luminosity functions (i.e., one-point statistics) and angular correlation functions (i.e., two-point statistics) of high redshift galaxies.
The luminosity function represents the volume density of galaxies as a function of the luminosity.
Since galaxies form in dark matter halos, the luminosity function is related to the dark matter mass function and baryonic physics of galaxy formation.
Studying the shape and evolution of the luminosity function in the high redshift universe allows us to obtain key insights into the star formation and feedback processes.

Great progress has been made in determining luminosity functions of high redshift galaxies, especially in the rest-frame ultraviolet (UV), which is redshifted to the optical wavelength at $z\sim4-7$ easily accessible from ground-based telescopes.
Since the time-averaged unobscured star formation rate (SFR) of galaxies is proportional to the luminosity of galaxies in the rest-frame UV, the UV luminosity function provides us with a measure of how quickly galaxies grow with cosmic time.
Analyses of galaxies in deep blank fields including the \textit{Hubble} Ultra Deep field (HUDF) have resulted in identifying $\sim20,000$ galaxy candidates at $z\sim2-10$ down to the absolute UV magnitude of $M_\m{UV}\sim-17$ mag (e.g., \citealt{2010ApJ...725L.150O}, \citealt{2013ApJ...763L...7E}; \citealt{2013ApJ...768..196S}; \citealt{2013MNRAS.432.2696M}; \citealt{2015ApJ...803...34B,2019ApJ...880...25B,2021arXiv210207775B}; \citealt{2015ApJ...810...71F}; \citealt{2016MNRAS.456.3194P}; \citealt{2017ApJ...838...29M}). 
In addition, the gravitational lensing by galaxy clusters has allowed us to probe even fainter galaxies and constrain the faint-end slope of the UV luminosity function (e.g., \citealt{2015ApJ...799...12I,2018ApJ...854...73I}; \citealt{2015ApJ...808..104O,2018ApJ...855..105O}, \citealt{2015ApJ...814...69A,2018MNRAS.479.5184A}; \citealt{2016ApJ...823L..40C}; \citealt{2016ApJ...832...56A}; \citealt{2016MNRAS.459.3812M}, \citealt{2017ApJ...843..129B}), although the impact of magnification uncertainties should be correctly considered (see \citealt{2017ApJ...843..129B}, \citealt{2018MNRAS.479.5184A}).

Investigating the bright-end of the luminosity function is also important. 
Previously the luminosity function is thought to follow the Schechter function \citep{1976ApJ...203..297S}, which is derived from the shape of the halo mass function \citep{1974ApJ...187..425P} with several modifications.
The Schechter function has an exponential cutoff at the bright end, which is possibly attributed to several different mechanisms such as heating from an active galactic nucleus (AGN; e.g., \citealt{2004MNRAS.347.1093B}; \citealt{2004ApJ...608...62S}; \citealt{2004ApJ...600..580G}; \citealt{2006MNRAS.365...11C}; \citealt{2006MNRAS.370..645B}), inefficiency of gas cooling in massive dark matter haloes due to virial shock heating (e.g., \citealt{1977ApJ...215..483B}; \citealt{1977MNRAS.179..541R}; \citealt{1977ApJ...211..638S}; \citealt{2003ApJ...599...38B}), and dust obscuration which becomes substantial for the most luminous galaxies (e.g., \citealt{1996ApJ...457..645W}; \citealt{2000ApJ...544..218A}; \citealt{2005ApJ...619L..59M}, \citealt{2020MNRAS.493.2059B}).
However, recent studies based on wide area surveys have reported an overabundance of objects at the bright end of UV luminosity functions beyond the Schechter function \citep[bright end excess, e.g.,][]{2014MNRAS.440.2810B,2015MNRAS.452.1817B,2017MNRAS.466.3612B,2020MNRAS.493.2059B,2017ApJ...851...43S,2019ApJ...883...99S,2018PASJ...70S..10O,2018ApJ...863...63S,2018ApJ...867..150M,2020MNRAS.494.1771A,2020MNRAS.494.1894M,2021arXiv210613813F}.
These studies suggest that the bright-end ($\leq-23$ mag) of the luminosity function is contributed by faint quasars or AGNs, at least at $z\sim4-7$ \citep{2018PASJ...70S..10O,2018ApJ...863...63S,2020MNRAS.494.1771A}.
In addition, \citet{2018PASJ...70S..10O} calculate the galaxy UV luminosity function that is estimated by the subtraction of the AGN contribution, and report that the galaxy luminosity function still shows a bright end excess beyond the Schechter function at $z\sim4-7$.
They claim that this bright end excess implies inefficient mass quenching (e.g., the AGN feedback, virial shock heating) in these high redshift galaxies, or significant number of merging or gravitationally-lensed galaxies at the bright end \citep[see also; e.g.,][]{2014MNRAS.440.2810B,2015ApJ...803...34B}.

Together with studying luminosity functions, the clustering analysis with the angular correlation function is important to understand the connection between galaxies and their dark matter halos.
The galaxy-dark matter halo connection is investigated with the weak lensing analysis \citep[e.g.,][]{2012ApJ...744..159L,2015ApJ...806....2M,2015MNRAS.449.1352C}, the abundance matching/empirical model \citep[e.g.,][]{2013ApJ...770...57B,2019MNRAS.488.3143B,2013MNRAS.428.3121M,2018MNRAS.477.1822M,2015ApJ...814...95F}, and the clustering analysis \citep[e.g.,][]{2016ApJ...821..123H,2018PASJ...70S..11H,2017ApJ...841....8I,2020ApJ...904..128I,2018ApJ...853...69C,2018MNRAS.481.4885Q,2020MNRAS.494..804C}.
Since the weak lensing analysis cannot be applied at $z\gtrsim2$ due to the limited number of the background galaxies and their lower image quality with current observational datasets (but see also \citealt{2021arXiv210315862M}), the clustering analysis is a crucial tool for estimating the dark matter halo mass of high redshift galaxies.
Many studies have investigated the dark matter halos of high redshift galaxies as a function of their redshifts and UV luminosities \citep[e.g.,][]{2004ApJ...611..685O,2005ApJ...635L.117O,2009A&A...498..725H,2011ApJ...737...92S,2013ApJ...774...28B,2014ApJ...793...17B,2016ApJ...821..123H,2018ApJ...859...84H,2018MNRAS.477.3760H}.
These studies reveal that the more UV luminous galaxies reside in more massive halos.

Recently, \citet{2018ApJ...859...84H} have identified a tight relation between the ratio of the SFR to the dark matter accretion rate, $SFR/\dot{M}_\m{h}$, and the halo mass, $M_\m{h}$, over $z\sim4-7$, suggesting the existence of a fundamental relation between the galaxy growth and its dark matter halo assembly.
This redshift-independent relation indicates that the star formation efficiency does not significantly change at $z\sim4-7$, and star formation activities are regulated by the dark matter mass assembly.
Several studies show that this redshift-independent $SFR/\dot{M}_\m{h}-M_\m{h}$ relation can reproduce the UV luminosity functions  at $z\gtrsim4$ \citep[e.g.,][]{2015ApJ...813...21M,2018PASJ...70S..11H,2018ApJ...868...92T,2021arXiv210207775B} and the trend of the redshift evolution of the cosmic SFR density \citep[e.g.,][]{2015ApJ...813...21M,2018PASJ...70S..11H,2018ApJ...868...92T,2018ApJ...855..105O}, a.k.a the cosmic star formation history or the Lilly-Madau plot (e.g., \citealt{1996ApJ...460L...1L}; \citealt{1996MNRAS.283.1388M}; \citealt{1997AJ....113....1S}; \citealt{1999ApJ...519....1S}; \citealt{2015ApJ...803...34B}; see review by \citealt{2014ARA&A..52..415M}).
As discussed in \citet{2018PASJ...70S..11H}, this suggests a simple picture that the evolution of the cosmic SFR density is primarily driven by the steep increase of the number density of halos (and galaxies) due to the structure formation to $z\sim4-2$, and the decrease of the accretion rate from $z\sim2$ to $z\sim0$ due to the cosmic expansion.
However, the $SFR/\dot{M}_\m{h}-M_\m{h}$ relation is only constrained at $z\sim4-7$, and it is not known whether the relation evolves from $z\sim4$ to $z\sim1-3$ or not, where the cosmic SFR density reaches its peak.

\begin{deluxetable*}{ccccccccc}
\tablecaption{HSC-SSP Data Used for the $z\sim4-7$ Selection}
\tablehead{\colhead{Field} & \colhead{R.A.} & \colhead{Decl.} & \colhead{Area} & \colhead{$g$} & \colhead{$r$} & \colhead{$i$} & \colhead{$z$} & \colhead{$y$}  \\
\colhead{(1)}& \colhead{(2)}& \colhead{(3)}& \colhead{(4)} &  \colhead{(5)} & \colhead{(6)} & \colhead{(7)} & \colhead{(8)} & \colhead{(9)}}
\startdata
\multicolumn{9}{c}{UltraDeep (UD)}\\
UD-SXDS & 02:18:23.26 & $-$04:52:51.40 & 1.3 & 27.15 & 26.68 & 26.57 & 26.09 & 25.27\\
UD-COSMOS & 10:00:23.43 & $+$02:12:39.11 & 1.3 & 26.85 & 26.58 & 26.75 & 26.56 & 25.90\\
\hline
\multicolumn{9}{c}{Deep (D)}\\
D-XMM-LSS & 02:25:26.62 & $-$04:20:10.79 & 2.2 & 26.83 & 26.18 & 25.87 & 25.73 & 24.47\\
D-COSMOS & 10:00:33.67 & $+$02:10:07.02 & 4.9 & 26.61 & 26.37 & 26.32 & 26.02 & 25.15\\
D-ELAIS-N1 & 16:10:56.49 & $+$54:58:13.69 & 5.4 & 26.71 & 26.34 & 26.13 & 25.73 & 24.81\\
D-DEEP2-3 & 23:28:17.72 & $-$00:15:57.55 & 5.1 & 26.78 & 26.38 & 25.98 & 25.73 & 24.98\\
\hline
\multicolumn{9}{c}{Wide (W)}\\
W-W02 & 02:15:36.65 & $-$04:03:28.04 & 33.3 & 26.43 & 25.94 & 25.69 & 25.03 & 24.22\\
W-W03 & 09:23:02.23 & $+$00:36:45.81 & 66.2 & 26.20 & 25.84 & 25.76 & 25.17 & 24.36\\
W-W04 & 13:21:04.83 & $-$00:12:07.68 & 72.2 & 26.41 & 25.99 & 25.86 & 25.17 & 24.33\\
W-W05 & 21:26:59.14 & $+$01:41:02.30 & 86.6 & 26.18 & 25.81 & 25.61 & 25.01 & 24.25\\
W-W06 & 15:38:28.05 & $+$43:18:51.64 & 28.4 & 26.44 & 26.05 & 25.78 & 25.09 & 24.15\\
W-W07 & 14:17:03.01 & $+$52:30:29.70 & 0.9 & 26.60 & 25.88 & 25.79 & 24.98 & 24.00\\
\hline
Total Area & & & 307.9 & & & 
\enddata
\tablecomments{(1) Field name.
(2) Right ascension.
(3) Declination.
(4) Effective area in $\m{deg^2}$.
(5)-(9) $5\sigma$ limiting magnitude measured in $1.\carcsec5$ diameter circular apertures in the $g$-, $r$-, $i$-, $z$-, and $y$-bands.
\redc{These limiting magnitudes are not corrected to total.}
}
\label{tab_limitmag}
\end{deluxetable*}

In this work, we present new measurements of the rest-frame UV luminosity functions at $z\sim4-7$ and clustering at $z\sim2-6$ based on wide and deep optical images obtained in the Hyper-Suprime Cam Subaru Strategic Program (HSC-SSP) survey (\citealt{2018PASJ...70S...4A}, see also \citealt{2012SPIE.8446E..0ZM,2018PASJ...70S...1M,2018PASJ...70S...2K,2018PASJ...70S...3F}) and the CFHT large area U-band deep survey (CLAUDS; \citealt{2019MNRAS.489.5202S}).
This paper is one in a series of papers from twin programs dedicated to scientific results on high redshift galaxies based on the HSC-SSP survey data.
One program is our luminous Lyman break galaxy (LBG) or dropout galaxy studies, named Great Optically Luminous Dropout Research Using Subaru HSC (GOLDRUSH; \citealt{2018PASJ...70S..10O}, \citealt{2018PASJ...70S..11H}, \citealt{2018PASJ...70S..12T}).
The other program is high redshift Ly$\alpha$ emitter (LAE) studies using HSC narrowband filters, Systematic Identification of LAEs for Visible Exploration and Reionization Research Using Subaru HSC \citep[SILVERRUSH;][]{2018PASJ...70S..13O,2018PASJ...70S..14S,2018PASJ...70S..15S,2018PASJ...70S..16K,2018ApJ...859...84H,2019ApJ...883..142H,2018PASJ...70...55I,2019ApJ...879...28H,2019arXiv190600173K,2021ApJ...911...78O}.
Our new LBG catalogs are made public on our project webpage\footnote{http://cos.icrr.u-tokyo.ac.jp/rush.html} \redc{or zenodo.\footnote{https://doi.org/10.5281/zenodo.5512721}}

This paper is organized as follows.
We show the observational datasets in Section \ref{ss_data} and describe sample selections in Section \ref{ss_selection}.
The results of the UV luminosity functions and clustering analysis are presented in Sections \ref{ss_LF} and \ref{ss_clustering}, respectively.
We discuss our results in Section \ref{ss_discussion}, and summarize our findings in Section \ref{ss_summary}.
Throughout this paper we use the Planck cosmological parameter sets of the TT, TE, EE+lowP+lensing+ext result \citep{2016A&A...594A..13P}:
$\Omega_\m{m}=0.3089$, $\Omega_\Lambda=0.6911$, $\Omega_\m{b}=0.049$, $h=0.6774$, and $\sigma_8=0.8159$.
We define $r_\m{200}$ that is the radius in which the mean enclosed density is 200 times higher than the mean cosmic density.
To define the halo mass, we use $M_\m{200}$ that is the dark matter and baryon mass enclosed in $r_\m{200}$.
Note that this definition is the same as \citet{2016ApJ...821..123H} but different from the one in \citet{2018PASJ...70S..11H} who use the total dark matter mass without baryons.
We assume the \citet{1955ApJ...121..161S} initial mass function (IMF).
All magnitudes are in the AB system \citep{1983ApJ...266..713O}.

\section{Observational Datasets}\label{ss_data}

\subsection{Subaru/HSC Data}\label{ss_data_HSC}

\begin{deluxetable*}{cccc}
\setlength{\tabcolsep}{-1pt}
\vspace{-0.5cm}
\tablecaption{Selection Criteria for Our Catalog Construction}
\tablehead{\colhead{Parameter} & \colhead{Value} & \colhead{Band} & \colhead{Comment}}  \\
\startdata
\verb|isprimary| & True & --- & Object is a primary one with no deblended children \\
\verb|pixelflags_edge| & False & $grizy$ & Locate within images \\
\verb|pixelflags_saturatedcenter| & False & $grizy$ &  None of the central $3 \times 3$ pixels of an object is saturated \\
\verb|pixelflags_bad| & False & $grizy$ & None of the pixels in the footprint of an object is labelled as bad \\
\verb|mask_brightstar_any| & False & $grizy$ & None of the pixels in the footprint of an object is close to bright sources \\
\verb|mask_brightstar_ghost15| & -99 & $grizy$ & None of the pixels in the footprint of an object is close to the ghost masks \\
\verb|sdsscentroid_flag| & False & $ri$ for $g$-drop & Object centroid measurement has no problem \\
 & False & $iz$ for $r$-drop & --- \\
 & False & $zy$ for $i$-drop & --- \\
 & False & $y$ for $z$-drop &  --- \\
\verb|cmodel_flag| & False & $gri$ for $g$-drop & Cmodel flux measurement has no problem \\
& False & $riz$ for $r$-drop & --- \\
& False & $izy$ for $i$-drop &  --- \\
& False & $zy$ for $z$-drop &  --- \\
\verb|merge_peak| & True & $ri$ for $g$-drop & Detected in $r$ and $i$. \\
& False/True & $g$/$iz$ for $r$-drop &  Undetected in $g$ and detected in $r$ and $i$. \\
& False/True & $gr$/$zy$ for $i$-drop & Undetected in $g$ and $r$, and detected in $z$ and $y$.  \\
& False/True & $gri$/$y$ for $z$-drop & Undetected in $g$, $r$ and $i$, and detected in $y$.  \\
\verb|blendedness_abs_flux| & $<0.2$ & $ri$ for $g$-drop & The target photometry is not significantly affected by neighbors.\\
& $<0.2$ & $iz$ for $r$-drop & --- \\
& $<0.2$ & $zy$ for $i$-drop & --- \\
& $<0.2$ & $y$ for $z$-drop & --- \\
\verb|inputcount_value| & $\geq3$/$\geq5$ & $gr$/$izy$ & The number of exposures is equal to or larger than 3/5 in $gr$/$izy$.
\enddata
\label{tab_flag}
\end{deluxetable*}

We use the internal S18A data release product taken in the HSC-SSP survey \citep{2018PASJ...70S...4A} from 2014 March to 2018 January, which is basically identical to the version of the Public Data Release 2 \citep{2019PASJ...71..114A}.\footnote{https://hsc.mtk.nao.ac.jp/ssp/data-release}
The HSC-SSP survey obtains deep optical imaging data with the five broadband filters, $g$, $r$, $i$, $z$, and $y$ \citep{2018PASJ...70...66K}, which are useful to select $z\sim4-7$ galaxies with the dropout selection technique.
The HSC-SSP survey has three layers, the UltraDeep, Deep, and Wide, with different combinations of area and depth.
Total effective survey areas of the data we use are $\sim3$, $\sim18$, and $\sim288$ $\m{deg^2}$ for the UltraDeep, Deep, and Wide layers, respectively (Table \ref{tab_limitmag}).
Here we define the effective survey area as area where the number of visits in $g$, $r$, $i$, $z$, and $y$-bands are equal to or larger than threshold values after masking interpolated, saturated, or bad pixels, cosmic rays, and bright source halos \citep{2018PASJ...70S...7C}.
The applied flags and threshold values are summarized in Table \ref{tab_flag}.
In addition to these flags, we mask some regions that are affected by bright source halos or ghosts of bright sources but not flagged.

The HSC data are reduced by the HSC-SSP collaboration with {\tt hscPipe} \citep{2018PASJ...70S...5B} that is the HSC data reduction pipeline based on the Large Synoptic Survey Telescope (LSST) pipeline \citep{2008arXiv0805.2366I,2010SPIE.7740E..15A}.
{\tt hscPipe} performs all the standard procedures including bias subtraction, flat fielding with dome flats, stacking, astrometric and photometric calibrations, flagging, source detections and measurements, and construction of a multiband photometric catalog. 
The astrometric and photometric calibration are based on the data of Panoramic Survey Telescope and Rapid Response System (Pan-STARRS) 1 imaging survey \citep{2013ApJS..205...20M,2012ApJ...756..158S,2012ApJ...750...99T}.
PSFs are calculated in {\tt hscPipe}, and typical full width at half maximum (FWHM) of the PSFs are $0.\carcsec6-0.\carcsec9$.

We use forced photometry, which allows us to measure fluxes in multiple bands with a consistent aperture defined in a reference band.
The reference band is $i$ by default and is switched to $z$ ($y$) for sources with no detection in the $i$ ($z$) and bluer bands. 
In previous studies based on the S16A data release \citep[e.g.,][]{2018PASJ...70S..10O,2018PASJ...70S..11H,2018PASJ...70S..12T}, the CModel magnitude \citep{2004AJ....128..502A} was used to measure total fluxes and colors of sources.
However, we have found that some objects in the S18A data release have unnaturally bright CModel magnitudes compared to their aperture magnitudes, as also reported in \citet{2020PASJ...72...86H}.
Thus in this paper, we instead use magnitudes measured with a fixed $2\arcsec$-diameter aperture after aperture correction, {\tt convolvedflux\_0\_20}, for measuring total fluxes and colors of sources.
\redc{The aperture correction factor is calculated in each band assuming the point spread function.}
Among several magnitudes with different aperture sizes and corrections calculated with {\tt hscPipe}, we have found that this magnitude provides the best match to the CModel magnitude in the S16A data release \redc{with the typical difference less than 5\%}.
Limiting magnitudes and source detections are evaluated with magnitudes measured in a $1.\carcsec5$-diameter aperture, $m_\m{aper}$.
All the magnitudes are corrected for Galactic extinction \citep{1998ApJ...500..525S}.
We measure the $5\sigma$ limiting magnitudes which are defined as the $5\sigma$ levels of sky noise in a $1.\carcsec5$-diameter aperture. 
The sky noise is calculated from fluxes in sky apertures which are randomly placed on the images in the reduction process.
The limiting magnitudes measured in $g$, $r$, $i$, $z$, and $y$-bands are presented in Table \ref{tab_limitmag}.

We select isolated or cleanly deblended sources from detected source catalogs available on the database that are provided by the HSC-SSP survey team. 
We require that none of the central $3 \times 3$ pixels are saturated, and there are no bad pixels in their footprint, like the definition of the effective area described above.
We also require that there are no problems in measuring the CModel fluxes in the $gri$ images for $g$-dropouts, in the $riz$ images for $r$-dropouts, in the $izy$ images for $i$-dropouts, and in the $zy$ images for $z$-dropouts, except for the problem of unnaturally bright magnitudes described above.
In addition, we remove sources if there are any problems in measuring their centroid positions in the $ri$ images for $g$-dropouts, in the $iz$ images for $r$-dropouts, in the $zy$ images for $i$-dropouts, 
and in the $y$ image for $z$-dropouts.  
To remove severely blended sources, we apply a blendedness parameter threshold of $b<0.2$ in the $ri$, $iz$, and $zy$-bands at $z\sim4$, $5$, and $6$, respectively
These selection criteria are summarized in Table \ref{tab_flag}.

\subsection{CLAUDS Data}\label{ss_data_CLAUDS}

In the UltraDeep and Deep layers of the HSC-SSP survey, deep $U$-band images are taken in CLAUDS \citep[][]{2019MNRAS.489.5202S}.
These $U$-band images are useful to select $z\sim2-3$ galaxies by using the $U$-dropout or BX/BM selection techniques \citep[e.g.,][]{2003ApJ...592..728S,2004ApJ...604..534S,2004ApJ...607..226A}.
The $U$-band images are obtained with two filters, $u$ and $u^*$, because CFHT updated the MegaCam filter set by replacing the old $u^*$-filter by the new $u$-filter during the CLAUDS observing campaign (2014B to 2016B).
Specifically, we have deep $u^*$-band images in the UD-SXDS, UD-COSMOS, and D-XMM-LSS fields, and deep $u$-band images in D-COSMOS, D-DEEP2-3 and D-ELAIS-N1 fields. 
\citet{2019MNRAS.489.5202S} describe the data reduction and procedures for making combined CLAUDS+HSC-SSP catalogs in detail.
The $5\sigma$ depths of the $u^*$ and $u$-band images are typically 27.9 and 27.5 mag, respectively, sufficiently deep to select $z\sim2-3$ galaxies.

\subsection{Spectroscopic Data}\label{ss_spec}

We carried out spectroscopic follow-up observations for sources in our dropout catalogs at $z\sim4-7$ with DEep Imaging Multi-Object Spectrograph (DEIMOS; \citealt{2003SPIE.4841.1657F}) on the Keck Telescope on 2018 August 11 (S18B-014, PI: Y. Ono), AAOmega+2dF (\citealt{2006SPIE.6269E..0GS}; \citealt{2002MNRAS.333..279L}) on the Anglo-Australian Telescope (AAT) from December 31 2018 to January 3 2019 (A/2018B/03, PI: Y. Ono), and the Faint Object Camera and Spectrograph (FOCAS: \citealt{2002PASJ...54..819K}) on the Subaru Telescope on 2019 May 13 (S19A-006, PI: Y. Ono).

In the DEIMOS observations, we used the 600ZD grating with the GG455 filter.
The spectroscopic observations were made in multi-object slit mode.
We used a total of two masks.
Slit widths were 0.\carcsec8, and the integration time was 3600-6000 seconds per each mask.
The DEIMOS spectra were reduced with the {\tt spec2d} IDL pipeline developed by the DEEP2 Redshift Survey Team (\citealt{2012MNRAS.419.3018C}; \citealt{2013ApJS..208....5N}).
Wavelength calibrations was conducted by using the arc lamp emission lines.
The spectral resolutions in an FWHM based on the widths of night-sky emission lines were $\sim3.7\ \m{\AA}$.
Flux calibration was achieved with data of a standard star G191B2B.
The details of the DEIMOS observations will be presented in Ono et al. in prep.

In the AAOmega+2dF observations, we used the X5700 dichroic beam splitter, the 580V grating with the central wavelength at 4821 {\AA} in the blue channel, and the 385R grating with the central wavelength at 7251 {\AA} in the red channel.
This configuration covered a wavelength range of $3800-8800 \m{\AA}$ with a resolution of $R\sim1400$. 
We used a total of four masks, two covering the UD-COSMOS field and two covering  the UD-SXDS field.
The integration time is 1800-7800 seconds per each mask, although weather conditions were not excellent.
The spectra are reduced in the standard manner by using the OzDES pipeline \citep{2015MNRAS.452.3047Y,2017MNRAS.472..273C,2020MNRAS.496...19L}.

In the FOCAS observations, we used the VPH900 grism with the SO58 order-cut filter.
The spectroscopic observations were made in the multi-object slit mode.
We used a total of two masks.
Slit widths were 0.\carcsec8, and the integration time was 7200 seconds per each mask. 
The FOCAS data were reduced with the {\tt focasred} pipeline.
Wavelength calibrations was conducted by using night-sky emission lines.
The spectral resolution in an FWHM of FOCAS VPH900 based on the night-sky lines was $\sim5.7\ \m{\AA}$.
Flux calibration was performed with data of a standard star BD+28d4211.
The details of the FOCAS observations will be presented in Ono et al. in prep.

\redc{A total of 55 dropout candidates were targeted in these observations.
Target priorities were determined by apparent magnitudes, and brighter sources were assigned higher priorities.
The apparent magnitude range of the targets was $19-25$ mag and most of them were $21-24$ mag.}

\begin{figure*}
\centering
\begin{minipage}{0.325\hsize}
\begin{center}
\includegraphics[width=0.99\hsize, bb=7 18 325 348]{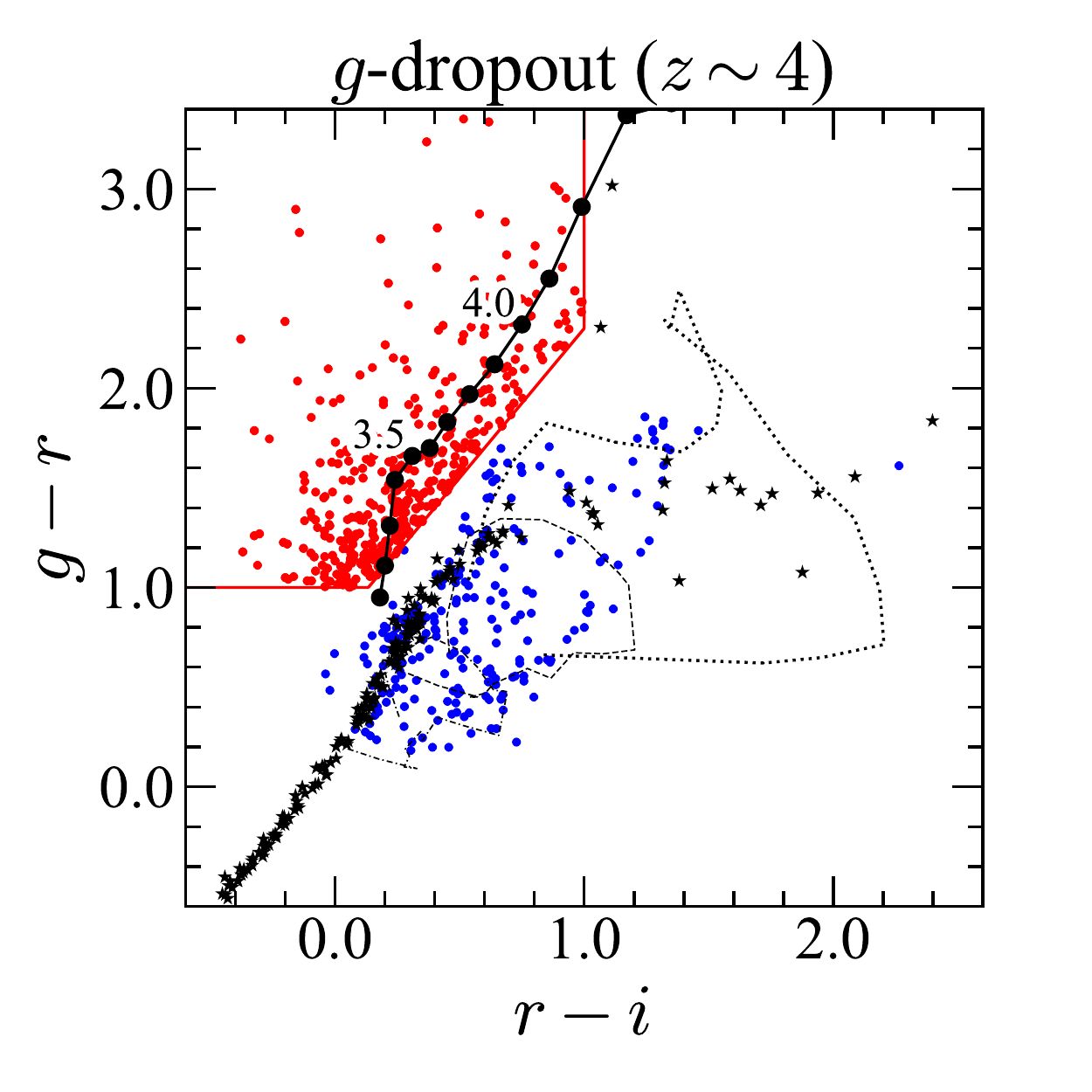}
\end{center}
\end{minipage}
\begin{minipage}{0.325\hsize}
\begin{center}
\includegraphics[width=0.99\hsize, bb=7 18 325 348]{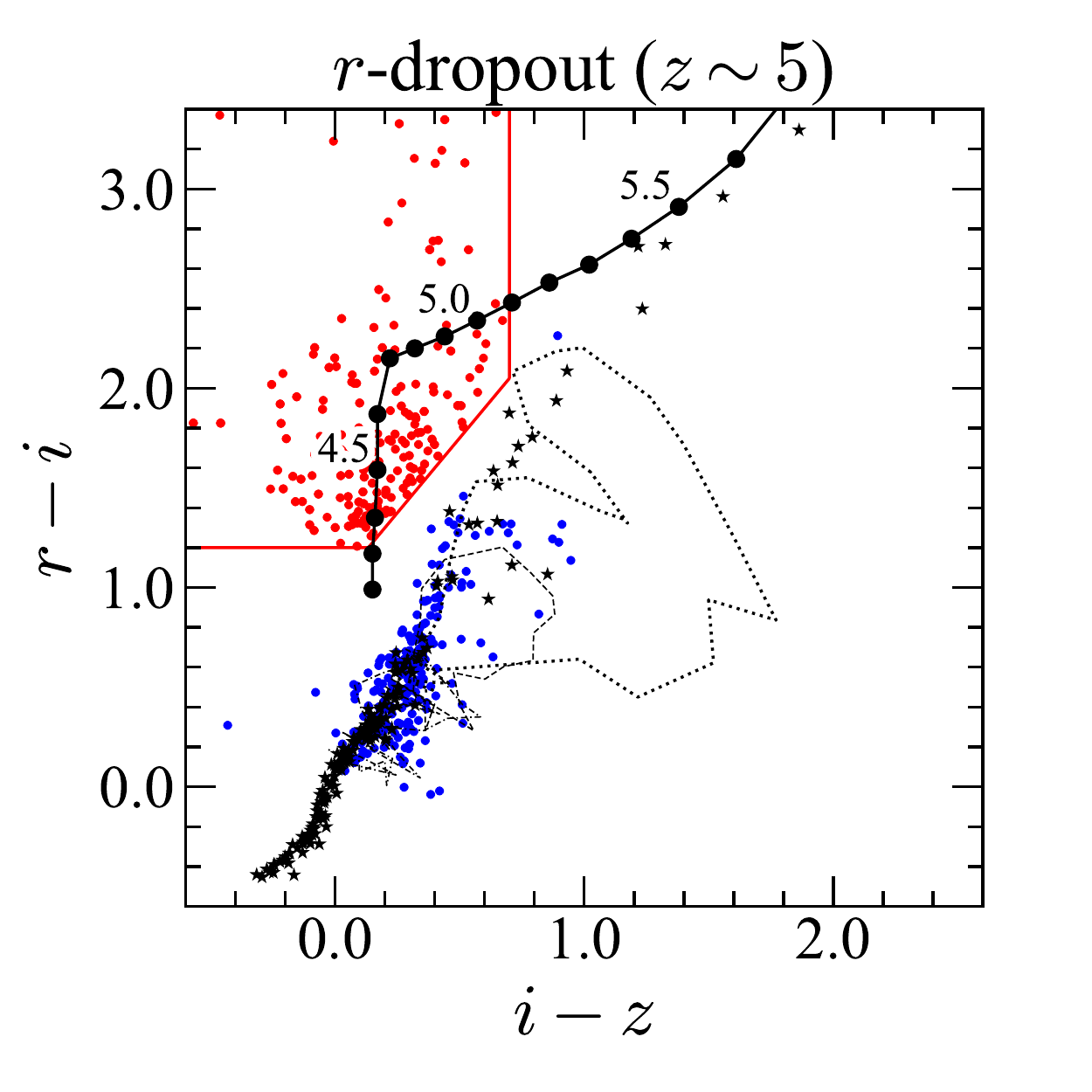}
\end{center}
\end{minipage}
\begin{minipage}{0.325\hsize}
\begin{center}
\includegraphics[width=0.99\hsize, bb=7 18 325 348]{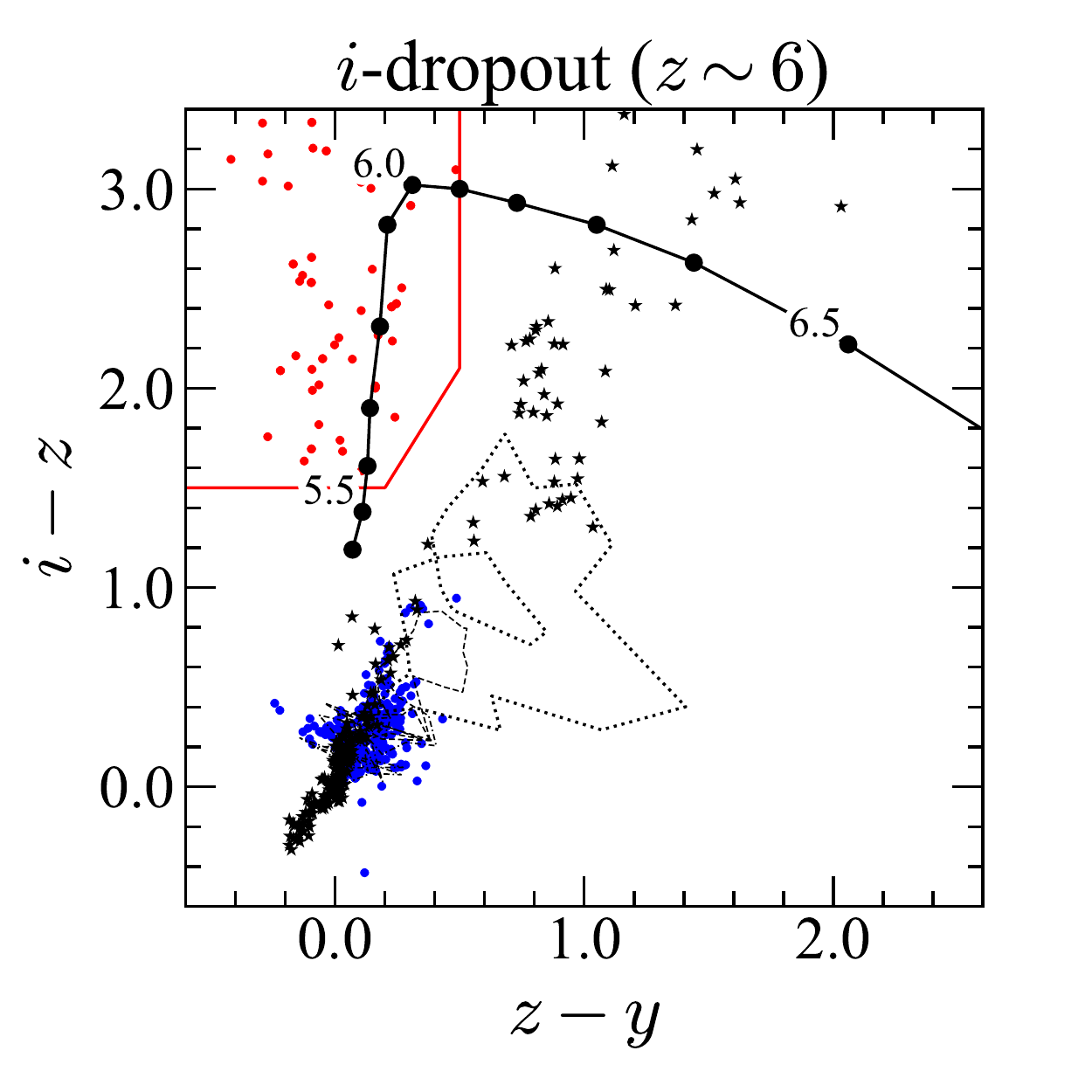}
\end{center}
\end{minipage}
\caption{Two-color diagrams to select dropout sources.
The left, middle, and right panels show two-color diagrams to select $g$-dropout ($z\sim4$), $r$-dropout ($z\sim5$), and $i$-dropout ($z\sim6$) sources, respectively.
The red lines indicate color criteria we use to select dropout sources (Equations (\ref{eq_z4_1})-(\ref{eq_z6_3})), and the red circles are spectroscopically-identified sources in the UltraDeep layers (for the left panel) and in all layers (for the middle and right panels).
The solid black lines are colors of star forming galaxies expected from the \citet{2003MNRAS.344.1000B} model as a function of redshift.
\redc{As model parameters, we adopt the \citet{1955ApJ...121..161S} IMF, an age of $70$ Myr after the initial star formation, metallicity of $Z / Z_\odot = 0.2$, and the \citet{2000ApJ...533..682C} dust extinction law with reddening of $E(B-V) = 0.16$ (see Section \ref{ss_spec_source}).}
The circles on the line show their redshifts with an interval of $\Delta z=0.1$.
The blue circles are $z=0-3$ sources spectroscopically identified in the UD-COSMOS region.
The dotted, dashed, and dot-dashed lines are, respectively, typical spectra of elliptical, Sbc, and irregular galaxies \citep{1980ApJS...43..393C} redshifted from $z=0$ to $z=2$ (for the left panel) and to $z=3$ (for the middle and right panels).
The black stars indicate Galactic stars taken from \citet{1983ApJS...52..121G} and L /T dwarfs from \citet{2004AJ....127.3553K}.
}
\label{fig_2color}
\end{figure*}

In addition to the observations described above, we include results of our observations with the Inamori Magellan Areal Camera and Spectrograph (IMACS; \citealt{2011PASP..123..288D}) on the Magellan I Baade telescope in 2007 -- 2011 (PI: M. Ouchi). 
The IMACS observations were carried out on 2007 November 11--14, 2008 November 29--30, December 1--2, December 18--20,  2009 October 11--13, 2010 February 8--9, July 9--10,  and 2011 January 3--4. 
In these observations, main targets were high redshift LAE candidates found in the deep Subaru Suprime-Cam narrowband images obtained in the SXDS \citep{2008ApJS..176..301O,2010ApJ...723..869O} and COSMOS fields \citep{2007ApJS..172..523M,2009ApJ...696..546S}.
High redshift dropout galaxy candidates selected from deep broadband images in these two fields \citep{2008ApJS..176....1F,2007ApJS..172...99C} were also observed as mask fillers. 
The data were reduced with the Carnegie Observatories System for MultiObject Spectroscopy ({\sc cosmos}) pipeline.\footnote{http://code.obs.carnegiescience.edu/cosmos}

\section{Sample Selection}\label{ss_selection}

\subsection{Source Selection at $z\sim4-7$}\label{ss_selection_z47}

From the source catalogs made in Section \ref{ss_data_HSC},  we construct $z\sim4-7$ dropout candidate catalogs based on the Lyman break color selection technique (e.g., \citealt{1996ApJ...462L..17S}; \citealt{2002ARA&A..40..579G}).
As shown in Figure \ref{fig_2color}, galaxy candidates can be selected based on their $gri$, $riz$, $izy$, and $zy$ colors at $z\sim4$, $5$, $6$, and $7$, respectively.

First, to identify secure sources, we select sources whose signal-to-noise (SN) ratios are higher than $5$ within $1.\carcsec5$-diameter apertures in the $i$-band for $g$-dropouts and in the $z$-band for $r$-dropouts.
\redc{For $i$-dropouts, we select sources with signal-to-noise ratios higher than $5$ and $4$ in the $z$ and $y$-bands, respectively, because the $y$-band images are relatively shallow.}
For the $z$-dropouts, we select sources with signal-to-noise ratios higher than $5$ in the $y$-band.
We then select dropout galaxy candidates by using their broadband spectral energy distribution (SED) colors. 
Like our previous studies \citep{2018PASJ...70S..10O,2018PASJ...70S..11H,2018PASJ...70S..12T}, we adopt the following color criteria:\\
$g$-dropouts ($z\sim4$)
\begin{eqnarray}
(g-r &>& 1.0) \land \label{eq_z4_1}\\
(r-i &<& 1.0) \land \label{eq_z4_2}\\
(g-r &>& 1.5(r-i)+0.8),\label{eq_z4_3}
\end{eqnarray}
$r$-dropouts ($z\sim5$)
\begin{eqnarray}
(r-i &>& 1.2) \land \label{eq_z5_1}\\
(i-z &<& 0.7) \land \label{eq_z5_2}\\
(r-i &>& 1.5(i-z)+1.0),\label{eq_z5_3}
\end{eqnarray}
$i$-dropouts ($z\sim6$)
\begin{eqnarray}
(i-z &>& 1.5) \land \label{eq_z6_1}\\
(z-y &<& 0.5) \land \label{eq_z6_2}\\
(i-z &>& 2.0(z-y)+1.1),\label{eq_z6_3}
\end{eqnarray}
$z$-dropouts ($z\sim7$)
\begin{eqnarray}
z-y &>& 1.6. \label{eq_z7_1}
\end{eqnarray}
\redc{As shown in Figure \ref{fig_2color}, these color criteria are set to avoid low redshift galaxies and stellar contaminants.
Although only the $z-y$ color is used in the $z$-dropout selection, we can efficiently select $z\sim7$ sources with this strict color criterion, as shown in previous studies similarly using the $z-y$ color \citep[e.g.,][]{2009ApJ...706.1136O}.}

To remove foreground interlopers, we exclude sources with continuum detections at $>2\sigma$ levels in the $g$-band for $r$-dropouts, in the $g$- or $r$-bands for $i$-dropouts, and in the $g$, $r$, or $i$-bands for $z$-dropouts, using the $1.\carcsec5$ diameter aperture magnitudes.
Since our $z$-dropout candidates are detected only in $y$-band images, we carefully check coadd and  single epoch observation images of the selected candidates to remove spurious sources and moving objects.

\begin{deluxetable*}{cccccccc}
\tablecaption{Number of Sources in Galaxy Samples Used in This Work.}
\tablehead{\colhead{Field} & \colhead{BM$^\dagger$} & \colhead{BX$^\dagger$} & \colhead{$U$-drop$^\dagger$} & \colhead{$g$-drop$^*$} & \colhead{$r$-drop$^*$} & \colhead{$i$-drop$^*$} & \colhead{$z$-drop$^*$}  \\
\colhead{}& \colhead{$z\sim1.7$}& \colhead{$z\sim2.2$}& \colhead{$z\sim3$} &  \colhead{$z\sim4$} & \colhead{$z\sim5$} & \colhead{$z\sim6$} & \colhead{$z\sim7$} }
\startdata
\multicolumn{8}{c}{UltraDeep (UD)}\\
UD-SXDS & $49184$ & $20292$ & $36769$ & $15282$ & $1517$ & $59$ & $8$\\
UD-COSMOS & $60726$ & $26139$ & $59838$ & $10067$ & $2760$ & $212$ & $27$\\
\hline
\multicolumn{8}{c}{Deep (D)}\\
D-XMM-LSS & $286685$ & $110755$ & $185555$ & $6919$ & $1237$ & $5$ & $1$\\
D-COSMOS & $192809$ & $81284$ & $160028$ & $32602$ & $6439$ & $114$ & $15$\\
D-ELAIS-N1 & $172950$ & $85458$ & $172735$ & $40815$ & $3947$ & $61$ & $9$\\
D-DEEP2-3 & $173450$ & $81541$ & $165561$ & $28725$ & $3808$ & $131$ & $13$\\
\hline
\multicolumn{8}{c}{Wide (W)}\\
W-W02 & \nodata & \nodata & \nodata & $150775$ & $8034$ & $158$ & $26$\\
W-W03 & \nodata & \nodata & \nodata & $357845$ & $29758$ & $549$ & $52$\\
W-W04 & \nodata & \nodata & \nodata & $550136$ & $35440$ & $410$ & $53$\\
W-W05 & \nodata & \nodata & \nodata & $473160$ & $35105$ & $757$ & $60$\\
W-W06 & \nodata & \nodata & \nodata & $164936$ & $11166$ & $110$ & $27$\\
W-W07 & \nodata & \nodata & \nodata & $4982$ & $148$ & $1$ & $1$\\
\hline
Total($z$) & $935804$ & $405469$ & $780486$ & $1836244$ & $139359$ & $2567$ & $292$\\
\hline
Total & \multicolumn{7}{c}{4100221}
\enddata
\tablenotetext{}{$^\dagger$Selected in C. Liu et al. in prep.}
\tablenotetext{}{$^*$Selected in This work.}
\label{tab_galnum}
\end{deluxetable*}

\begin{figure*}
\centering
\begin{minipage}{0.49\hsize}
\begin{center}
\includegraphics[width=0.49\hsize, bb=15 15 350 360]{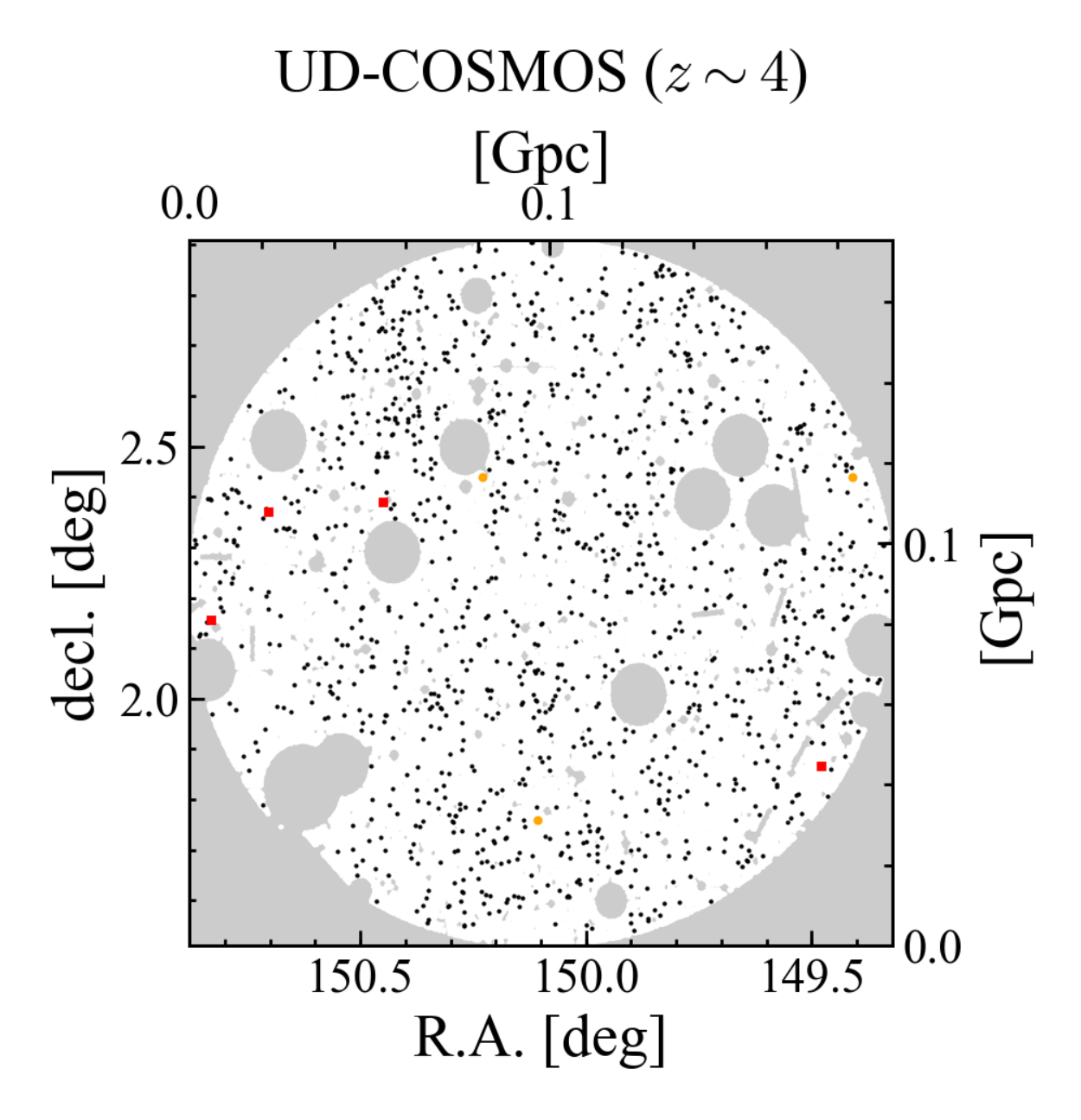}
\includegraphics[width=0.49\hsize, bb=15 0 345 320]{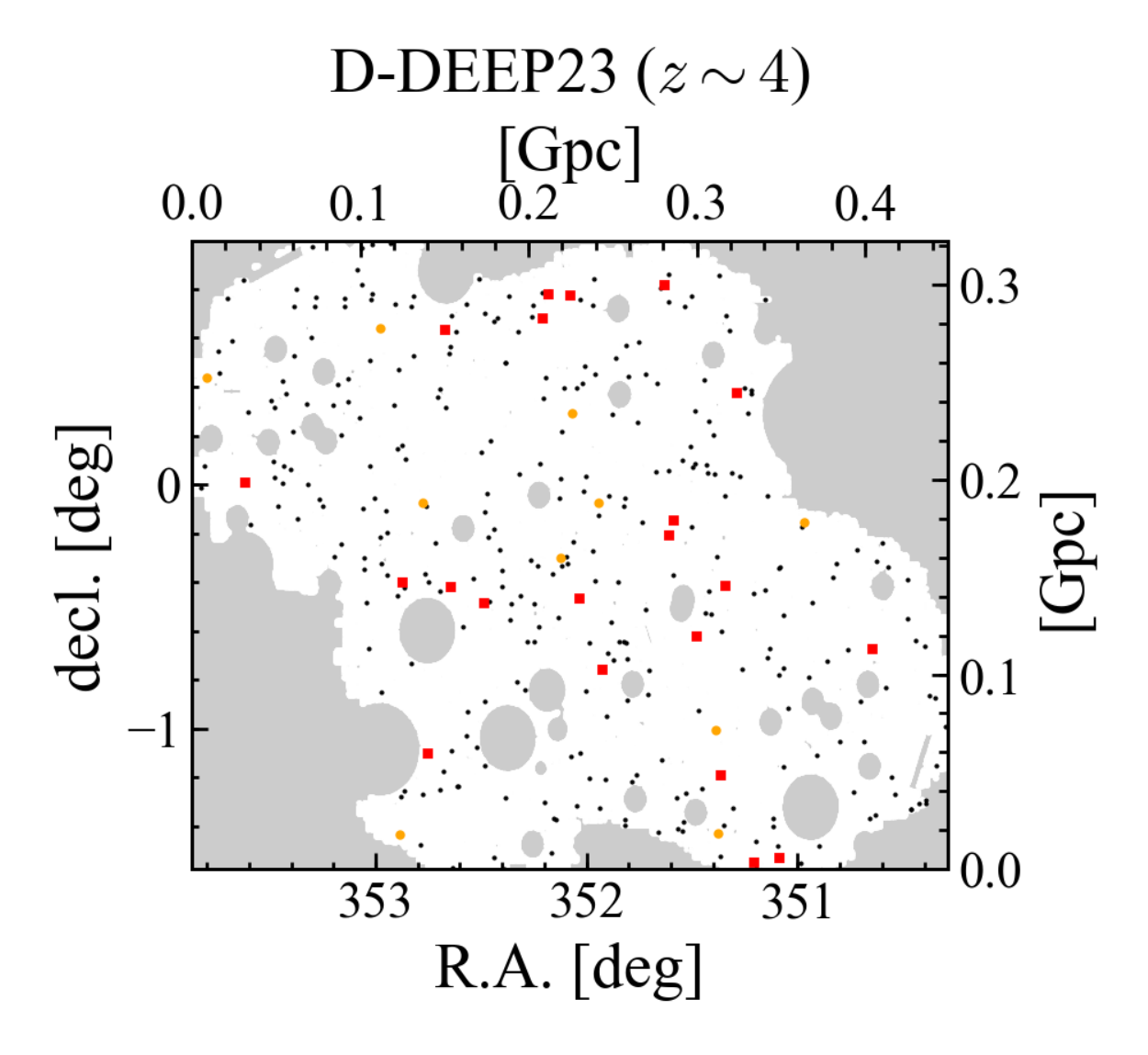}
\end{center}
\end{minipage}
\begin{minipage}{0.49\hsize}
\begin{center}
\includegraphics[width=0.49\hsize, bb=15 15 350 360]{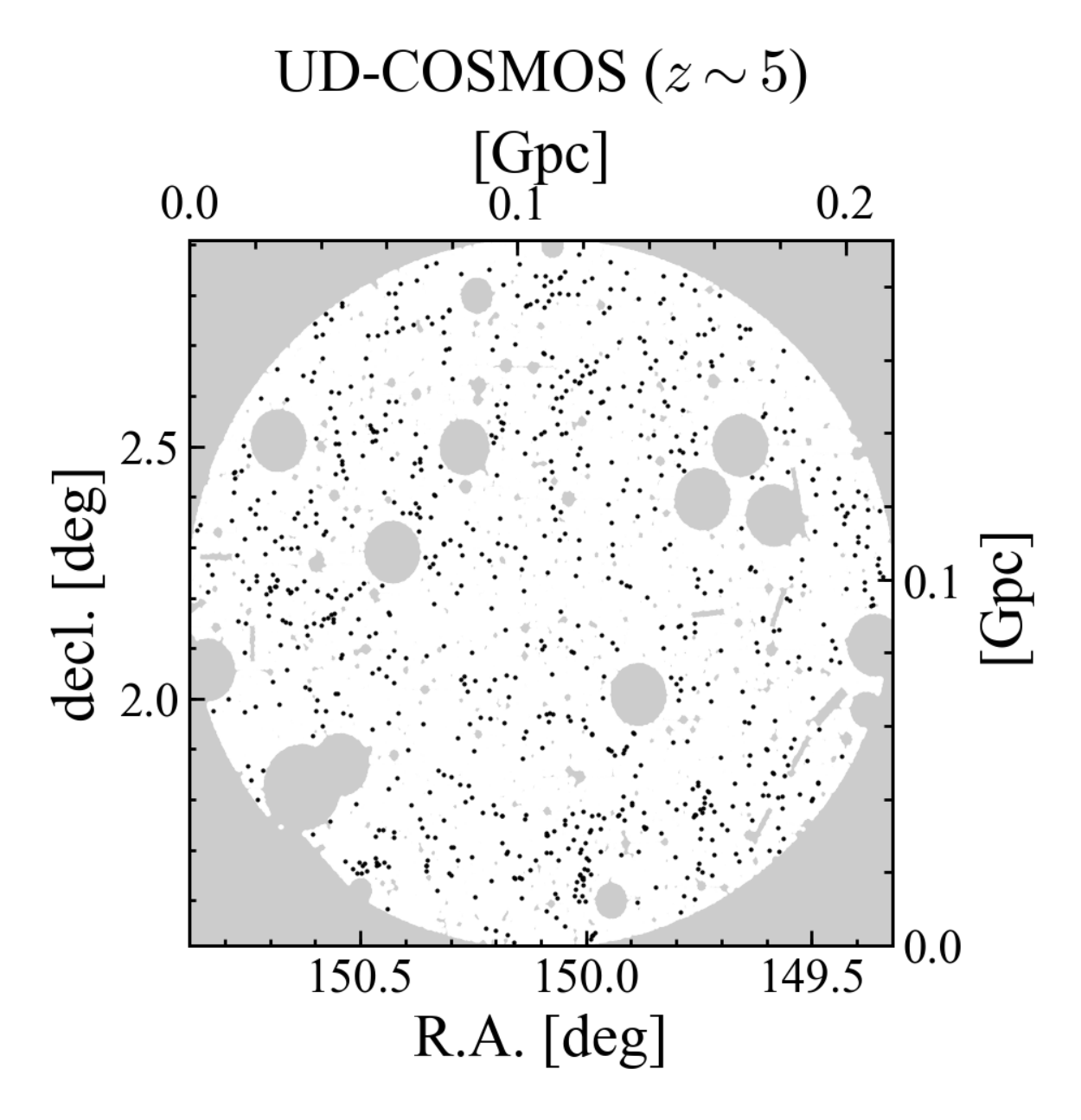}
\includegraphics[width=0.49\hsize, bb=15 0 345 320]{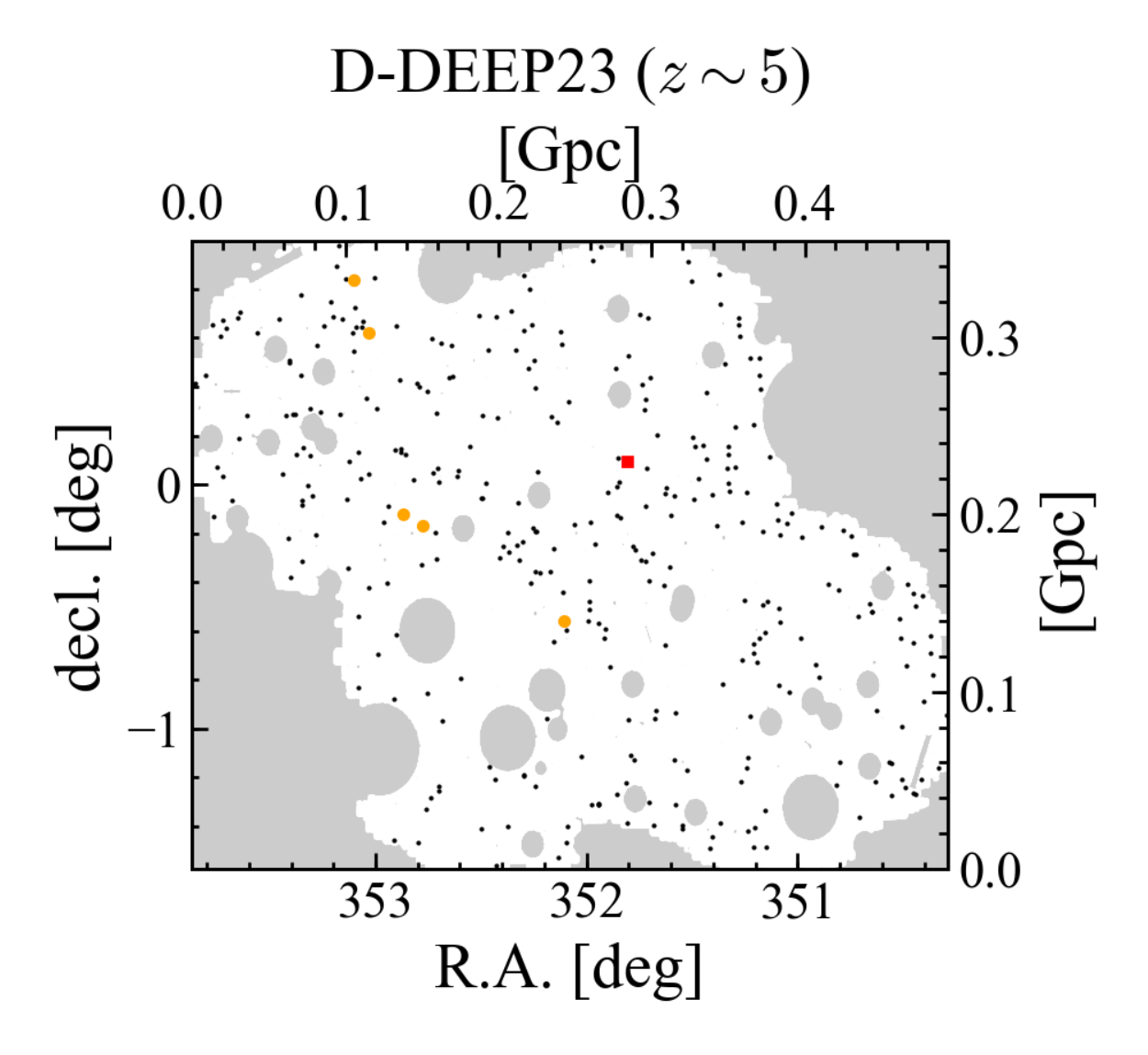}
\end{center}
\end{minipage}
\begin{minipage}{0.49\hsize}
\begin{center}
\includegraphics[width=0.99\hsize, bb=15 10 705 505]{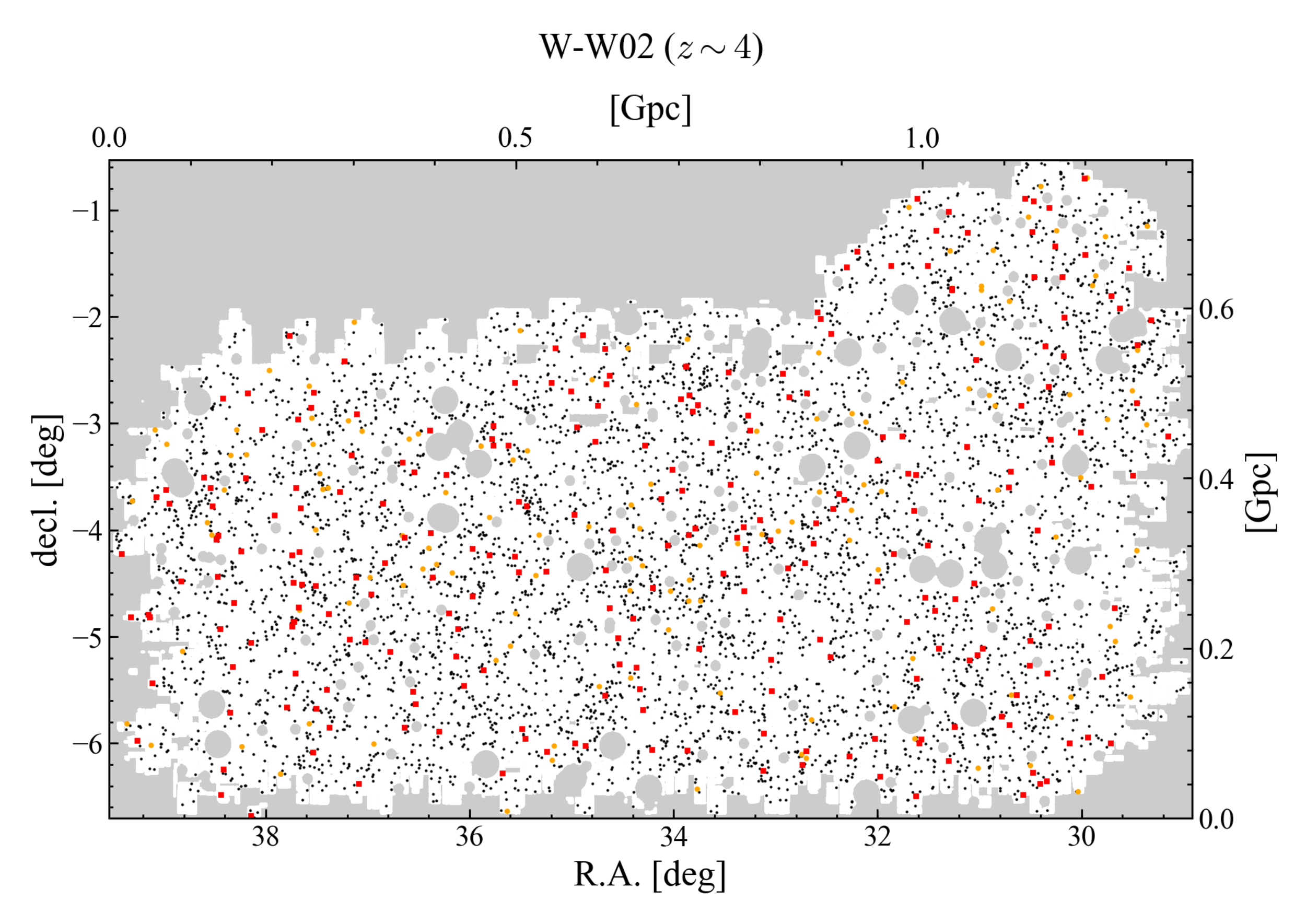}
\end{center}
\end{minipage}
\begin{minipage}{0.49\hsize}
\begin{center}
\includegraphics[width=0.99\hsize, bb=15 10 705 505]{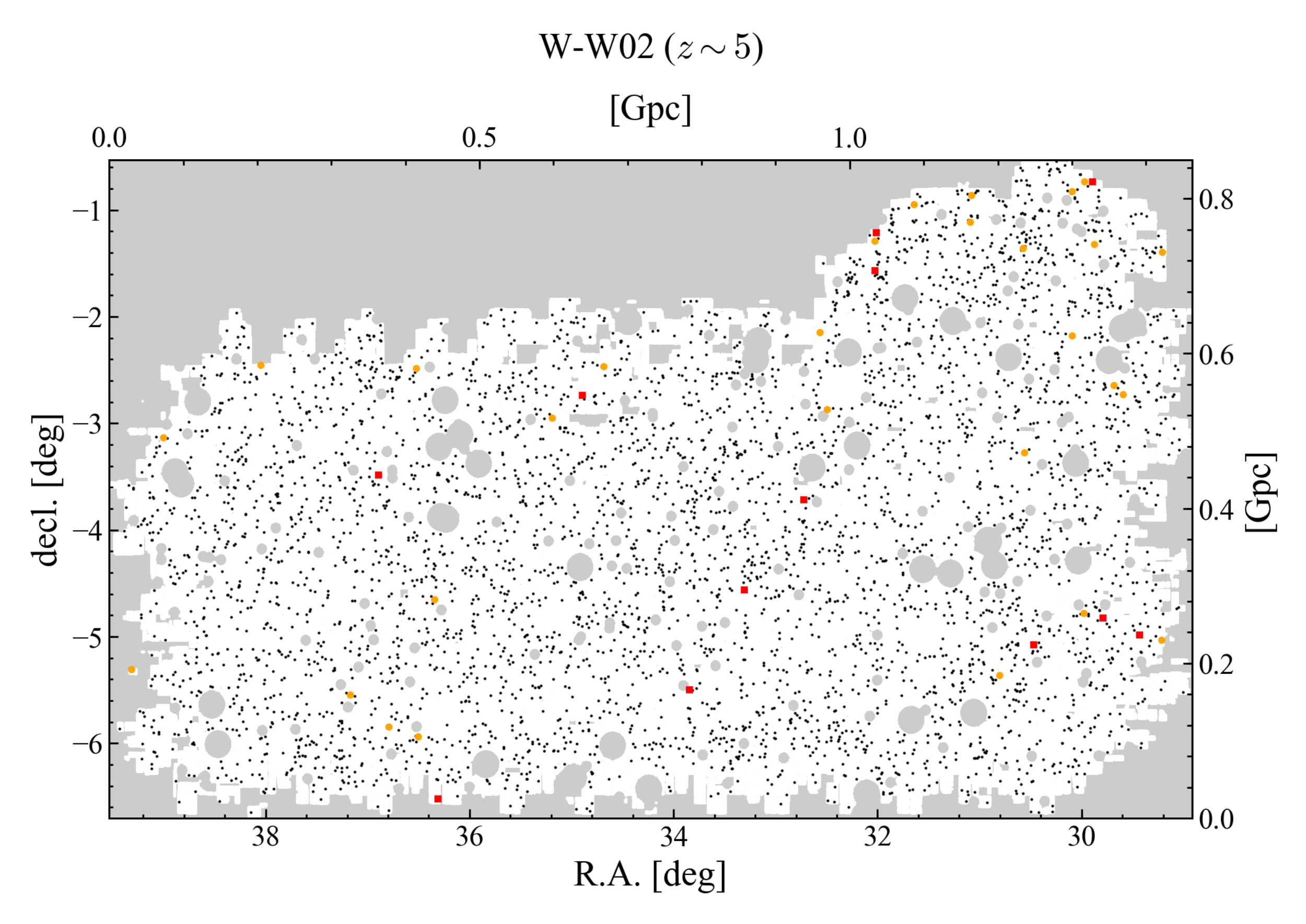}
\end{center}
\end{minipage}
\\
\vspace{0.2cm}
\begin{minipage}{0.49\hsize}
\begin{center}
\includegraphics[width=0.49\hsize, bb=15 15 350 360]{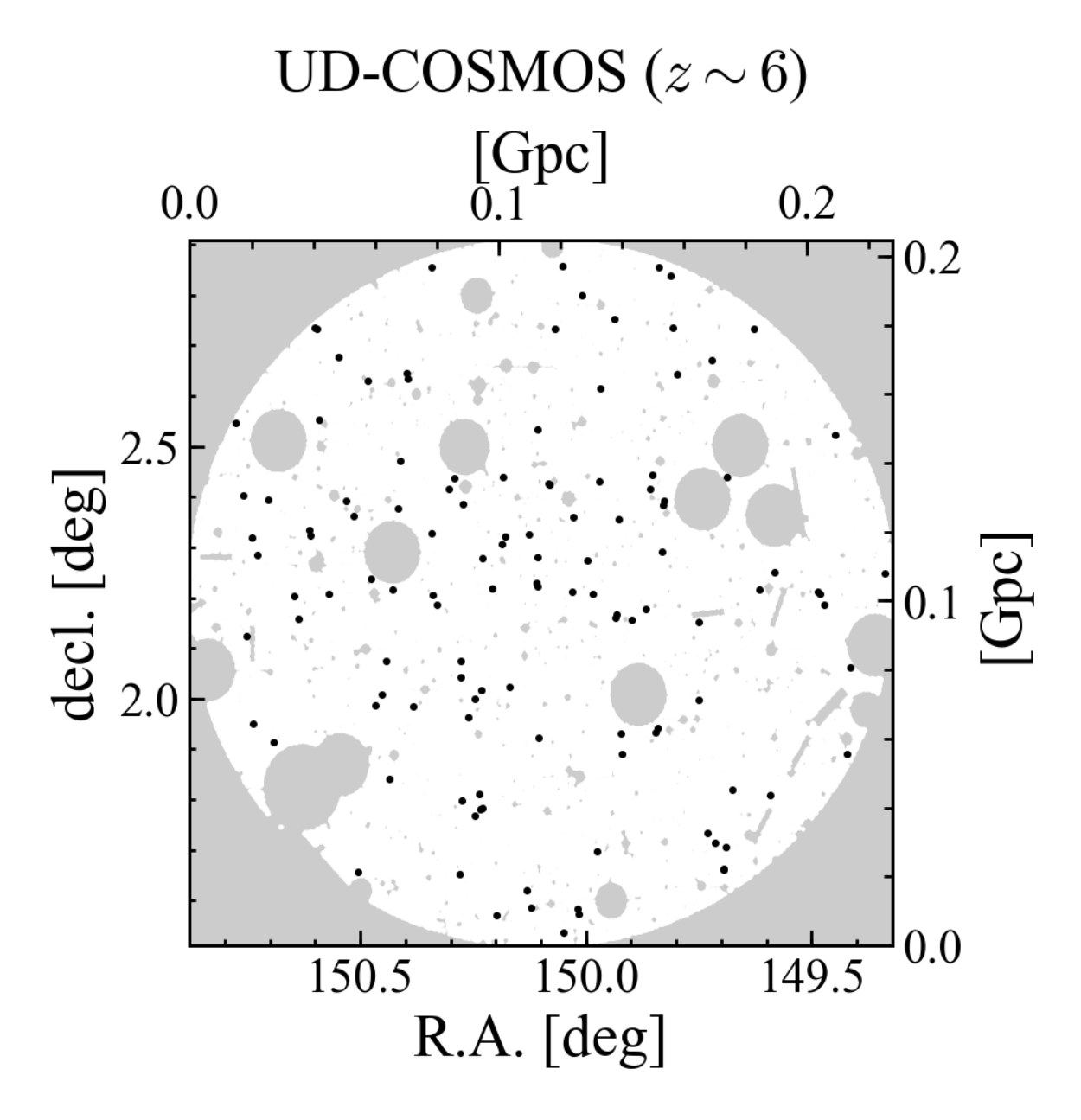}
\includegraphics[width=0.49\hsize, bb=15 0 345 320]{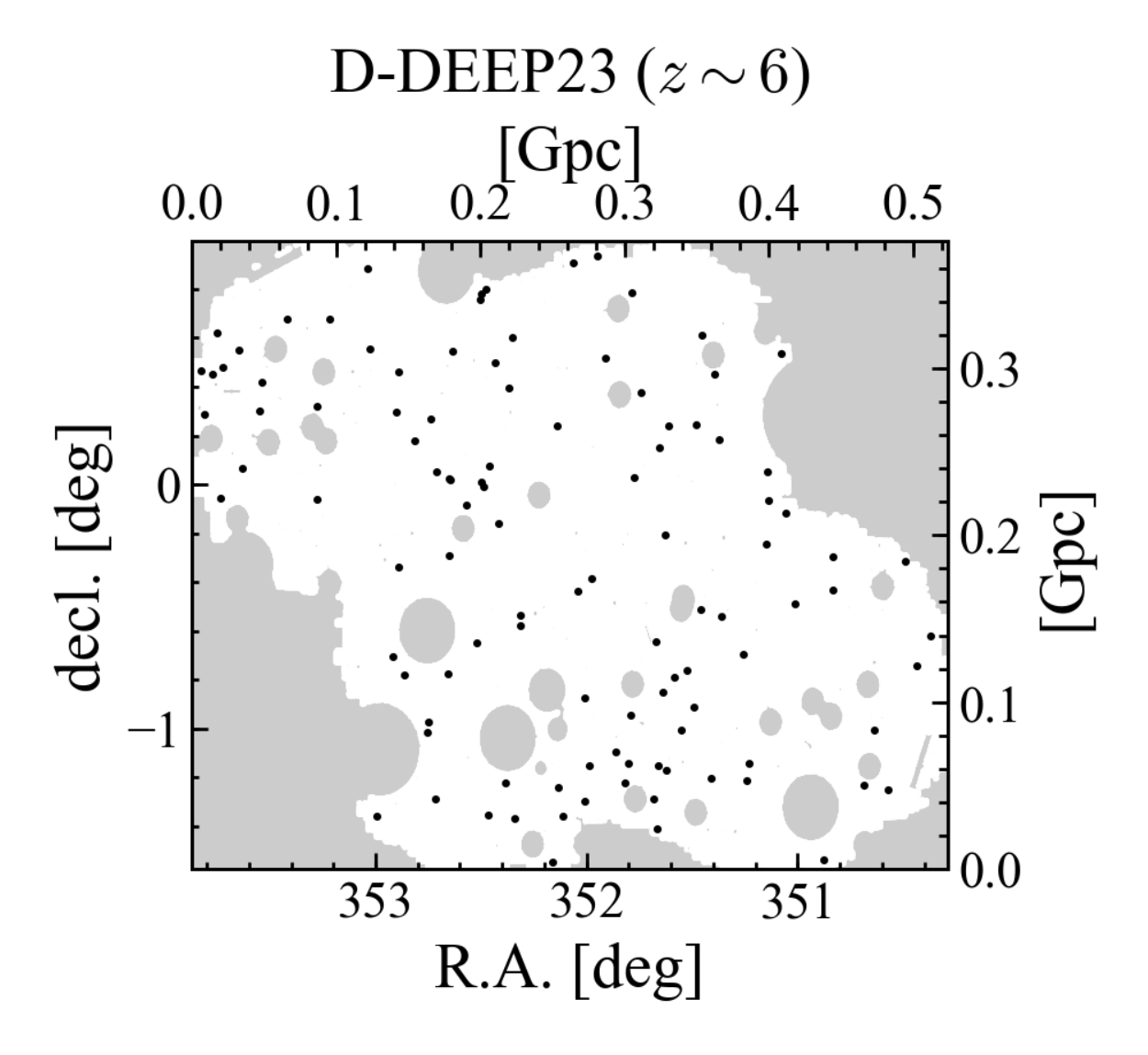}
\end{center}
\end{minipage}
\begin{minipage}{0.49\hsize}
\begin{center}
\includegraphics[width=0.49\hsize, bb=15 15 350 360]{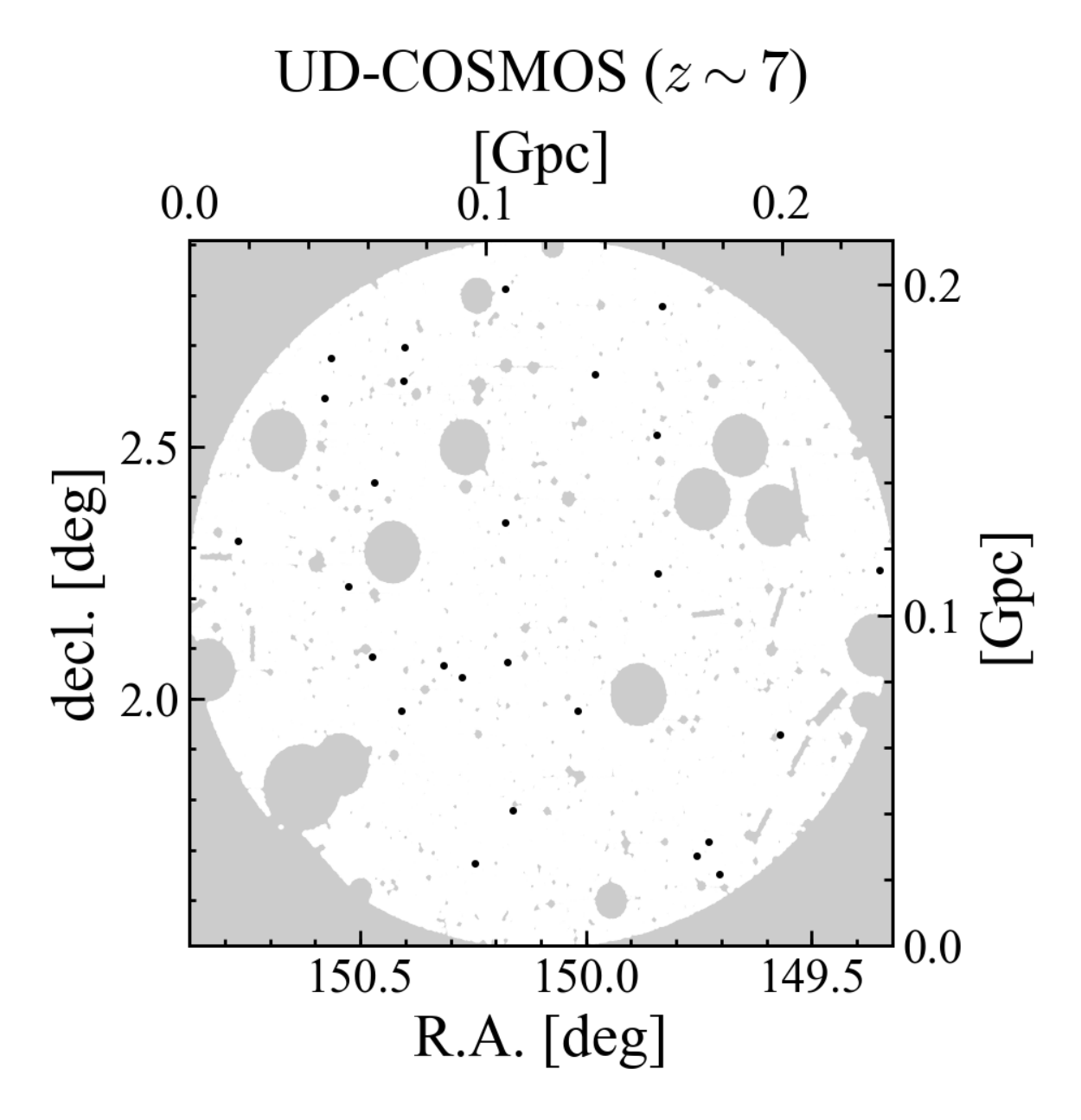}
\includegraphics[width=0.49\hsize, bb=15 0 345 320]{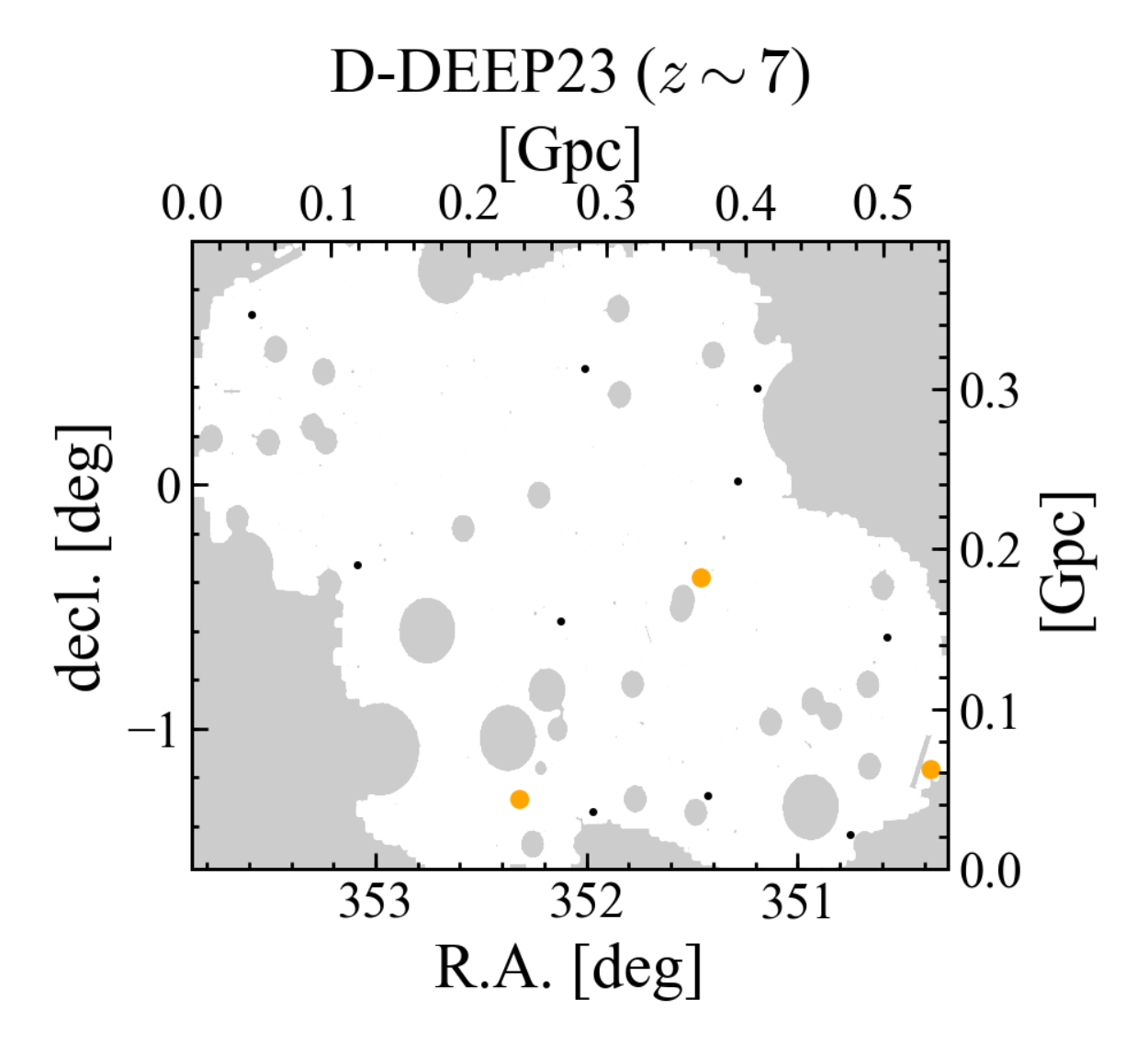}
\end{center}
\end{minipage}
\begin{minipage}{0.49\hsize}
\begin{center}
\includegraphics[width=0.99\hsize, bb=15 10 705 505]{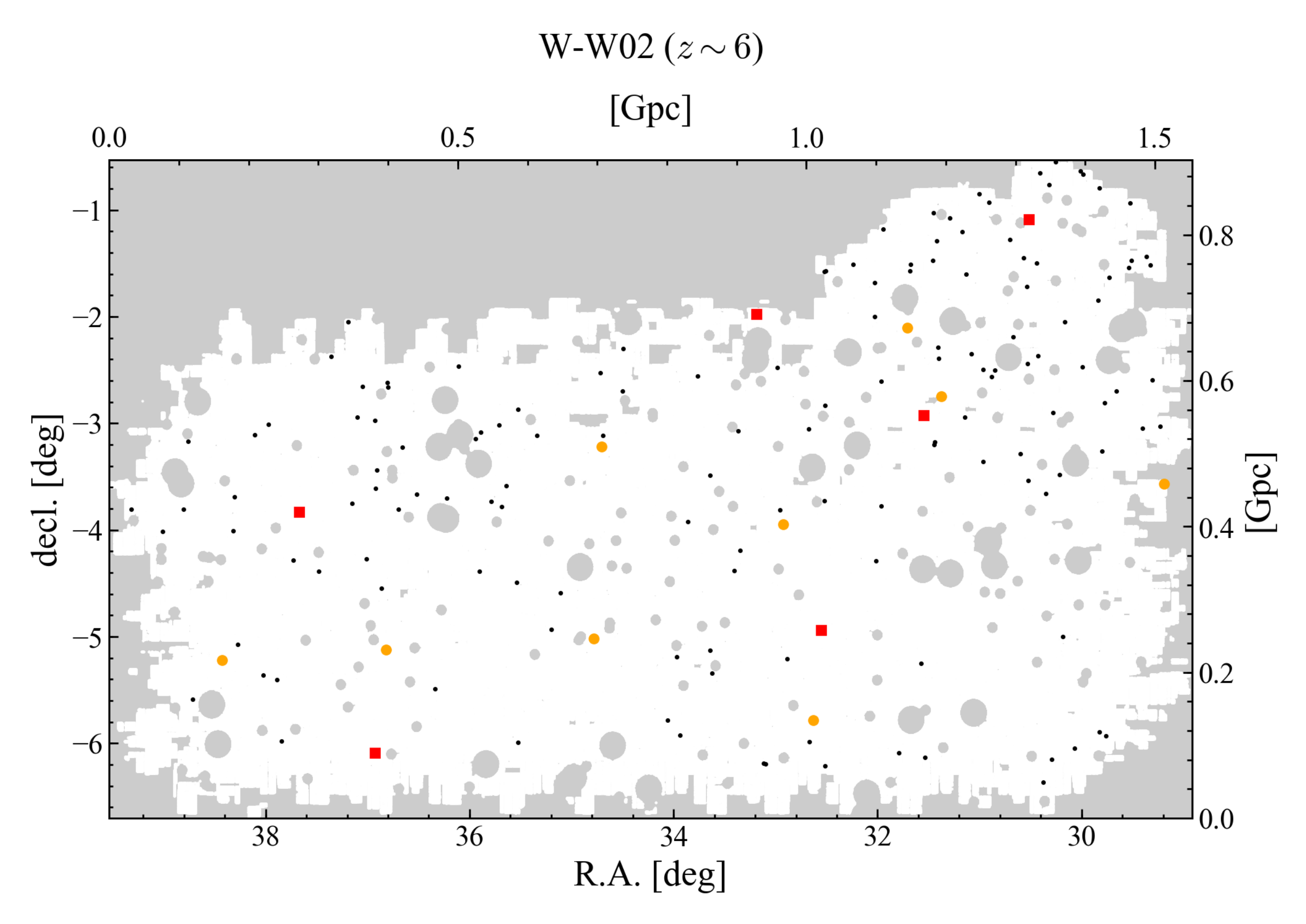}
\end{center}
\end{minipage}
\begin{minipage}{0.49\hsize}
\begin{center}
\includegraphics[width=0.99\hsize, bb=15 10 705 505]{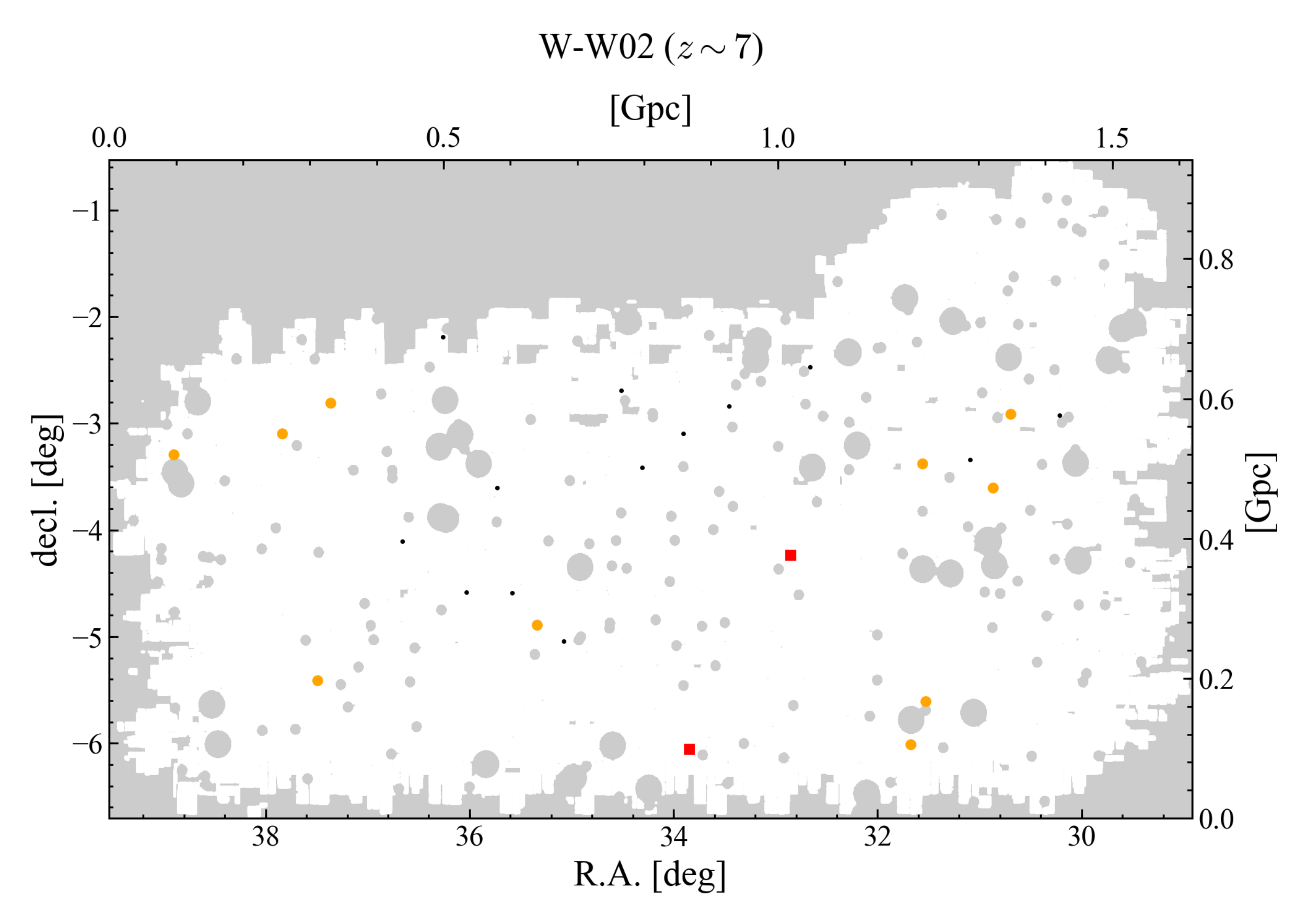}
\end{center}
\end{minipage}
\caption{Examples of sky distributions of dropout sources.
The red squares, orange circles, and black dots are, respectively, the positions of sources whose magnitudes are $i<22.0$, $22.0-22.5$, and $22.5-25.0\ (24.0)$ mag for $z\sim4$ sources, $z<22.5$, $22.5-23.0$, and $23.0-25.5\ (24.5)$ mag for $z\sim5$ sources, and $y<23.0$, $23.0-23.5$, and $23.5-25.5\ (25.0)$ mag for $z\sim6$ and $7$ sources, in the UltraDeep (Deep and Wide) layers.
The scale on the map is marked in degrees and in the projected distance (comoving Gpc).}
\label{fig_map}
\end{figure*}

Using the selection criteria described above, we select a total of 1,978,462 dropout candidates at $z\sim4-7$, consisting of 1,836,244 $g$-dropouts, 139,359 $r$-dropouts, 2,567 $i$-dropouts, and $292$ $z$-dropouts.
Our sample is selected from the $307.9\ \m{deg^2}$ wide area data corresponding to a $5.92\ \mathrm{Gpc^3}$ survey volume, and is the largest sample of the high-redshift ($z\gtrsim$4) galaxy population to date.
Especially, combined with $z\sim2-3$ galaxies selected later in Section \ref{ss_selection_z23}, we have a total of $4,100,221$ galaxies at $z\sim2-7$, which is the largest among high redshift galaxy studies.
Table \ref{tab_galnum} summarizes the number of dropout candidates in each field, and Figure \ref{fig_map} shows examples of sky distributions of the dropouts.
The differences in the numbers of the selected candidates mainly come from the differences in the survey areas and depths.

\begin{figure*}
\centering
\includegraphics[width=0.99\hsize, bb=21 4 914 216]{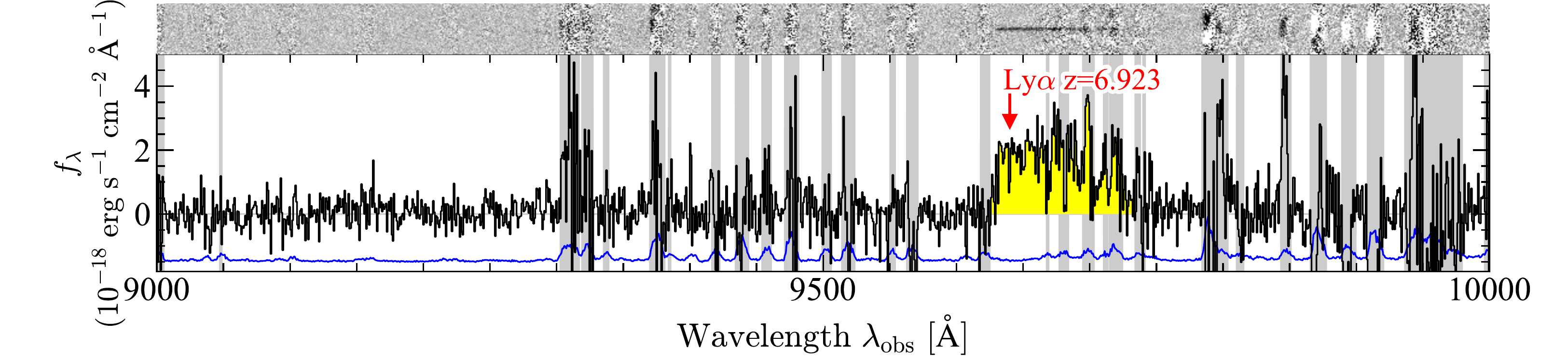}
\includegraphics[width=0.99\hsize, bb=21 4 914 240]{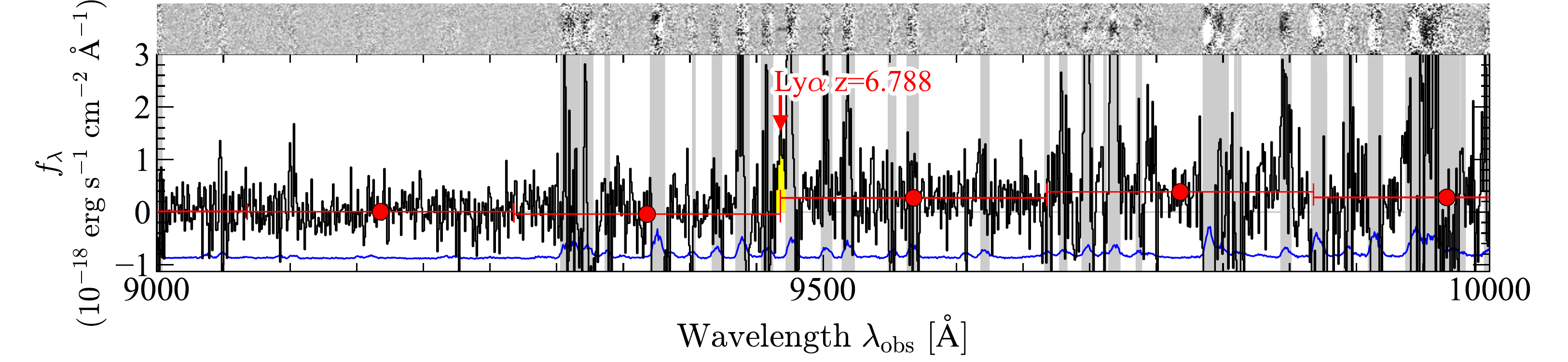}
\includegraphics[width=0.99\hsize, bb=21 4 914 240]{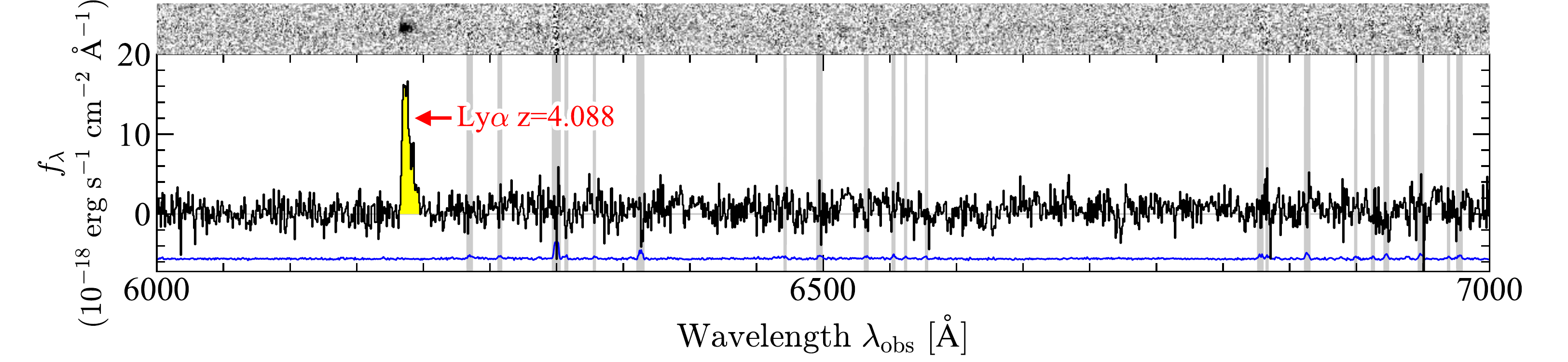}
\includegraphics[width=0.99\hsize, bb=21 4 914 240]{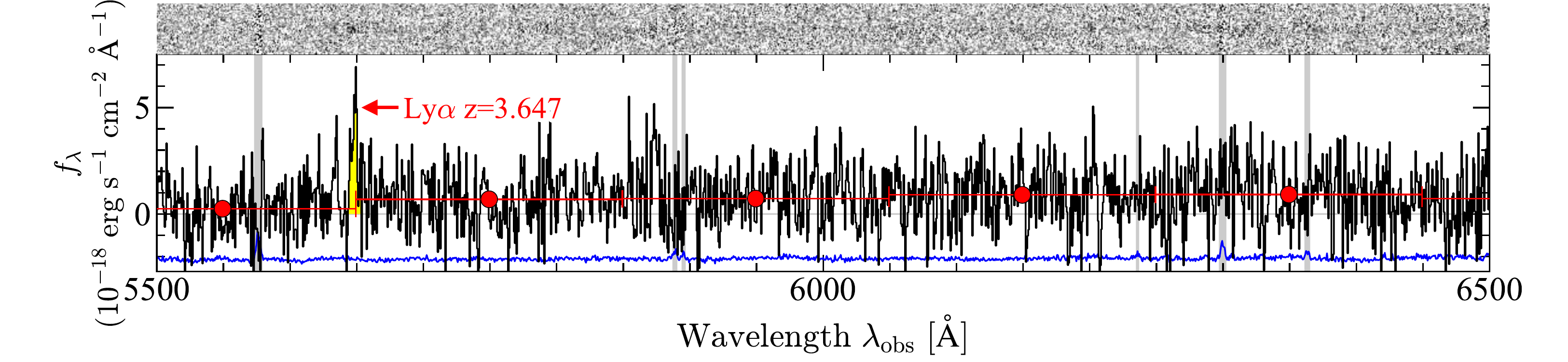}
\caption{Spectra of HSC J160953$+$532821 at $z=6.923$, HSC J161207$+$555919 at $z=6.788$, HSC J020834$-$021239 at $z=4.088$, and HSC J022552$-$054439 at $z=3.647$ from top to bottom, obtained in our spectroscopic follow-up observations.
In each figure, the top panel shows the two-dimensional spectrum (black is positive) and the bottom panel shows the one-dimensional spectrum. 
In the top panel, our dropout galaxy is located at the center in the spatial direction, and the spatial range is $\pm3\arcsec$.
In the bottom panel, the black line indicates the spectrum of the object, and the blue line shows the sky spectrum with an arbitrary normalization.
For the sources with weak Ly$\alpha$ emission, we also plot the averaged spectra over 200 $\m{\AA}$ bins with red filled circles to show the Lyman break features.
}
\label{fig_spec}
\end{figure*}

\subsection{Spectroscopically-Identified Sources}\label{ss_spec_source}

In our samples, a total of 46 sources are identified as $z>3$ objects through our spectroscopic follow-up observations with DEIMOS, AAOmega+2dF, and FOCAS (Section \ref{ss_spec}).
Redshifts are determined based on the Ly$\alpha$ line and/or Lyman break.
Figure \ref{fig_spec} shows examples of spectroscopically identified sources, HSC J160953$+$532821, HSC J161207$+$555919, HSC J020834$-$021239, and HSC J022552$-$054439.
HSC J160953$+$532821 is a faint quasar at $z=6.923$ with a UV magnitude of $M_\m{UV}=-22.7$ mag, confirmed in our FOCAS observations.
This source is also identified in \citet{2019ApJ...883..183M}.
An FWHM of its Ly$\alpha$ line is $\sim2500\ \m{km\ s^{-1}}$, and a rest-frame Ly$\alpha$ equivalent width (EW) is $EW_\m{Ly\alpha}^0\sim600\ \m{\AA}$.
HSC J161207$+$555919 is a bright ($M_\m{UV}=-23.0$ mag) galaxy at $z=6.788$ with a narrow Ly$\alpha$ emission line of $<300\ \m{km\ s^{-1}}$ and $EW_\m{Ly\alpha}^0\sim3\ \m{\AA}$, confirmed in our FOCAS observations.
This source is identified as a narrow-line quasar in \citet{2019ApJ...883..183M}.
The Lyman break feature can be seen in the spectrum, whose redshift is consistent with that derived from the Ly$\alpha$ emission line.
HSC J020834$-$021239 and HSC J022552$-$054439 are bright galaxies at $z=4.088$ and $3.647$ identified in our DEIMOS observations, with $M_\mathrm{UV}=-22.7$ and $-21.8$ mag, FWHMs of 250 and 180 $\m{km\ s^{-1}}$, and $EW_\m{Ly\alpha}^0\sim40$ and $9\ \m{\AA}$, respectively.

In addition, we incorporate results of our spectroscopic observations for high redshift galaxies with Magellan/IMACS. 
We also check spectroscopic catalogs in other studies (\citealt{2003A&A...405L..19C}, \citealt{2008ApJS..176..301O}, \citealt{2008ApJ...675.1076S}, \citealt{2009AJ....137.3541W}, \citealt{2010AJ....139..906W}, \citealt{2012MNRAS.422.1425C}, \citealt{2012ApJ...760..128M}, \citealt{2012ApJ...755..169M}, \citealt{2013A&A...559A..14L}, \citealt{2013AJ....145....4W}, \citealt{2015ApJS..218...15K}, \citealt{2016ApJS..227...11B}, \citealt{2016ApJ...825L...7H}, \citealt{2016ApJ...828...26M}, \citealt{2016ApJS..225...27M}, \citealt{2016ApJ...826..114T}, \citealt{2016ApJ...819...24W}, \citealt{2017ApJ...845L..16H}, \citealt{2017ApJ...846..134J}, \citealt{2017ApJ...841..111M}, \citealt{2017A&A...600A.110T}, \citealt{2017AJ....153..184Y}, \citealt{2018ApJ...858...77H}, \citealt{2018PASJ...70S..35M}, \citealt{2018ApJS..237....5M}, \citealt{2018PASJ...70S..10O}, \citealt{2018A&A...613A..51P}, \citealt{2018A&A...619A.147P}, \citealt{2018PASJ...70S..15S}, \citealt{2019ApJ...883..183M}, \citealt{2020ApJ...896...93H}, \citealt{2020ApJ...902..117H}, \citealt{2020ApJ...891..177Z}, \citealt{2021MNRAS.502.6044E}, and \citealt{2021A&A...647A.150G}). 
We adopt their classifications between galaxies and AGNs in their catalogs, which are mostly based on apparent AGN features such as broad emission lines. 
For the catalogs of the VIMOS VLT Deep Survey (VVDS; \citealt{2013A&A...559A..14L}) and the VIMOS Ultra Deep Survey (VUDS; \citealt{2017A&A...600A.110T}), we take into account sources whose reliabilities of the redshift determinations are $>70-75${\%}, i.e., sources with redshift reliability flags of 2, 3, 4, 9, 12, 13, 14, and 19.  
Here we focus on sources with spectroscopic redshifts of $z_{\rm spec} > 3$ in these catalogs.

In total, 1037 dropouts in our sample have been spectroscopically identified at $z_{\rm spec}\geq3$ in our observations and previous studies, including 770 galaxies and 267 AGNs.
These sources are listed in Table \ref{tab_speccat}, and the redshift distributions of the sources are shown in Figure \ref{fig_redshift_hist}.
Figure \ref{fig_2color} shows the distributions of the spectroscopically identified sources at $z_{\rm spec} > 3$ in the two-color diagrams. 
We also plot sources in the UD-COSMOS field with spectroscopic redshifts of $z_{\rm spec} < 3$ as foreground interlopers.
In addition, the tracks of model spectra of young star-forming galaxies that are produced with the stellar population synthesis code {\sc galaxev} \citep{2003MNRAS.344.1000B} are shown. 
As model parameters, the \citet{1955ApJ...121..161S} IMF, an age of $70$ Myr after the initial star formation, and metallicity of $Z / Z_\odot = 0.2$ are adopted. 
We use the \citet{2000ApJ...533..682C} dust extinction law with reddening of $E(B-V) = 0.16$. 
The IGM absorption is considered following the prescription of \citet{1995ApJ...441...18M}. 
The colors of the spectroscopically identified galaxies are broadly consistent with those expected from the model spectra.

\subsection{Selection Completeness and Redshift Distribution}\label{ss_completeness}

The selection completeness and redshift distributions of dropout candidates at $z\sim4-7$ are estimated based on results of Monte Carlo simulations in \citet{2018PASJ...70S..10O}.
\citet{2018PASJ...70S..10O} run a suite of Monte Carlo simulations with an input mock catalog of high redshift galaxies with the size distribution of \citet{2015ApJS..219...15S}, the Sersic index of $n=1.5$, the intrinsic ellipticities of $0.0$--$0.8$.
To produce galaxy SEDs, the stellar population synthesis model of {\sc galaxev} \citep{2003MNRAS.344.1000B} is used with the \citet{1955ApJ...121..161S} IMF, a constant rate of star formation, age of $25$ Myr, metallicity of $Z/Z_\odot = 0.2$, and the \citet{2000ApJ...533..682C} dust extinction ranging from $E(B-V) = 0.0$--$0.4$, corresponding to the UV spectral slope of $-3\lesssim\beta_\m{UV}\lesssim-1$.
We do not include Ly$\alpha$ emission in the galaxy SEDs because the line fluxes are typically not significant compared to the continuum in the broadband fluxes.
The IGM attenuation is taken into account by using the prescription of \citet{1995ApJ...441...18M}. 
Different simulations are carried out for the Wide, Deep, and UltraDeep layers by using the SynPipe software \citep{2018PASJ...70S...6H}, which utilizes GalSim v1.4 \citep{2015A&C....10..121R} and the HSC pipeline {\tt hscpipe} \citep{2018PASJ...70S...5B}.
We insert large numbers of artificial sources into HSC images.
Then we select high redshift galaxy candidates with the same selection criteria, and calculate the selection completeness as a function of magnitude and redshift, $C(m,z)$, averaged over UV spectral slope $\beta_\m{UV}$ weighted with the $\beta_\m{UV}$ distribution of \citet{2014ApJ...793..115B}.  
Then the obtained completeness is scaled based on the limiting magnitudes to correct for differences in depths of the S18A data in this study and S16A data used in \citet{2018PASJ...70S..10O}.

\begin{figure}
\centering
\includegraphics[width=0.99\hsize, bb=30 16 633 316]{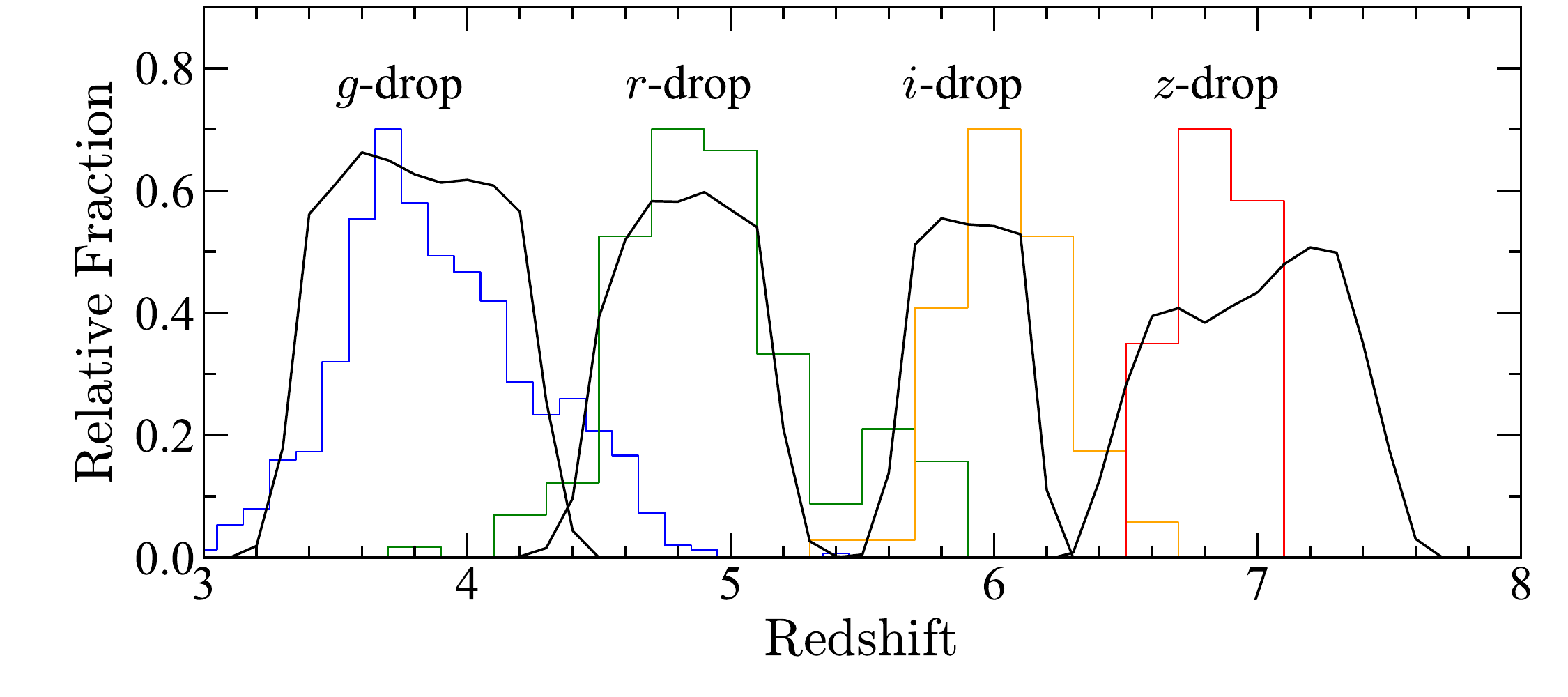}
\caption{Selection completeness estimates for our $z\sim4$, $5$, $6$, and $7$ samples.
The black curves correspond to the results of the Monte Carlo calculations in Section \ref{ss_completeness} averaged over the Wide, Deep, and UltraDeep layers.
Average redshifts of these samples are 3.8, 4.9, 5.9, and 6.9.
The histograms show redshift distributions of spectroscopically identified sources in the $z\sim4$ (blue), $z\sim5$ (green), $z\sim6$ (orange), and $z\sim7$ (red) samples.}
\label{fig_redshift_hist}
\end{figure}

Figure \ref{fig_redshift_hist} shows results of the selection completeness estimates as a function of redshift.  
The average redshift values are roughly ${\bar z}=3.8$ for $g$-dropouts, ${\bar z}=4.9$ for $r$-dropouts, ${\bar z}=5.9$  for $i$-dropouts, and  ${\bar z}=6.9$  for $z$-dropouts. 
In Figure \ref{fig_redshift_hist}, we also show the redshift distributions of the spectroscopically identified galaxies in our samples (Section \ref{ss_spec_source}). 
The redshift distributions of the spectroscopically identified sources are broadly consistent with the results of our selection completeness simulations.
However, the distributions of the spectroscopically identified sources in the $g$-, $r$-, and $i$-dropout samples appear to be shifted toward slightly higher redshifts compared to the simulation results.
This is probably because the spectroscopically identified sources are biased to ones with strong Ly$\alpha$ emission in the $g$- and $r$-dropout samples, and ones identified in the SHELLQs project searching for $z\sim6-7$ quasars \citep[e.g.,][]{2016ApJ...828...26M} in the $i$-dropout sample.
In particular, the redshift distribution of the spectroscopically identified $r$-dropouts has a secondary peak at $z\sim5.7$, which is caused by $z=5.7$ Ly$\alpha$ emitters found in the Subaru Suprime-Cam and HSC narrowband surveys in the literature \citep[e.g.,][]{2008ApJS..176..301O,2018PASJ...70S..15S}.
Another possible reason is the systematic uncertainty of the IGM attenuation model in the simulations.
The model of \citet{1995ApJ...441...18M} predicts lower redshift dropouts than \citet{2014MNRAS.442.1805I}, which may explain the discrepancy.
For the $z$-dropout sample, the redshift distribution of the spectroscopically identified sources is shifted to the lower redshift, because of the increasing fraction of the neutral hydrogen at $z>7$, which resonantly scatters Ly$\alpha$ photons.

\subsection{Contamination}\label{ss_contami} 

Some foreground objects such as red galaxies at low redshifts can satisfy our color selection criteria \redc{due to photometric scatters}, although intrinsically they do not enter the color selection window. 
This happens especially in the Wide and Deep layers, whose limiting magnitudes are relatively shallow.
We evaluate contamination fractions in our dropout samples with the following three methods.

The first method is the one using estimates of photometric redshifts.
We use photometric redshifts estimated with the MIZUKI code \citep{2015ApJ...801...20T,2018PASJ...70S...9T}. 
Here a foreground interloper is defined as a source whose 95\% upper bound of the photometric redshift is less than $z=2$.
We derive the fraction of the foreground interlopers as a function of the $i$- ($z$-) band magnitude in our $z\sim4$ ($z\sim5$) dropout samples in the Wide and Deep layers.
\redc{The derived contamination fractions are presented in Table \ref{tab_contami}.}
In the UltraDeep layer, the fractions of the interlopers in our $z\sim4-5$ samples are negligibly small ($<10\%$) at $m_\m{UV}\geq25$ mag.
\redcc{The contamination fraction becomes higher for brighter sources at $\lesssim24$ mag, because the number density of high redshift galaxies decreases to brighter magnitudes, while that of foreground interlopers (e.g., low redshift passive galaxies) does not significantly change \citep[e.g.,][]{2013ApJ...777...18M}.}
We do not derive the contamination fraction of the $z\sim6-7$ sources with the MIZUKI code, because accuracies of the photometric redshifts are not high due to the limited number of available bands redder than the Lyman break.
\redcc{In the UD-COSMOS field}, we also check photometric redshifts in the COSMOS 2015 catalog \citep{2016ApJS..224...24L} that are determined with multi-band photometric data including near-infrared images such as {\it Spizter}/IRAC, useful to eliminate stellar contaminants \citep{2018ApJ...863...63S}.
\redcc{We match our dropout sources with the COSMOS 2015 catalog  within a $1\arcsec$ search radius of the object coordinate.}
\redcc{For bright $z\sim4-5$ sources detected with $>10\sigma$ significance levels in the HSC $i$-band, whose flux measurements are reliable, typically less than $20\%$ of them are classified as foreground interlopers, consistent with the estimates of the MIZUKI code.
For fainter sources at $z\sim4-5$, about $\sim70\%$ of them are classified as $z>3$ galaxies.
The contamination rate for $z\sim6-7$ sources is less than $40\%$, although the number of sources is small and their flux measurements have relatively large uncertainties due to their faintness compared to the bright $z\sim4-5$ sources.}
\redcc{Stellar templates are also fitted for the sources in the COSMOS 2015 catalog, and only 3.5\% of them are classified as stars with $\chi^2_\m{star}<\chi^2_\m{galaxy}$ (see \citealt{2016ApJS..224...24L}).
We also check the COSMOS {\it Hubble} ACS catalog \citep{2007ApJS..172..219L}, and only 3.7\% of them are point sources with the SExtractor \citep{1996A&AS..117..393B} stellarity parameters of $>0.9$ (1=star; 0=galaxy).
These analyses indicate that the fraction of the stellar contamination is negligibly small.}

\setlength{\tabcolsep}{0.1cm}
\begin{deluxetable}{ccccc}
\tablecaption{\redc{Estimated Contamination Fraction for the $z\sim4$ and $z\sim5$ Samples in UltraDeep (UD), Deep (D), and Wide (W) Layers}}
\tablehead{
\colhead{Redshift} & \colhead{$m_\m{UV}$} & \colhead{Fraction (UD)} & \colhead{Fraction (D)} & \colhead{Fraction (W)}}
\startdata
$z\sim4$  & 22.0 & \nodata & $0.69^{+0.06}_{-0.06}$ & $0.60^{+0.07}_{-0.07}$ \\
 & 23.0 & $0.33^{+0.15}_{-0.15}$ & $0.64^{+0.03}_{-0.03}$ & $0.61^{+0.03}_{-0.03}$ \\
 & 24.0 & $0.12^{+0.06}_{-0.06}$ & $0.18^{+0.02}_{-0.02}$ & $0.20^{+0.02}_{-0.02}$ \\
 & 24.5 & $0.05^{+0.04}_{-0.04}$ & $0.07^{+0.01}_{-0.01}$ & $0.11^{+0.01}_{-0.01}$ \\
 & 25.0 & $0.02^{+0.02}_{-0.02}$ & $0.04^{+0.01}_{-0.01}$ & $0.08^{+0.01}_{-0.01}$ \\
 & 25.5 & $0.02^{+0.02}_{-0.02}$ & $0.03^{+0.01}_{-0.01}$ & $0.05^{+0.01}_{-0.01}$ \\
 & 26.0 & $0.03^{+0.01}_{-0.01}$ & $0.03^{+0.01}_{-0.01}$ & $0.04^{+0.01}_{-0.01}$ \\
 & 26.5 & $0.02^{+0.03}_{-0.03}$ & $0.02^{+0.03}_{-0.03}$ & $0.04^{+0.19}_{-0.19}$ \\
$z\sim5$ & 23.0 & \nodata & $0.55^{+0.11}_{-0.11}$ & $0.45^{+0.10}_{-0.10}$ \\
 & 24.0 & $0.02^{+0.21}_{-0.21}$ & $0.35^{+0.04}_{-0.04}$ & $0.35^{+0.04}_{-0.04}$ \\
 & 24.5 & $0.03^{+0.09}_{-0.09}$ & $0.13^{+0.03}_{-0.03}$ & $0.15^{+0.03}_{-0.03}$ \\
 & 25.0 & $0.03^{+0.05}_{-0.05}$ & $0.08^{+0.02}_{-0.02}$ & $0.09^{+0.02}_{-0.02}$ \\
 & 25.5 & $0.04^{+0.03}_{-0.03}$ & $0.07^{+0.02}_{-0.02}$ & $0.08^{+0.02}_{-0.02}$ \\
 & 26.0 & $0.06^{+0.04}_{-0.04}$ & $0.06^{+0.03}_{-0.03}$ & $0.06^{+0.06}_{-0.06}$ \\
 & 26.5 & $0.06^{+0.11}_{-0.11}$ & $0.06^{+0.11}_{-0.11}$ & \nodata 
\enddata
\tablecomments{\redcc{These contamination fractions are estimated based on the photometric redshift analysis (the first method in Section \ref{ss_contami}).}}
\label{tab_contami}
\end{deluxetable}

The second method is the one using the spectroscopic redshift catalog created in Section \ref{ss_spec}.
We estimate the contamination fraction in the $z\sim4$ sample using the results in our AAOmega+2dF spectroscopy and the VVDS spectroscopy, which target a sufficient number of bright $z\sim4$ sources, and whose target selections are not significantly biased to low or high redshift sources.
\redc{Foreground interlopers are identified based on continuum emission bluewards of the expected wavelength of the Lyman break, and/or rest-frame optical emission lines.}
At $z\gtrsim5$, we cannot derive robust contamination fractions because of the small number of spectroscopically confirmed sources in the AAOmega+2dF and VVDS data.
Nonetheless, we find that a total of 80 sources from our $z\sim6-7$ dropout samples ($y=21.0-25.6$ mag) are spectroscopically identified in the entire spectroscopic catalog, and all of them are at $z_\m{spec}>5$, although it is possible that the actual contamination rate is higher than inferred from these numbers due to various biases including the publication bias and the fact that spectroscopic observations usually prioritize the most promising candidates.

The third method is a simulation with shallower data, in the same manner as \citet{2018PASJ...70S..10O}.
We use a shallower dataset whose depth is comparable with that of the Wide layer in the UD-COSMOS field, the Wide-layer-depth COSMOS data. 
We assume that the UD-COSMOS data are sufficiently deep and the contamination fraction in our dropout selections is small. 
First, we select objects that do not satisfy our selection criteria at each redshift from the UD-COSMOS catalog. 
Then, we find the closest source in the Wide-layer-depth COSMOS catalog that matches within a $1\arcsec$ search radius of the object coordinate.
If the objects satisfy our selection criteria for the Wide-layer dropout, we regard them as foreground interlopers, and calculate their number densities.  
Based on comparisons between the surface number densities of interlopers and those of the selected dropouts, we estimate the fractions of foreground interlopers \redc{that satisfy our color selection criteria due to the photometric scatters}. 
The estimated contamination fractions are \redcc{$\sim0-40\%$ for sources with $24-25$ mag at $z\sim4-5$,} comparable to those in \citet{2018PASJ...70S..10O}.
For the $z \sim 6-7$ dropout samples, we cannot estimate the surface number densities of interlopers by adopting this method, because the number densities of such sources in the shallower depth COSMOS field data are too small due to the limited survey area of the UD-COSMOS field.

We find that the three methods above give the contamination fractions consistent with each other within their uncertainties at $z\sim4-5$.
As the contamination fractions used in derivations of luminosity functions later, we adopt the fractions determined based on the photometric redshifts, given their high accuracies compared to those of the other methods.
For the $z\sim4-5$ samples in the UltraDeep layer and the $z\sim6-7$ samples, we assume that contamination fractions are negligibly small based on the results of the photometric and spectroscopic redshifts.

\subsection{Source Selection at $z\sim2-3$}\label{ss_selection_z23}

In addition to the $z\sim4-7$ catalogs, we also use catalogs of galaxy candidates at $z\sim2-3$ to study clustering properties.
We use BM, BX, and $U$-dropout galaxy catalogs at $z\sim1.7$, $2.2$, and $3$ constructed in C. Liu et al. in prep.
Here we briefly describe our source selection at $z\sim2-3$.
The $z\sim2-3$ galaxy candidates are selected from a combined CLAUDS+HSC-SSP catalog made by \citet{2019MNRAS.489.5202S} based on {\tt hscpipe}.
Note that the HSC-SSP data in the combined CLAUDS+HSC-SSP catalog is based on the S16A internal data release, which is different from the S18A data release product we use for the $z\sim4-7$ selection.
As described in Section \ref{ss_data_CLAUDS}, the U-band images are obtained with two filters, the $u$ and $u^*$-bands.
For the $u^*$-band filter, we adopt color criteria same as \cite{2009A&A...498..725H}, who select $z\sim3$ dropout galaxies with the similar filter set to this study:\\
$U$-dropouts ($z\sim3$)
\begin{eqnarray}\label{eq_ugr_uS}
  (g-r   &<& 1.2)\land \\
  (u^*-g &>& 0.9)\land \\
  (u^*-g &>& 1.5 (g-r) + 0.75). 
\end{eqnarray}
For the BX and BM galaxies, we adopt the following color criteria, respectively:\\
BX ($z\sim2.2$)
\begin{eqnarray}\label{eq_BX_uS}
  (g-r   &>& -0.5)\land \\
  (u^*-g &>& 2.57 (g-r) + 0.21)\land\\
  (u^*-g &>& 0.42 (g-r) + 0.54)\land\\
  ((u^*-g &<& 0.9) \lor (u^*-g < 1.5 (g-r) + 0.75)),
\end{eqnarray}
BM ($z\sim1.7$)
\begin{eqnarray}\label{eq_BM_uS}
  (g-r   &>& -0.5)\land \\
  (u^*-g &>& -2.5 (g-r) + 0.98)\land\\
  (u^*-g &>& 0.93 (g-r) + 0.12)\land\\
  (u^*-g &<& 0.42 (g-r) + 0.54).
\end{eqnarray}
For the $ugr$ filter set, we define selection criteria by comparing the positions of stars in the $ugr$ and $u^*gr$ diagrams:\\
$U$-dropouts ($z\sim3$)
\begin{eqnarray}\label{eq_ugr_u}
  (g-r &<& 1.2)\land \\
  (u-g &>& 0.98)\land \\
  (u-g &>& 1.99 \times g-r + 0.68),
\end{eqnarray}
BX ($z\sim2.2$)
\begin{eqnarray}\label{eq_BX_u}
  (g-r   &>& -0.5)\land \\
  (u-g &>& 3.07 (g-r) + 0.14)\land\\
  (u-g &>& 0.42 (g-r) + 0.54)\land\\
  ((u-g &<& 0.9)\lor(u-g < 1.5 (g-r) + 0.75)).
\end{eqnarray}
Selection criteria of BM galaxies for the $ugr$ filter set are the same as those for the $u^*gr$ filter set.
Details of the selection are presented in C. Liu et al. in prep.
A total of 935,804, 405,469, and 780,486 galaxy candidates are selected at $z\sim1.7$, $2.2$, and $3$, respectively.
The number densities of the selected galaxies are comparable to previous studies.
The selection completeness and contamination fraction of these samples are described in C. Liu et al. in prep.

\section{UV Luminosity Function}\label{ss_LF}

\subsection{Dropout UV Luminosity Function}\label{ss_LFall}

\subsubsection{Derivation}

We derive the rest-frame UV luminosity functions of $z\sim4-7$ dropout sources by applying the effective volume method \citep{1999ApJ...519....1S}. 
Based on the results of the selection completeness simulations in Section \ref{ss_completeness}, we estimate the effective survey volume per unit area as a function of the apparent magnitude, 
\begin{equation}
V_{\rm eff} (m) = \int C(m,z) \frac{dV(z)}{dz} dz, 
\end{equation}
where $C(m,z)$ is the selection completeness, i.e., the probability that a galaxy with an apparent magnitude $m$ at redshift $z$ is detected and satisfies the selection criteria, and $dV(z)/dz$ is the differential comoving volume as a function of redshift.

The space number densities of the dropouts that are corrected for incompleteness and contamination effects are obtained by calculating
\begin{equation}
\psi(m) = \left[1-f_\m{cont}(m)\right]\frac{n_{\rm raw}(m)}{V_{\rm eff}(m)},
\end{equation}
where $n_{\rm raw}(m)$ is the surface number density of selected dropouts in an apparent magnitude bin of $m$, and $f_{\rm cont}(m)$ is the contamination fraction in the magnitude bin estimated in Section \ref{ss_contami}. 
The $1\sigma$ uncertainties are calculated by taking account of the Poisson confidence limits \citep{1986ApJ...303..336G} on the numbers of the sources. 
To calculate $1\sigma$ uncertainties of the space number densities of dropouts, we consider uncertainties of the surface number densities and the contamination fractions.

We convert the number densities of dropouts as a function of apparent magnitude, $\psi(m)$, into the UV luminosity functions, $\Phi[M_{\rm UV}(m)]$, which is the number densities of dropouts as a function of rest-frame UV absolute magnitude.
We calculate the absolute UV magnitudes of dropout samples from their apparent magnitudes using their averaged redshifts $\bar z$:   
\begin{equation}
M_{\rm UV} = m + 2.5 \log(1+\bar z) - 5 \log \left( \frac{d_{\rm L}(\bar z) }{ 10 \, {\rm pc}} \right) + (m_{\rm UV} - m),  
\end{equation}
where $d_{\rm L}$ is the luminosity distance in units of parsecs and $(m_{\rm UV} - m)$ is the $K$-correction term between the magnitude at rest-frame UV and the magnitude in the bandpass that we use. 
We define the UV magnitude, $m_\m{UV}$, as the magnitude at the rest-frame $1500\ \m{\AA}$.
For the apparent magnitude $m$, we use a magnitude in a band whose central wavelength is the nearest to the rest-frame wavelength of $1500\ \m{\AA}$, namely $i$-, $z$-, $y$-, and $y$-bands for $g$-, $r$-, $i$-, and $z$-dropouts, respectively.
We set the $K$-correction term to be $0$ by assuming that dropout galaxies have flat UV continua, i.e., constant $f_\nu$ in the rest-frame UV ($m_\m{UV}=m$). 
\redc{Note that this assumption does not have a significant impact on the calculated UV magnitudes.
If we vary the UV slope ($\beta_\m{UV}$) with $-2.5<\beta_\m{UV}<-1.5$ \citep{2014ApJ...793..115B}, the calculated UV magnitude differs only within 0.1 mag.
}

\subsubsection{Results}\label{ss_LFall_res}

\begin{figure*}
\centering
\includegraphics[width=0.95\hsize, bb=17 7 928 503]{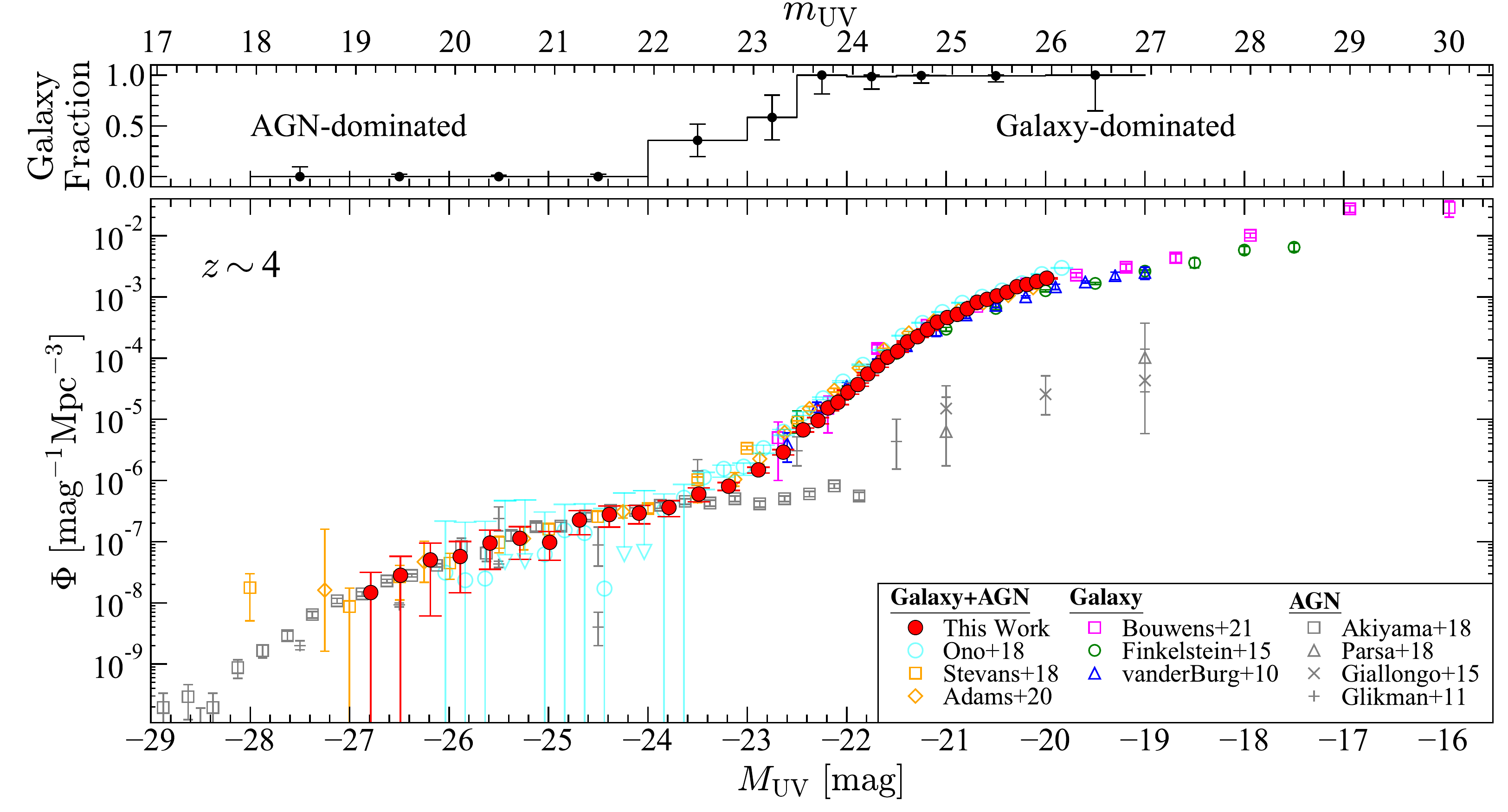}
\includegraphics[width=0.95\hsize, bb=17 7 928 513]{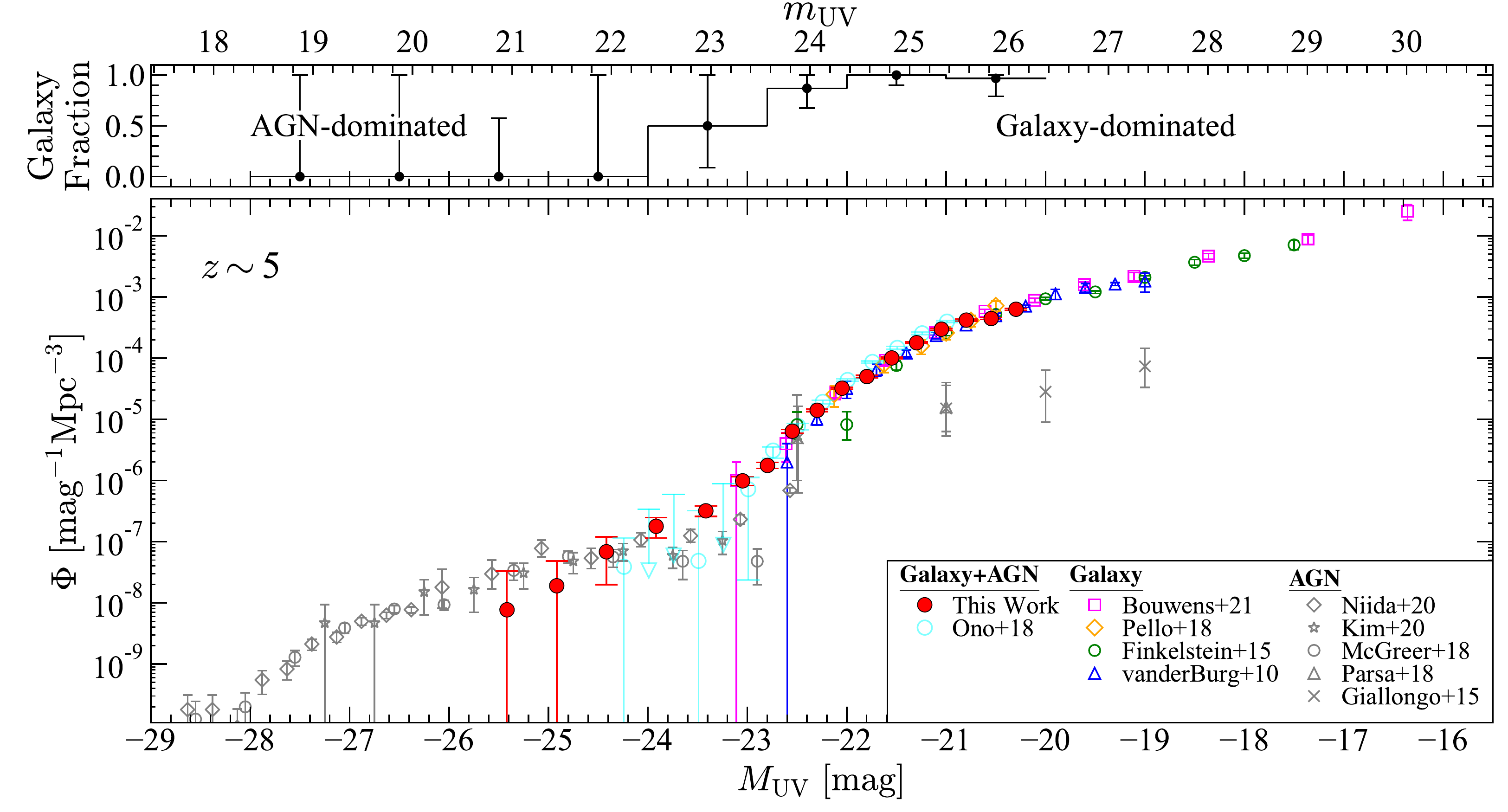}
\caption{Each bottom panel shows rest-frame UV luminosity functions of dropout sources (including galaxies and AGNs) at $z\sim4$ and $5$.
The red circles show our results based on the HSC-SSP survey data.
For comparison, we also show previous results for galaxies+AGNs in \citet[][cyan circles]{2018PASJ...70S..10O}, \citet[][orange squares]{2018ApJ...863...63S}, and \citet[][orange diamonds]{2020MNRAS.494.1771A}, galaxies in \citet[][magenta squares]{2021arXiv210207775B}, \citet[][green circles]{2015ApJ...810...71F}, \citet[][blue triangles]{2010A&A...523A..74V}, and \citet[][orange diamonds]{2018A&A...620A..51P}, and AGNs in \citet[][black squares]{2018PASJ...70S..34A}, \citet[][black triangles]{2018MNRAS.474.2904P}, \citet[][black crosses]{2015A&A...578A..83G}, \citet[][black pluses]{2011ApJ...728L..26G}, \citet[][black diamonds]{2020ApJ...904...89N}, and \redc{\citet[][black stars]{2020ApJ...904..111K}}.
Each top panel shows a fraction of galaxies in our dropout sample based on spectroscopic results.
For the denominator of the fraction, the sum of the numbers of galaxies and AGNs is used.
\redc{Note that this fraction is estimated based on various spectroscopic catalogs, and its uncertainty is discussed in Section \ref{ss_galLF_res}.}
}
\label{fig_LFall_1}
\end{figure*}

The top panel of Figure \ref{fig_LFall_1} shows our derived luminosity function for $g$-dropouts at $z \sim 4$ with previous studies.
Our measurements have smaller error bars compared to \citet{2018PASJ...70S..10O} because of the improved constraint on the contamination fraction (Section \ref{ss_contami}).
Our measurements agree well with previous studies of quasars at $M_\m{UV}\lesssim-24$ \citep[e.g.,][]{2018PASJ...70S..34A}, studies of galaxies at $M_\m{UV}\gtrsim-22$ \citep[e.g.,][]{2021arXiv210207775B,2015ApJ...810...71F}, and studies of galaxies and AGNs \citep{2018PASJ...70S..10O,2018ApJ...863...63S,2020MNRAS.494.1771A}.

\begin{figure*}
\centering
\includegraphics[width=0.95\hsize, bb=17 7 928 503]{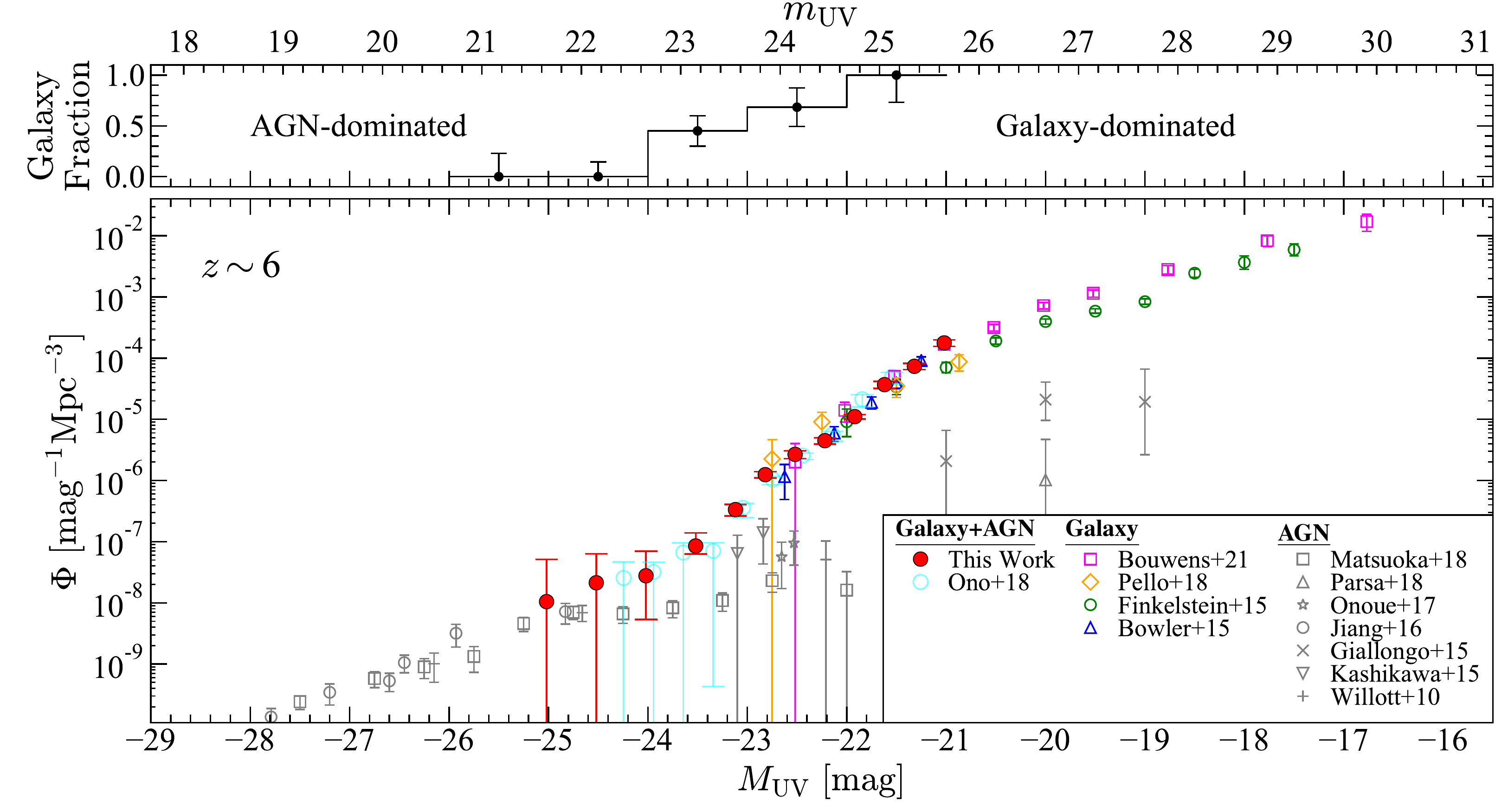}
\includegraphics[width=0.95\hsize, bb=17 7 928 513]{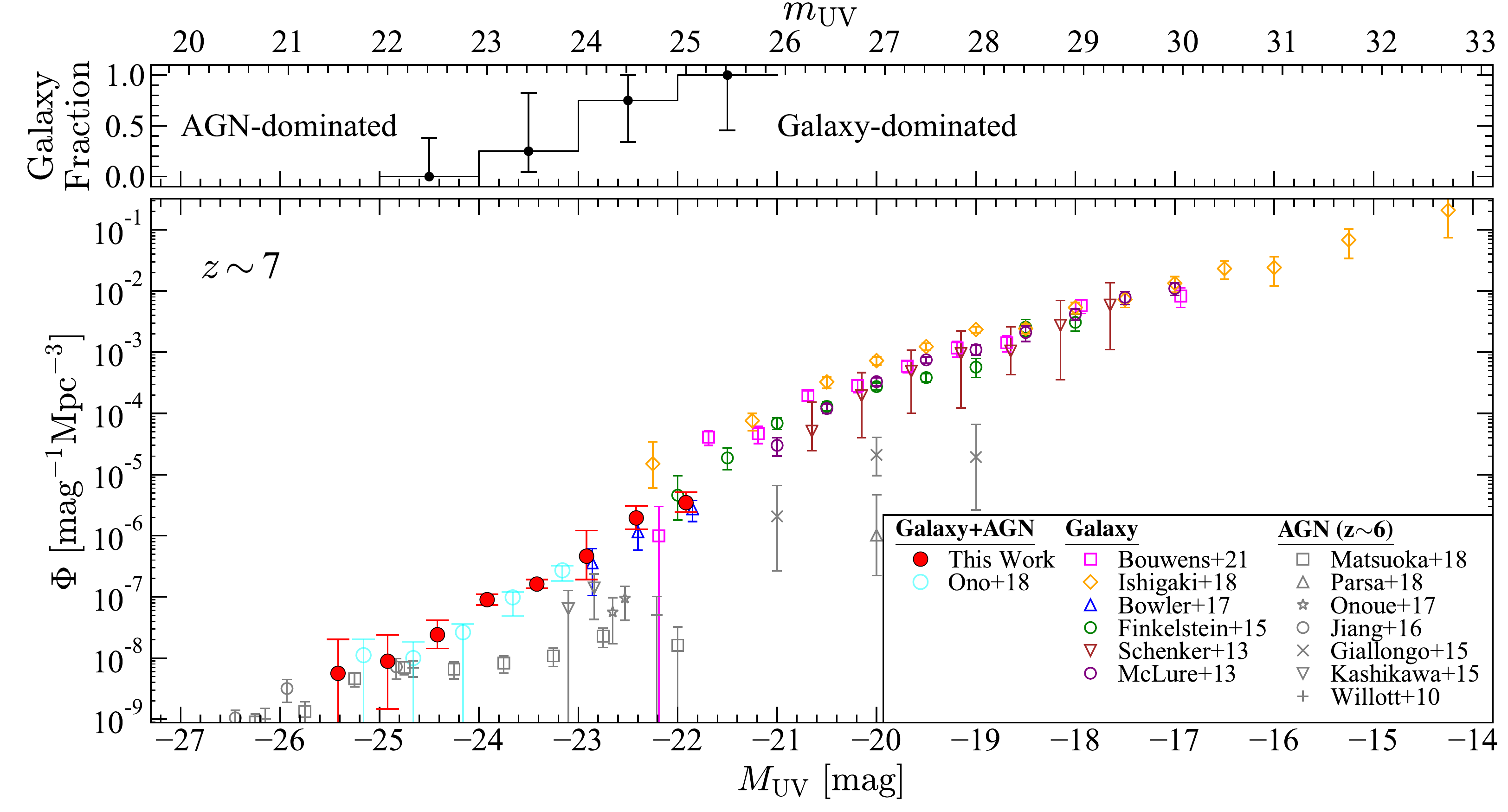}
\caption{Same as Figure \ref{fig_LFall_1} but at $z\sim6$ and $7$.
For comparison, we also show previous results for galaxies+AGNs in \citet[][cyan circles]{2018PASJ...70S..10O}, galaxies in \citet[][magenta squares]{2021arXiv210207775B}, \citet[][orange diamonds]{2018A&A...620A..51P}, \citet[][orange diamonds]{2018ApJ...854...73I}, \citet[][blue triangles]{2017MNRAS.466.3612B,2015MNRAS.452.1817B}, \citet[][green circles]{2015ApJ...810...71F}, \citet[][brown triangles]{2013ApJ...768..196S}, and \citet[][purple circles]{2013MNRAS.432.2696M}, and AGNs in \citet[][black squares]{2018ApJ...869..150M}, \citet[][black triangles]{2018MNRAS.474.2904P}, \citet[][black stars]{2017ApJ...847L..15O}, \citet[][black circles]{2016ApJ...833..222J}, \citet[][black crosses]{2015A&A...578A..83G}, \citet[][black triangles]{2015ApJ...798...28K}, and \citet[][black pluses]{2010AJ....139..906W}.
In the $z\sim7$ panel, the $z\sim6$ AGN luminosity functions are plotted because there are no results of the AGN luminosity function at $z\sim7$.
}
\label{fig_LFall_2}
\end{figure*}

\begin{deluxetable*}{cccccc}
\setlength{\tabcolsep}{1.5pt}
\tablecaption{Obtained Dropout (Galaxy+AGN) and Galaxy UV luminosity functions at $z\sim4$, $5$, $6$, and $7$.}
\tablehead{
\colhead{$M_\m{UV}$} & \colhead{$\Phi$} & \colhead{$\Phi_\m{galaxy}$$^\dagger$} & \colhead{$M_\m{UV}$} & \colhead{$\Phi$} & \colhead{$\Phi_\m{galaxy}$$^\dagger$} \\
\colhead{(mag)} & \colhead{($\m{Mpc^{-3}}\ \m{mag^{-1}}$)} & \colhead{($\m{Mpc^{-3}}\ \m{mag^{-1}}$)} &\colhead{(mag)} & \colhead{($\m{Mpc^{-3}}\ \m{mag^{-1}}$)} & \colhead{($\m{Mpc^{-3}}\ \m{mag^{-1}}$)}}
\startdata
\multicolumn{3}{c}{$z\sim4$} & \multicolumn{3}{c}{$z\sim5$} \\
$-26.79$ & $(1.47^{+1.67}_{-1.47})\times10^{-8}$ & \nodata & $-25.42$ & $(7.70^{+25.20}_{-7.70})\times10^{-9}$ & \nodata \\
$-26.49$ & $(2.82^{+2.95}_{-2.82})\times10^{-8}$ & \nodata & $-24.92$ & $(1.90^{+2.92}_{-1.90})\times10^{-8}$ & \nodata \\
$-26.19$ & $(5.03^{+4.45}_{-4.42})\times10^{-8}$ & \nodata & $-24.42$ & $(6.84^{+5.14}_{-4.85})\times10^{-8}$ & $(2.58^{+14.68}_{-2.58})\times10^{-9}$ \\
$-25.89$ & $(5.72^{+4.32}_{-4.25})\times10^{-8}$ & \nodata & $-23.92$ & $(1.78^{+0.68}_{-0.64})\times10^{-7}$ & $(4.73^{+10.22}_{-4.73})\times10^{-8}$ \\
$-25.59$ & $(9.48^{+6.03}_{-5.96})\times10^{-8}$ & \nodata & $-23.42$ & $(3.19^{+0.65}_{-0.61})\times10^{-7}$ & $(1.57^{+2.48}_{-1.57})\times10^{-7}$ \\
$-25.29$ & $(1.13^{+0.63}_{-0.62})\times10^{-7}$ & \nodata & $-23.05$ & $(9.87^{+1.60}_{-1.60})\times10^{-7}$ & $(6.22^{+4.19}_{-4.19})\times10^{-7}$ \\
$-24.99$ & $(9.75^{+4.92}_{-4.80})\times10^{-8}$ & \nodata & $-22.80$ & $(1.76^{+0.23}_{-0.19})\times10^{-6}$ & $(1.27^{+0.67}_{-0.66})\times10^{-6}$ \\
$-24.69$ & $(2.26^{+0.96}_{-0.96})\times10^{-7}$ & \nodata & $-22.55$ & $(6.39^{+0.43}_{-0.43})\times10^{-6}$ & $(5.21^{+2.28}_{-2.28})\times10^{-6}$ \\
$-24.39$ & $(2.77^{+1.04}_{-1.04})\times10^{-7}$ & $(1.11^{+2.74}_{-1.11})\times10^{-8}$ & $-22.30$ & $(1.40^{+0.07}_{-0.07})\times10^{-5}$ & $(1.24^{+0.43}_{-0.43})\times10^{-5}$ \\
$-24.09$ & $(2.92^{+0.97}_{-0.97})\times10^{-7}$ & $(4.30^{+5.65}_{-4.30})\times10^{-8}$ & $-22.05$ & $(3.22^{+0.13}_{-0.13})\times10^{-5}$ & $(2.97^{+0.80}_{-0.80})\times10^{-5}$ \\
$-23.79$ & $(3.63^{+1.06}_{-1.06})\times10^{-7}$ & $(9.22^{+9.32}_{-9.22})\times10^{-8}$ & $-21.80$ & $(5.00^{+0.22}_{-0.22})\times10^{-5}$ & $(4.78^{+1.10}_{-1.10})\times10^{-5}$ \\
$-23.49$ & $(6.01^{+1.54}_{-1.54})\times10^{-7}$ & $(2.17^{+1.83}_{-1.83})\times10^{-7}$ & $-21.55$ & $(1.01^{+0.04}_{-0.04})\times10^{-4}$ & $(1.00^{+0.20}_{-0.20})\times10^{-4}$ \\
$-23.19$ & $(8.05^{+1.17}_{-1.17})\times10^{-7}$ & $(3.63^{+2.30}_{-2.30})\times10^{-7}$ & $-21.30$ & $(1.79^{+0.06}_{-0.06})\times10^{-4}$ & $(1.77^{+0.38}_{-0.38})\times10^{-4}$ \\
$-22.89$ & $(1.49^{+0.16}_{-0.16})\times10^{-6}$ & $(8.07^{+4.28}_{-4.28})\times10^{-7}$ & $-21.05$ & $(2.97^{+0.11}_{-0.11})\times10^{-4}$ & $(2.93^{+0.71}_{-0.71})\times10^{-4}$ \\
$-22.64$ & $(2.91^{+0.30}_{-0.30})\times10^{-6}$ & $(1.97^{+0.92}_{-0.92})\times10^{-6}$ & $-20.80$ & $(4.18^{+0.14}_{-0.14})\times10^{-4}$ & $(4.09^{+1.14}_{-1.14})\times10^{-4}$ \\
$-22.44$ & $(6.72^{+0.51}_{-0.51})\times10^{-6}$ & $(5.67^{+2.01}_{-2.01})\times10^{-6}$ & $-20.55$ & $(4.45^{+0.17}_{-0.17})\times10^{-4}$ & $(4.31^{+1.50}_{-1.50})\times10^{-4}$ \\
$-22.29$ & $(9.54^{+1.11}_{-1.11})\times10^{-6}$ & $(9.24^{+4.10}_{-4.10})\times10^{-6}$ & $-20.30$ & $(6.29^{+0.24}_{-0.24})\times10^{-4}$ & $(6.13^{+2.52}_{-2.52})\times10^{-4}$ \\
$-22.19$ & $(1.53^{+0.15}_{-0.15})\times10^{-5}$ & $(1.53^{+0.61}_{-0.61})\times10^{-5}$ & \multicolumn{3}{c}{$z\sim6$} \\
$-22.09$ & $(1.90^{+0.16}_{-0.16})\times10^{-5}$ & $(1.89^{+0.68}_{-0.68})\times10^{-5}$ & $-25.02$ & $(1.05^{+4.11}_{-1.05})\times10^{-8}$ & \nodata \\
$-21.99$ & $(2.77^{+0.20}_{-0.20})\times10^{-5}$ & $(2.75^{+0.92}_{-0.92})\times10^{-5}$ & $-24.52$ & $(2.13^{+4.21}_{-2.13})\times10^{-8}$ & \nodata \\
$-21.89$ & $(3.70^{+0.24}_{-0.24})\times10^{-5}$ & $(3.66^{+1.13}_{-1.13})\times10^{-5}$ & $-24.02$ & $(2.77^{+4.19}_{-2.23})\times10^{-8}$ & $(6.00^{+9.95}_{-6.00})\times10^{-9}$ \\
$-21.79$ & $(5.52^{+0.30}_{-0.30})\times10^{-5}$ & $(5.44^{+1.58}_{-1.58})\times10^{-5}$ & $-23.52$ & $(8.51^{+5.38}_{-2.25})\times10^{-8}$ & $(3.76^{+2.97}_{-2.05})\times10^{-8}$ \\
$-21.69$ & $(7.49^{+0.36}_{-0.36})\times10^{-5}$ & $(7.38^{+1.89}_{-1.89})\times10^{-5}$ & $-23.12$ & $(3.34^{+0.72}_{-0.72})\times10^{-7}$ & $(1.80^{+1.09}_{-1.09})\times10^{-7}$ \\
$-21.59$ & $(1.04^{+0.04}_{-0.04})\times10^{-4}$ & $(1.03^{+0.23}_{-0.23})\times10^{-4}$ & $-22.82$ & $(1.24^{+0.15}_{-0.14})\times10^{-6}$ & $(7.59^{+4.14}_{-4.14})\times10^{-7}$ \\
$-21.49$ & $(1.29^{+0.05}_{-0.05})\times10^{-4}$ & $(1.28^{+0.26}_{-0.26})\times10^{-4}$ & $-22.52$ & $(2.67^{+0.39}_{-0.39})\times10^{-6}$ & $(1.81^{+0.96}_{-0.96})\times10^{-6}$ \\
$-21.39$ & $(1.84^{+0.06}_{-0.06})\times10^{-4}$ & $(1.82^{+0.34}_{-0.34})\times10^{-4}$ & $-22.22$ & $(4.48^{+0.53}_{-0.53})\times10^{-6}$ & $(3.46^{+1.78}_{-1.78})\times10^{-6}$ \\
$-21.29$ & $(2.24^{+0.07}_{-0.07})\times10^{-4}$ & $(2.23^{+0.38}_{-0.38})\times10^{-4}$ & $-21.92$ & $(1.10^{+0.09}_{-0.09})\times10^{-5}$ & $(9.58^{+4.82}_{-4.82})\times10^{-6}$ \\
$-21.19$ & $(2.93^{+0.08}_{-0.08})\times10^{-4}$ & $(2.92^{+0.49}_{-0.49})\times10^{-4}$ & $-21.62$ & $(3.69^{+0.48}_{-0.48})\times10^{-5}$ & $(3.55^{+1.80}_{-1.80})\times10^{-5}$ \\
$-21.09$ & $(3.88^{+0.17}_{-0.17})\times10^{-4}$ & $(3.86^{+0.67}_{-0.67})\times10^{-4}$ & $-21.32$ & $(7.35^{+0.85}_{-0.85})\times10^{-5}$ & $(7.35^{+4.05}_{-4.05})\times10^{-5}$ \\
$-20.99$ & $(4.61^{+0.19}_{-0.19})\times10^{-4}$ & $(4.58^{+0.81}_{-0.81})\times10^{-4}$ & $-21.02$ & $(1.77^{+0.21}_{-0.21})\times10^{-4}$ & $(1.77^{+1.22}_{-1.22})\times10^{-4}$ \\
$-20.89$ & $(5.19^{+0.20}_{-0.20})\times10^{-4}$ & $(5.16^{+0.93}_{-0.93})\times10^{-4}$ & \multicolumn{3}{c}{$z\sim7$} \\
$-20.79$ & $(6.39^{+0.22}_{-0.22})\times10^{-4}$ & $(6.35^{+1.16}_{-1.16})\times10^{-4}$ & $-25.42$ & $(5.64^{+14.65}_{-5.64})\times10^{-9}$ & \nodata \\
$-20.69$ & $(8.20^{+0.26}_{-0.26})\times10^{-4}$ & $(8.15^{+1.51}_{-1.51})\times10^{-4}$ & $-24.92$ & $(8.89^{+15.29}_{-7.41})\times10^{-9}$ & \nodata \\
$-20.59$ & $(9.14^{+0.27}_{-0.27})\times10^{-4}$ & $(9.08^{+1.72}_{-1.72})\times10^{-4}$ & $-24.42$ & $(2.41^{+1.75}_{-0.98})\times10^{-8}$ & $(5.00^{+24.81}_{-5.00})\times10^{-10}$ \\
$-20.49$ & $(1.04^{+0.03}_{-0.03})\times10^{-3}$ & $(1.03^{+0.20}_{-0.20})\times10^{-3}$ & $-23.92$ & $(9.02^{+2.13}_{-1.66})\times10^{-8}$ & $(1.31^{+2.46}_{-1.31})\times10^{-8}$ \\
$-20.39$ & $(1.20^{+0.03}_{-0.03})\times10^{-3}$ & $(1.19^{+0.24}_{-0.24})\times10^{-3}$ & $-23.42$ & $(1.62^{+0.30}_{-0.23})\times10^{-7}$ & $(4.39^{+6.02}_{-4.39})\times10^{-8}$ \\
$-20.29$ & $(1.47^{+0.04}_{-0.04})\times10^{-3}$ & $(1.46^{+0.31}_{-0.31})\times10^{-3}$ & $-22.92$ & $(4.63^{+7.53}_{-2.71})\times10^{-7}$ & $(1.83^{+3.62}_{-1.83})\times10^{-7}$ \\
$-20.19$ & $(1.60^{+0.04}_{-0.04})\times10^{-3}$ & $(1.59^{+0.36}_{-0.36})\times10^{-3}$ & $-22.42$ & $(1.95^{+1.13}_{-0.67})\times10^{-6}$ & $(1.06^{+1.20}_{-1.06})\times10^{-6}$ \\
$-20.09$ & $(1.79^{+0.04}_{-0.04})\times10^{-3}$ & $(1.78^{+0.44}_{-0.44})\times10^{-3}$ & $-21.92$ & $(3.47^{+1.70}_{-1.03})\times10^{-6}$ & $(2.75^{+2.77}_{-2.55})\times10^{-6}$ \\
$-19.99$ & $(2.03^{+0.05}_{-0.05})\times10^{-3}$ & $(2.02^{+0.54}_{-0.54})\times10^{-3}$ & \nodata & \nodata & \nodata 
\enddata
\tablenotetext{$\dagger$}{\redc{The galaxy luminosity function derived by using Equation (\ref{eq_LFgal}) with the spectroscopic galaxy fractions presented in Figures \ref{fig_LFall_1} and \ref{fig_LFall_2}.}}
\label{tab_UVLF}
\end{deluxetable*}

Our derived luminosity functions at $z\sim5$ and $6$ are shown in the bottom panel of Figure \ref{fig_LFall_1} and the top panel of Figure \ref{fig_LFall_2}, respectively.
Similar to the $z\sim4$ result, our measurements agree well with previous studies of quasars at $M_\m{UV}\lesssim-24$ mag \citep[e.g.,][]{2020ApJ...904...89N,2018ApJ...869..150M}, studies of galaxies at $M_\m{UV}\gtrsim-22$ mag \citep[e.g.,][]{2021arXiv210207775B,2015ApJ...810...71F}, and studies of galaxies and AGNs \citep{2018PASJ...70S..10O}.
At $z\sim7$, our derived luminosity function agrees with previous studies \citep[e.g.,][]{2018PASJ...70S..10O,2017MNRAS.466.3612B}, as shown in the bottom panel of Figure \ref{fig_LFall_2}.
Table \ref{tab_UVLF} summarizes our measurements of the luminosity functions at $z\sim4-7$.

These agreements clearly indicate that the dropout luminosity function is a superposition of the AGN luminosity function (dominant at $M_\m{UV}<-24$ mag) and the galaxy luminosity function (dominant at $M_\m{UV}>-22$ mag).
In our dropout selection, we probe redshifted Lyman break features of high redshift galaxies. 
However, high redshift AGNs also have similar Lyman break features. 
Thus it is expected that our dropout sample is composed of both galaxies and AGNs. 
Indeed, as described in Section \ref{ss_spec_source}, our dropout samples include both spectroscopically identified galaxies and AGNs. 
Based on our spectroscopic results and the literature, we derive the galaxy fraction which is the number of spectroscopically confirmed high redshift galaxies divided by the sum of the numbers of spectroscopically confirmed galaxies and AGNs.
The derived galaxy fractions for our $z\sim4-7$ samples in each magnitude bin are presented in Figures \ref{fig_LFall_1} and \ref{fig_LFall_2}. 
As shown in Figures \ref{fig_LFall_1} and \ref{fig_LFall_2}, the galaxy fractions at $z\sim4-7$ are about $0${\%} at $M_{\rm UV}<-24$ mag, but then increase with increasing magnitude and reach about $100${\%} at $M_{\rm UV}>-22$ mag. 
These results further suggest that our luminosity functions are dominated by AGNs at the bright end, and by galaxies at the faint end. 
The very wide area and deep depth of the HSC-SSP survey allow us to bridge the UV luminosity functions of high redshift galaxies and AGNs, both of which can be selected with redshifted Lyman break features \citep{2018PASJ...70S..10O}.  

\subsubsection{Fitting the Dropout Luminosity Functions}\label{ss_LFall_fit}

\begin{figure*}
\centering
\includegraphics[width=0.59\hsize, bb=9 2 909 368]{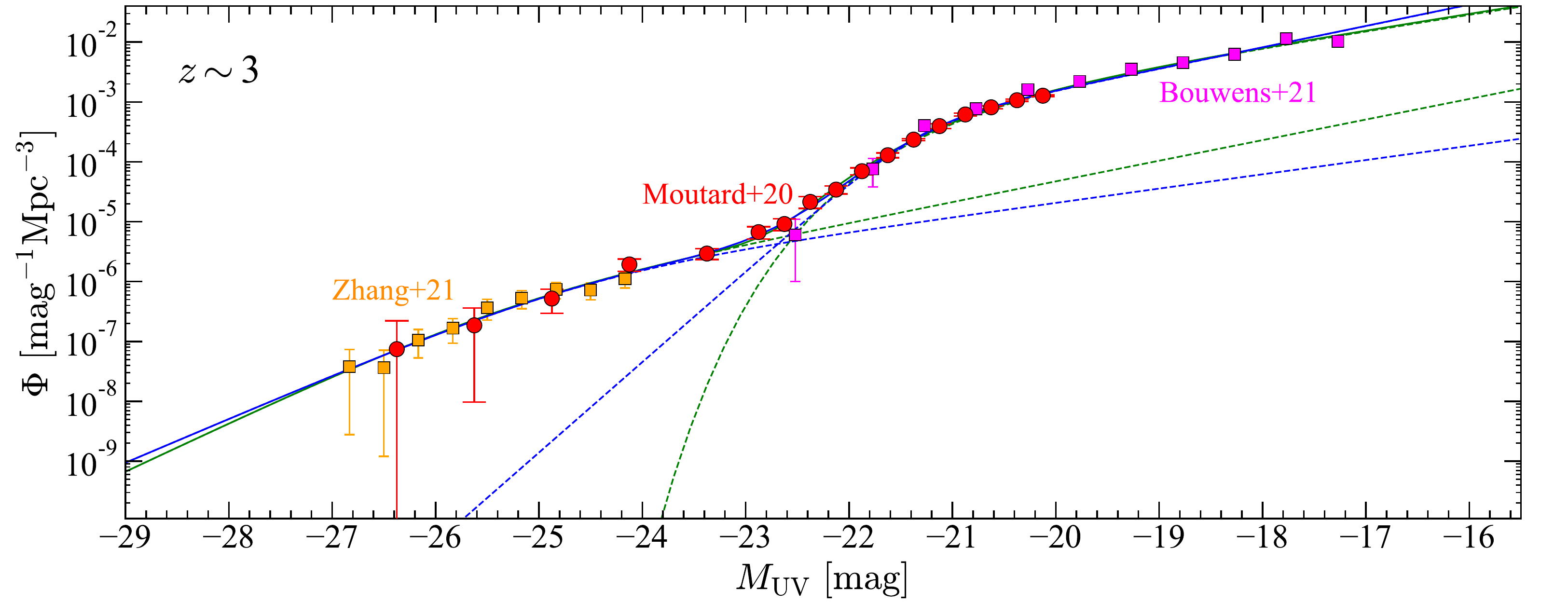}\\\includegraphics[width=0.59\hsize, bb=9 2 909 368]{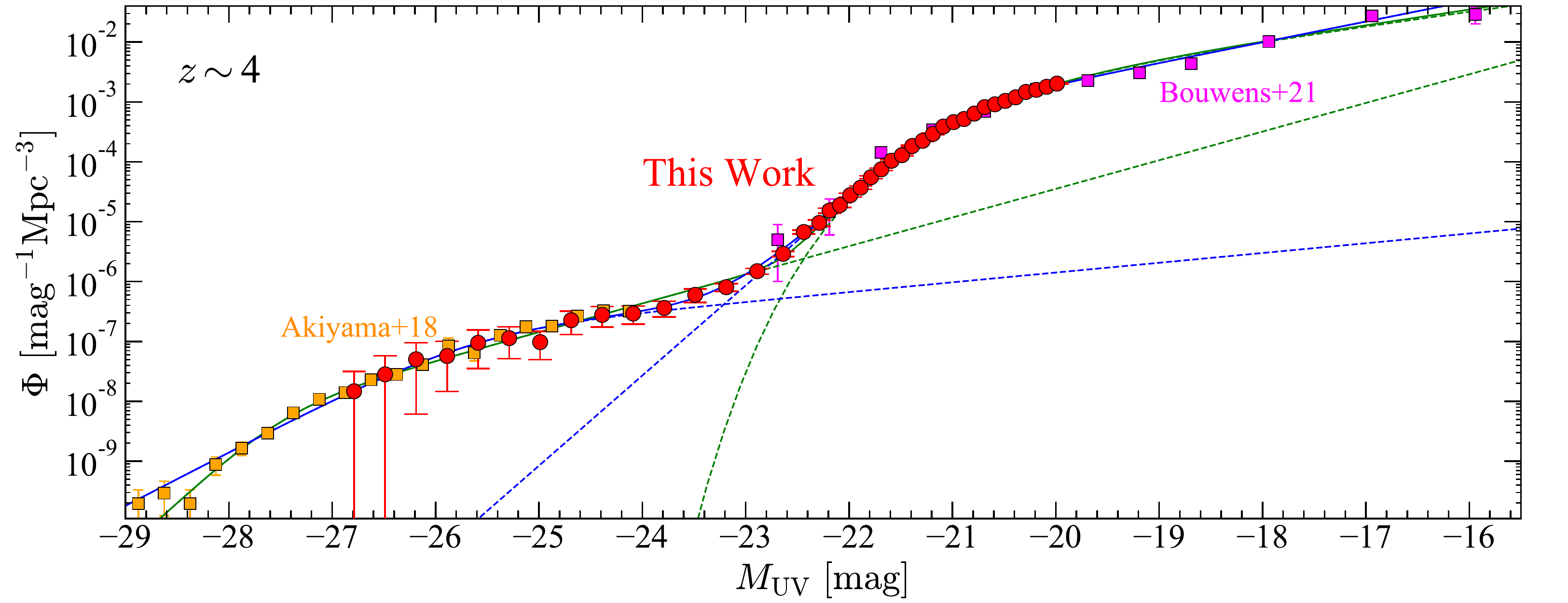}
\includegraphics[width=0.59\hsize, bb=9 2 909 368]{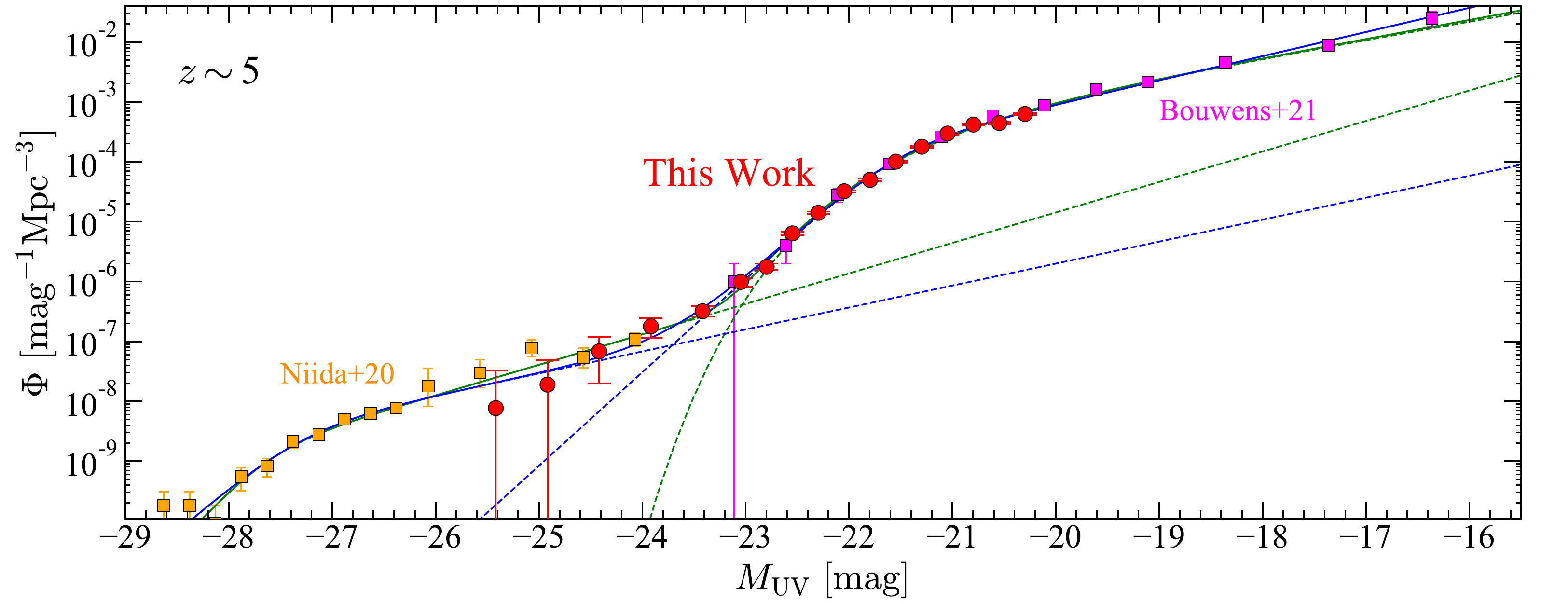}
\includegraphics[width=0.59\hsize, bb=9 2 909 368]{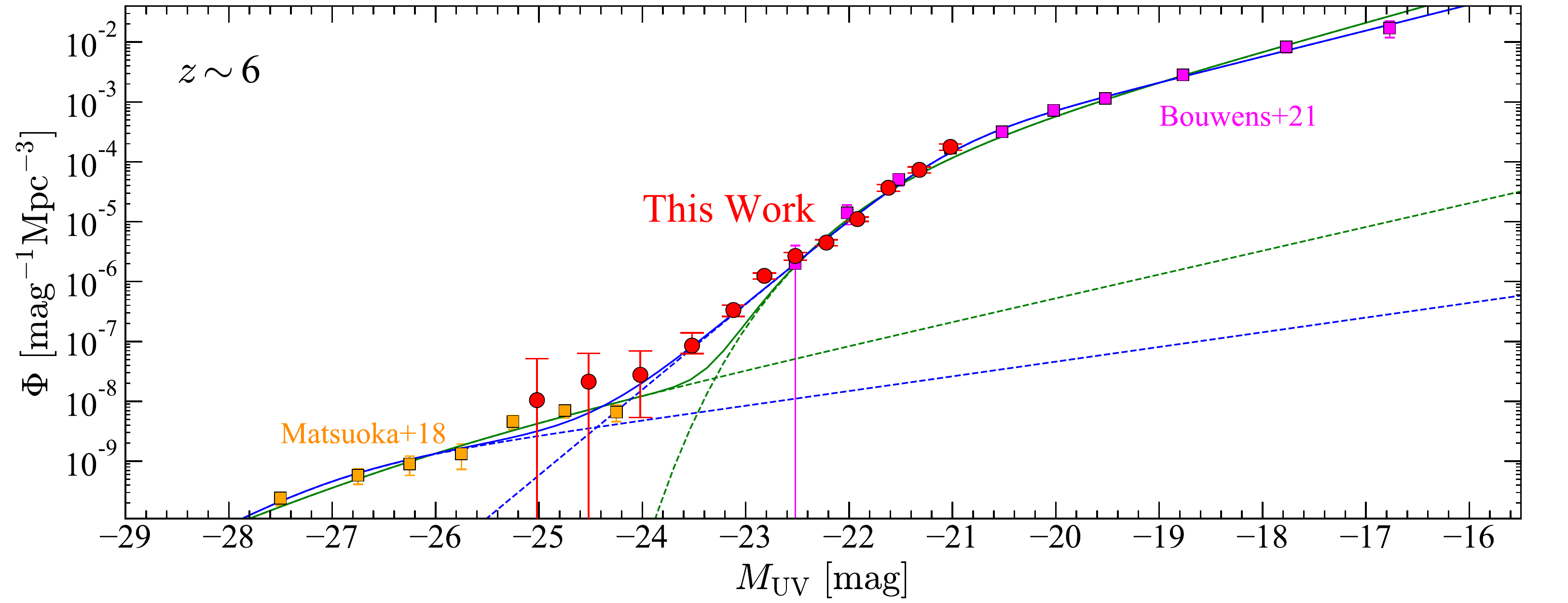}
\includegraphics[width=0.59\hsize, bb=9 2 909 368]{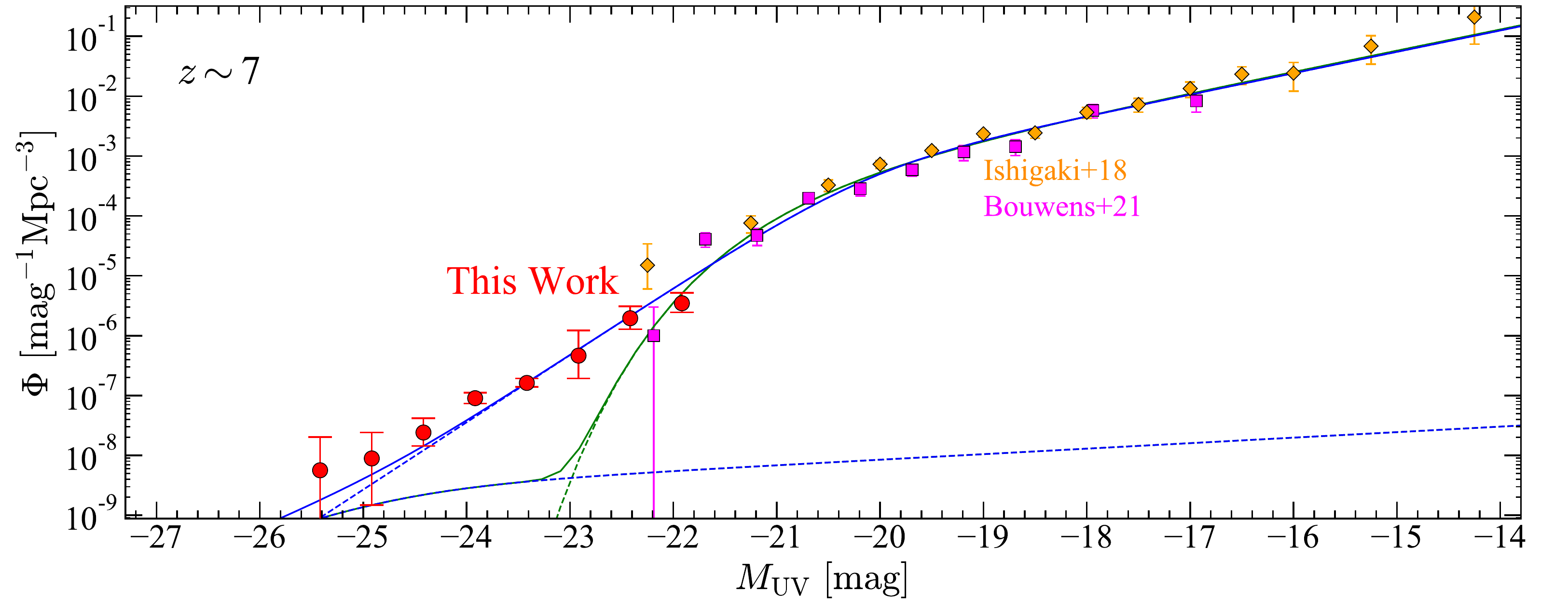}
\caption{Rest-frame UV luminosity functions of all rest-UV selected sources (galaxies+AGNs) at $z\sim3$, $4$, $5$, $6$, and $7$ from top to bottom.
The red circles show our results based on the HSC-SSP survey data at $z\sim4-7$ and results in \citet{2020MNRAS.494.1894M} at $z\sim3$.
The magenta squares and orange diamonds are results for galaxies taken from \citet{2021arXiv210207775B} and \citet{2018ApJ...854...73I}, respectively.
The orange squares are results for AGNs in \citet{2021arXiv210511497Z}, \citet{2018PASJ...70S..34A}, \citet{2020ApJ...904...89N}, and \citet{2018ApJ...869..150M} at $M_\m{UV}<-24$ mag.
The blue (green) lines show the best-fit DPL+DPL (DPL+Schechter) functions.
Note that at $z\sim7$, we fix the parameters of the AGN luminosity function to values in \citet{2018ApJ...869..150M} with decreasing the $\phi^*$ parameter by 0.7 dex following $\phi^*\propto10^{-0.7(z-6)}$ as assumed in \citet{2018ApJ...869..150M}.
}
\label{fig_LFall_wfit}
\end{figure*}

We investigate the shape of the UV luminosity functions of dropout sources (galaxies+AGNs) by fitting them with several functional forms.
Figure \ref{fig_LFall_wfit} shows our UV luminosity functions at $z\sim4-7$ with previous results of galaxies based on the {\it Hubble} data \citep{2021arXiv210207775B,2018ApJ...854...73I} and of quasars based on the HSC-SSP and Sloan Digital Sky Survey (SDSS) data \citep{2018PASJ...70S..34A,2020ApJ...904...89N,2018ApJ...869..150M}.
The combination of our results with the previous work reveals the UV luminosity functions in a very wide magnitude range of $-29 \lesssim M_{\rm UV} \lesssim -14$ mag, corresponding to the luminosity range of $0.002L^*_\m{UV}\lesssim L_\m{UV}\lesssim 2000L^*_\m{UV}$.
We also show luminosity functions at $z\sim3$ taken from \citet{2020MNRAS.494.1894M}, \citet{2021arXiv210207775B}, and \citet{2021arXiv210511497Z}.
As discussed in Section \ref{ss_LFall_res}, the dropout luminosity function is a superposition of the AGN luminosity function and the galaxy luminosity function.
Thus we simultaneously fit the AGN and galaxy luminosity functions.
For the AGN luminosity function, we fit with a double-power law (DPL) function that is widely used in studies of AGNs:
\begin{equation}\label{eq_dpl_L}
\phi(L) dL 
= \phi^* \left[ \left( \frac{L}{L^*} \right)^{-\alpha} + \left( \frac{L}{L^*} \right)^{-\beta}   \right]^{-1} \frac{dL}{L^*}, 
\end{equation}
where $\phi^*$ is the overall normalization, $L^*$ is the characteristic luminosity, and 
$\alpha$ and $\beta$ are the faint and bright-end power-law slopes, respectively. 
We define a DPL function as a function of absolute magnitude, $\Phi (M_{\rm UV})$, as $\phi(L)dL = \Phi (M_{\rm UV}) dM_{\rm UV}$, 
\begin{eqnarray}
&&\Phi(M_{\rm UV}) = \frac{\ln 10}{2.5} \phi^* \nonumber \\
&&\times \left[10^{0.4(\alpha+1)(M_{\rm UV} - M_{\rm UV}^*)} + 10^{0.4(\beta+1)(M_{\rm UV} - M_{\rm UV}^*)} \right]^{-1},
\label{eq_dpl}
\end{eqnarray}
where $M^*_{\rm UV}$ is the characteristic magnitude.
For the galaxy luminosity function, we fit with a DPL function or the Schechter function \citep{1976ApJ...203..297S}:
\begin{equation}
\phi(L) dL 
	= \phi^* \left( \frac{L}{L^*} \right)^\alpha 
		\exp \left( - \frac{L}{L^*} \right) d \left( \frac{L}{L^*} \right), 
\end{equation} 
where $\phi^*$, $L^*_{\rm UV}$, and $\alpha$ are  the overall normalization, the characteristic luminosity, and the faint power-law slope, respectively.
We define the Schechter function as a function of absolute magnitude, $\Phi(M_{\rm UV})$, as $\phi(L) dL = \Phi(M_{\rm UV}) dM_{\rm UV}$,
\begin{eqnarray}
\Phi(M_{\rm UV}) 
	&=& \frac{\ln 10}{2.5} \phi^* 10^{-0.4 (M_{\rm UV} - M_{\rm UV}^*) (\alpha +1)} \nonumber \\
	&& \times \exp \left( - 10^{-0.4 (M_{\rm UV} - M_{\rm UV}^*)} \right).
\label{eq_schechter}
\end{eqnarray}

Figure \ref{fig_LFall_wfit} shows the best-fit results in cases of the DPL+DPL and DPL+Schechter functions at $z\sim3-7$.
Note that in the fitting at $z\sim7$, we fix the parameters of the AGN luminosity function to values in \citet{2018ApJ...869..150M} with decreasing the $\phi^*$ parameter by 0.7 dex following $\phi^*\propto10^{-0.7(z-6)}$ as assumed in \citet{2018ApJ...869..150M}, because there are no measurements of the AGN luminosity function at $z\sim7$.
The luminosity functions in the very wide luminosity range of $-29 \lesssim M_{\rm UV} \lesssim -14$ mag are well fitted with either the DPL+DPL or DPL+Schechter functions, as partly shown at $z\sim4$ in previous studies \citep{2018ApJ...863...63S,2020MNRAS.494.1771A}, except for $z\sim7$ where the DPL+DPL functions provide a better fit.
Table \ref{tab_LFfit} summarizes the best-fit parameters and reduced $\chi^2$ values of the two cases.
We find that the best-fit parameters are roughly comparable to those of galaxy and AGN luminosity functions in previous studies \citep[e.g.,][]{2018PASJ...70S..10O,2018PASJ...70S..34A,2020ApJ...904...89N,2018ApJ...869..150M}.

\setlength{\tabcolsep}{0.1cm}
\begin{deluxetable*}{cccccccccccc}
\tabletypesize{\scriptsize}
\renewcommand{\arraystretch}{0.9}
\tablecaption{Fit Parameters for Luminosity Functions of All Rest-UV Selected Sources  (Galaxy+AGN)}
\tablehead{
\colhead{}& \colhead{} & \multicolumn{4}{c}{AGN Component} & & \multicolumn{4}{c}{Galaxy Component} & \colhead{}\\
\cline{3-6} \cline{8-11} \\ 
\colhead{Redshift} & \colhead{Fitted Function} & \colhead{$M_\m{UV}^*$} & \colhead{$\m{log}\phi^*$} & \colhead{$\alpha$} & \colhead{$\beta$} & & \colhead{$M_\m{UV}^*$} & \colhead{$\m{log}\phi^*$} & \colhead{$\alpha$} & \colhead{$\beta$} &\colhead{$\chi^2/\m{dof}$} \\
\colhead{}& \colhead{}& \colhead{($\m{Mpc^{-3}}$)}& \colhead{(mag)} &  \colhead{} & \colhead{} & & \colhead{($\m{Mpc^{-3}}$)}& \colhead{(mag)} &  \colhead{} & \colhead{} & \colhead{}}
\startdata
$z\sim3^*$ & DPL+DPL & $-24.59^{+0.25}_{-0.50}$ & $-5.74^{+0.13}_{-0.26}$ & $-1.59^{+0.41}_{-0.29}$ & $-2.84^{+0.41}_{-0.17}$ & & $-21.30^{+0.05}_{-0.05}$ & $-3.23^{+0.04}_{-0.04}$ & $-1.89^{+0.02}_{-0.02}$ & $-4.78^{+0.19}_{-0.23}$ & $52.6/29$\\
 & DPL+Schechter & $-25.31^{+1.01}_{-0.44}$ & $-6.11^{+0.50}_{-0.24}$ & $-1.86^{+0.43}_{-0.37}$ & $-3.04^{+0.19}_{-0.19}$ & & $-20.91^{+0.12}_{-0.08}$ & $-2.84^{+0.06}_{-0.05}$ & $-1.68^{+0.08}_{-0.04}$ & \nodata & $55.3/30$\\
$z\sim4$ & DPL+DPL & $-25.69^{+0.19}_{-0.10}$ & $-6.74^{+0.10}_{-0.06}$ & $-1.41^{+0.17}_{-0.09}$ & $-3.24^{+0.06}_{-0.07}$ & & $-20.99^{+0.03}_{-0.03}$ & $-3.00^{+0.02}_{-0.02}$ & $-1.86^{+0.03}_{-0.02}$ & $-4.77^{+0.05}_{-0.05}$ & $127.6/64$\\
 & DPL+Schechter & $-27.49^{+0.26}_{-0.08}$ & $-8.00^{+0.20}_{-0.06}$ & $-2.20^{+0.06}_{-0.03}$ & $-5.05^{+0.57}_{-0.24}$ & & $-20.49^{+0.03}_{-0.02}$ & $-2.52^{+0.02}_{-0.02}$ & $-1.59^{+0.03}_{-0.03}$ & \nodata & $115.0/65$\\
$z\sim5$ & DPL+DPL & $-27.32^{+0.76}_{-0.26}$ & $-8.35^{+0.47}_{-0.18}$ & $-1.92^{+0.31}_{-0.17}$ & $-4.77^{+0.61}_{-0.62}$ & & $-21.54^{+0.04}_{-0.04}$ & $-3.63^{+0.04}_{-0.03}$ & $-2.01^{+0.04}_{-0.03}$ & $-4.91^{+0.08}_{-0.08}$ & $100.9/36$\\
 & DPL+Schechter & $-27.67^{+1.47}_{-0.88}$ & $-8.71^{+0.89}_{-0.67}$ & $-2.27^{+0.48}_{-0.22}$ & $-5.92^{+0.66}_{-1.14}$ & & $-21.09^{+0.04}_{-0.03}$ & $-3.16^{+0.03}_{-0.03}$ & $-1.76^{+0.04}_{-0.03}$ & \nodata & $104.1/37$\\
$z\sim6$ & DPL+DPL & $-27.05^{+0.62}_{-0.44}$ & $-9.03^{+0.28}_{-0.19}$ & $-1.61^{+0.27}_{-0.23}$ & $-3.41^{+0.30}_{-0.40}$ & & $-21.03^{+0.09}_{-0.08}$ & $-3.52^{+0.09}_{-0.07}$ & $-2.08^{+0.07}_{-0.06}$ & $-4.57^{+0.09}_{-0.10}$ & $41.5/22$\\
 & DPL+Schechter & $-26.53^{+0.51}_{-0.74}$ & $-8.83^{+0.20}_{-0.41}$ & $-1.99^{+0.88}_{-0.36}$ & $-2.90^{+0.14}_{-0.21}$ & & $-21.22^{+0.15}_{-0.12}$ & $-3.65^{+0.16}_{-0.13}$ & $-2.19^{+0.08}_{-0.06}$ & \nodata & $97.4/23$\\
$z\sim7$ & DPL+DPL & $(-24.90)^\dagger$ & $(-8.49)\dagger$ & $(-1.23)\dagger$ & $(-2.73)\dagger$ & & $-20.12^{+0.21}_{-0.24}$ & $-3.05^{+0.15}_{-0.16}$ & $-1.89^{+0.10}_{-0.09}$ & $-3.81^{+0.10}_{-0.13}$ & $58.6/24$\\
 & DPL+Schechter & $(-24.90)\dagger$ & $(-8.49)\dagger$ & $(-1.23)\dagger$ & $(-2.73)\dagger$ & & $-20.49^{+0.12}_{-0.10}$ & $-3.14^{+0.10}_{-0.09}$ & $-1.88^{+0.07}_{-0.06}$ & \nodata & $105.2/25$
\enddata
\tablenotetext{$*$}{The $z\sim3$ values are based on our fitting for results in \citet{2020MNRAS.494.1894M}, \citet{2021arXiv210207775B}, and \citet{2021arXiv210511497Z}.}
\tablenotetext{$\dagger$}{The value in parenthesis is fixed to results in \citet{2018ApJ...869..150M} assuming the redshift evolution of $\phi^*\propto10^{-0.7(z-6)}$ to $z=6.8$.}
\label{tab_LFfit}
\end{deluxetable*}

\begin{figure*}
\centering
\includegraphics[width=0.95\hsize, bb=22 12 909 500]{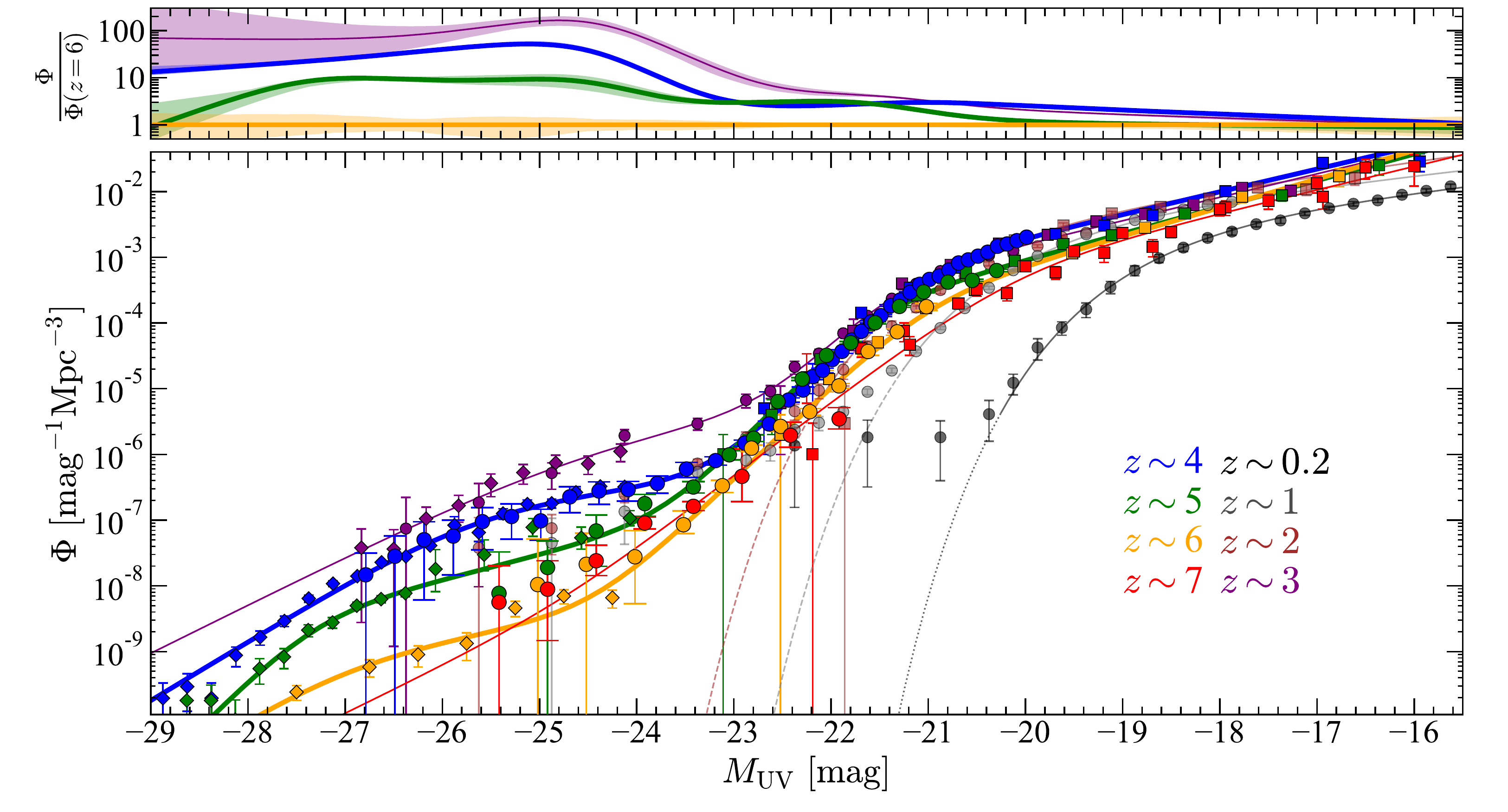}
\caption{Evolution of the rest-frame UV luminosity functions of all rest-UV selected sources (galaxies+AGNs) from $z\sim7$ to $z\sim0$.
The bottom panel shows the luminosity functions at $z\sim0-7$, and the black, grey, brown, purple, blue, green, orange, and red symbols show results at $z\sim0.2$, $1$, $2$, $3$, $4$, $5$, $6$, and $7$, respectively.
The circles at $z\sim4-7$ show our results based on the HSC-SSP survey data, and those at $z\sim0-3$ are taken from \citet{2020MNRAS.494.1894M} at $0.05<z<0.3$, $0.9<z<1.3$, $1.8<z<2.5$, and $2.5<z<3.5$.
The diamonds are results for AGNs in \citet{2021arXiv210511497Z}, \citet{2018PASJ...70S..34A}, \citet{2020ApJ...904...89N}, and \citet{2018ApJ...869..150M} at $z\sim3$, $4$, $5$, and $6$, respectively, and the squares show results for galaxies in \citet{2021arXiv210207775B} and \citet{2018ApJ...854...73I}.
The lines show the best-fit DPL+DPL functions at $z\sim3-7$ (this study), and Schechter functions at $z\sim0-2$ (\citealt{2020MNRAS.494.1894M}). 
The top panel shows ratios of the number densities at $z\sim3-5$ relative to those at $z\sim6$.
The shaded regions correspond to the $2\sigma$ uncertainties of the luminosity function parameters.
The number density of typical galaxies ($M_\m{UV}=M^*_\m{UV}\sim-21$ mag) increases only by a factor of $\sim3$ from $z\sim6$ to $3$, while the number density of typical quasars ($M_\m{UV}=M^*_\m{UV}\sim-27$ mag) significantly increases by a factor of $\sim100$.
}
\label{fig_LFall_evol}
\end{figure*}

\subsubsection{Redshift Evolution}\label{ss_LFall_evol}

Figure \ref{fig_LFall_evol} summarizes UV luminosity function estimates at $z\sim4-7$ in this work and the literature, and the best-fit DPL+DPL functions.
We also plot the rest-frame UV luminosity functions at $z\sim0-3$ from \citet{2020MNRAS.494.1894M}.
Like our $z\sim4-7$ results, the luminosity functions at $z\sim0-3$ also show number density excesses at the bright end compared to the Schechter functions, which are dominated by AGNs (see discussions in \citealt{2020MNRAS.494.1894M}).
Indeed, the number densities of bright sources ($M_\m{UV}\lesssim -23$ mag) at $z\sim3$ are comparable to the rest-frame UV luminosity function of spectroscopically identified AGNs in \citet{2021arXiv210511497Z}.
Interestingly, the number density of typical galaxies ($M_\m{UV}=M^*_\m{UV}\sim-21$ mag) increases only by a factor of $\sim3$ from $z\sim6$ to $3$, while the number density of typical quasars ($M_\m{UV}=M^*_\m{UV}\sim-27$ mag) significantly increases by a factor of $\sim100$, consistent with previous studies of the quasar luminosity functions \citep[e.g.,][]{2018ApJ...869..150M,2020ApJ...904...89N}.
This indicates a very rapid growth of AGNs in the first 1.5 Gyr.
If we extrapolate this evolution to higher redshift by assuming $\Phi\propto10^{-0.7(1+z)}$ \citep{2018ApJ...869..150M}, the number density of bright quasars ($M_\m{UV}\lesssim-26$ mag) will be very small ($\lesssim2\times10^{-10}\ \m{mag^{-1}\ \m{Mpc^{-3}}}$) at $z\geq7$. 
More specifically, the number density of typical quasars increases by a factor of 10 from $z\sim6$ to $5$, but increases only by a factor of 3 from $z\sim4$ to $3$, indicating the accelerated evolution of the quasar luminosity function at $z\sim3-6$ \citep{2020ApJ...904...89N}.

\subsection{Galaxy UV Luminosity Function}\label{ss_galLF}

\subsubsection{Derivation and Results}\label{ss_galLF_res}

\begin{figure}
\centering
\includegraphics[width=0.95\hsize, bb=9 7 350 206]{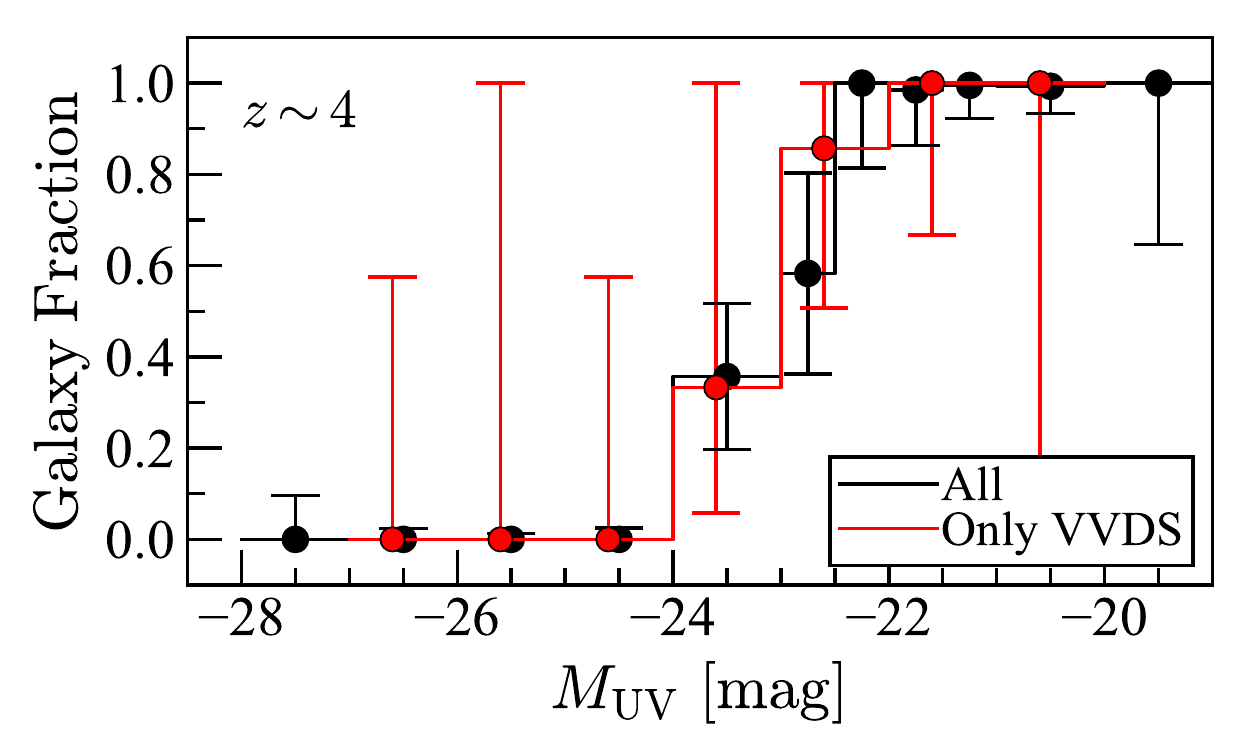}
\caption{\redc{Spectroscopic galaxy fraction at $z\sim4$.
The black circles are those estimated by using the all available spectroscopic catalogs (same as the top panel of Figure \ref{fig_LFall_1}), and the red circles are those estimated with the VVDS data \citep{2013A&A...559A..14L}.}
}
\label{fig_galFrac}
\end{figure}

We estimate the galaxy UV luminosity functions in a wide magnitude range by considering the contributions from AGNs in our dropout luminosity function measurements.  
To subtract the AGN contributions, we use the galaxy fraction estimates based on the spectroscopy  shown in Figures \ref{fig_LFall_1} and \ref{fig_LFall_2}.
We multiply the dropout (galaxy+AGN) UV luminosity functions by the spectroscopic galaxy fractions, $f_\m{galaxy}$, and obtain the galaxy luminosity functions, $\Phi_\m{galaxy}$:
\begin{equation}\label{eq_LFgal}
\Phi_\m{galaxy}(M_\m{UV})=f_\m{galaxy}(M_\m{UV})\Phi(M_\m{UV}).
\end{equation} 
\redc{Note that our galaxy fraction estimates are based on various spectroscopic catalogs, and it is possible that the estimates are biased due to a variety of different spectroscopic selection functions.
In order to check this possibility, we compares the estimated galaxy fractions at $z\sim4$ with those derived from the VVDS data \citep{2013A&A...559A..14L}, whose targets are purely selected based on their $i$-band magnitude.
As shown in Figure \ref{fig_galFrac}, the estimated galaxy fractions are consistent with each other  within the errors, indicating that our estimates are not significantly biased.
Future large spectroscopic surveys such as Subaru/Prime Focus Spectrograph (PFS) will allow us to acculately determine the spectroscopic galaxy fractions.}

\begin{figure*}
\centering
\includegraphics[width=0.6\hsize, bb=12 2 769 358]{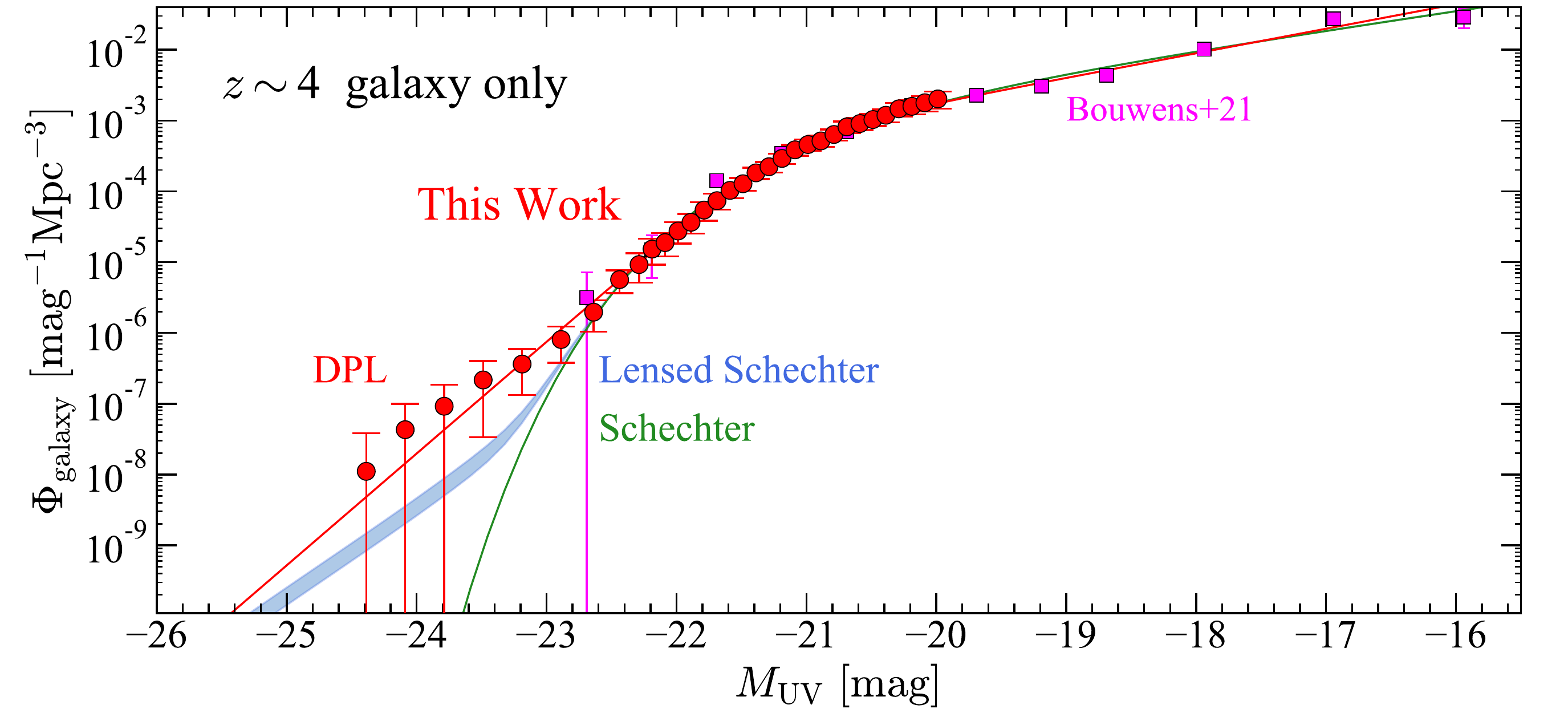}
\includegraphics[width=0.6\hsize, bb=12 2 769 368]{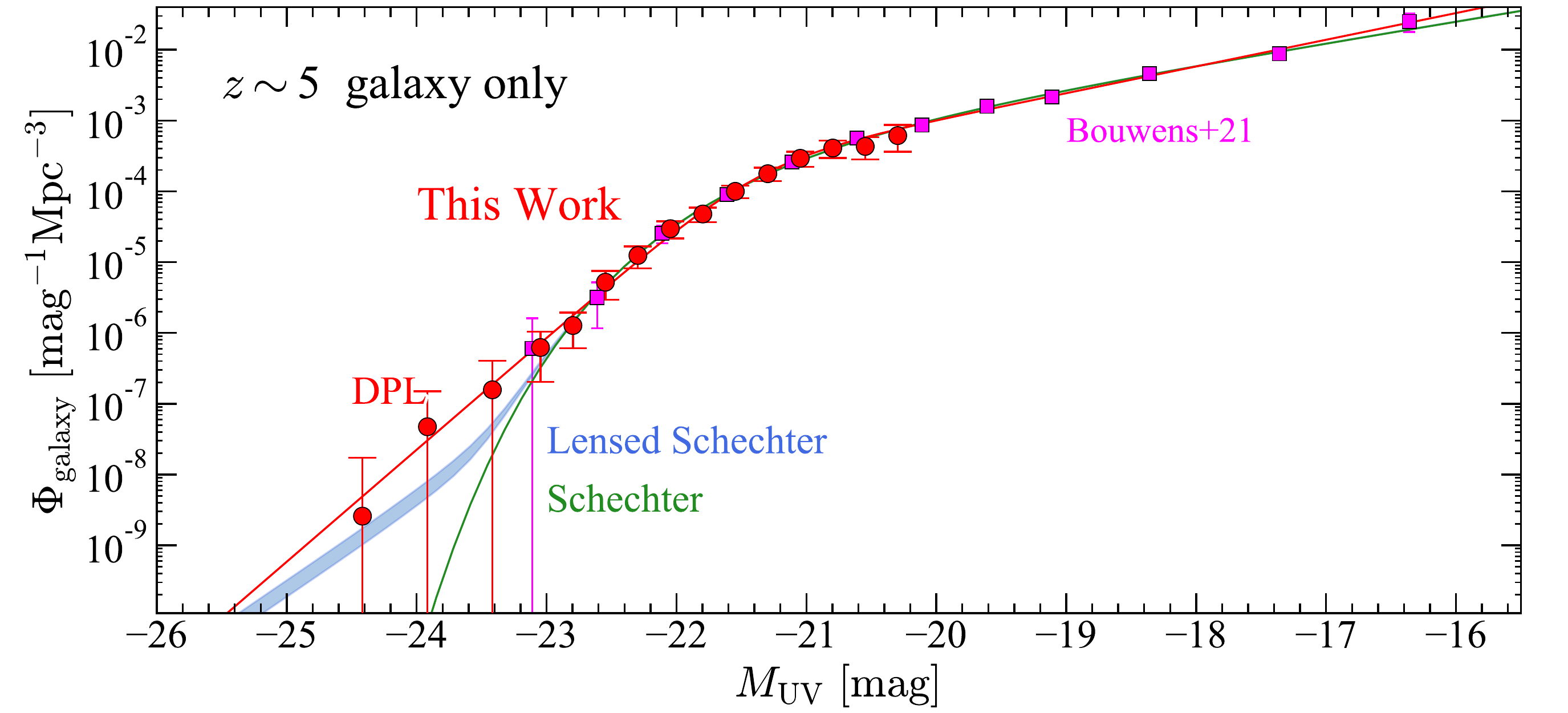}
\includegraphics[width=0.6\hsize, bb=12 2 769 358]{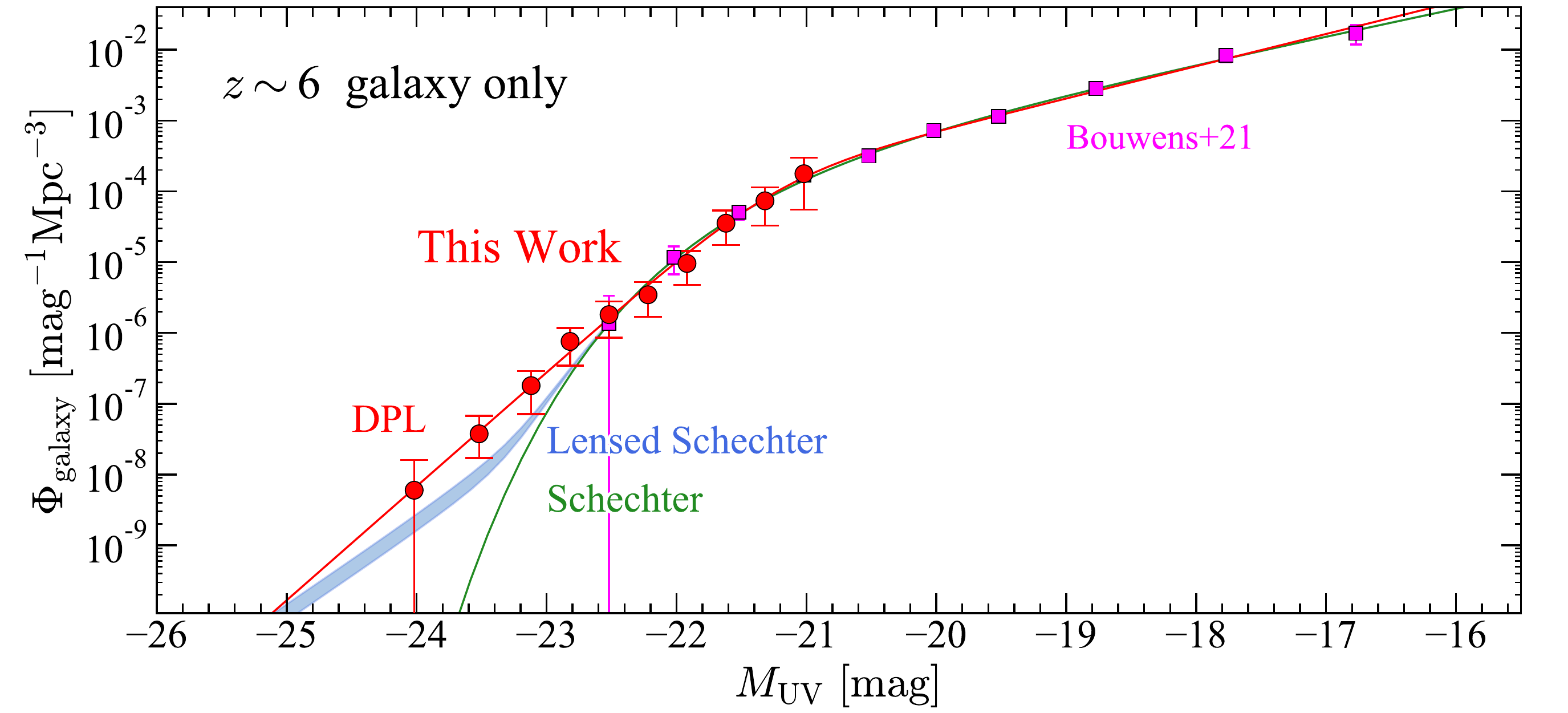}
\includegraphics[width=0.6\hsize, bb=12 2 769 368]{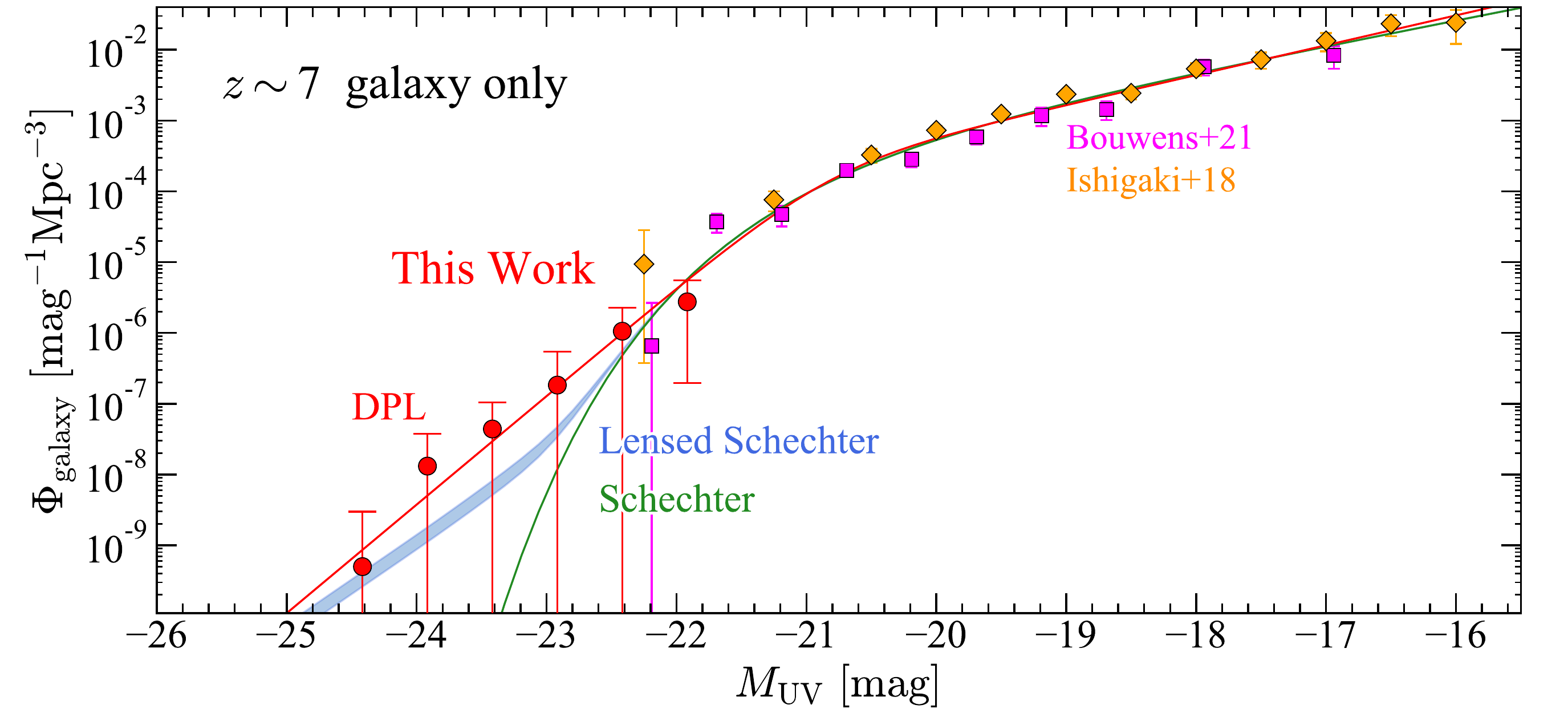}
\caption{Rest-frame UV luminosity functions of galaxies that take into account AGN fraction correction at $z\sim4$, $5$, $6$, and $7$ from top to bottom.
The red line is the best-fit DPL function, and the green line is the best-fit Schechter function without considering the lensing effect.
The blue shaded region corresponds to the lensed Schechter functions with the two cases of optical depth estimates (\citealt{2011ApJ...742...15T} and \citealt{2015MNRAS.450.1224B}). 
The red circles show our results based on the HSC-SSP survey data.
The magenta squares and orange diamonds are results for galaxies taken from \citet{2021arXiv210207775B} and \citet{2018ApJ...854...73I}, respectively.}
\label{fig_LFgal_wfit}
\end{figure*}

Figure \ref{fig_LFgal_wfit} and Table \ref{tab_UVLF} show our estimates of the galaxy UV luminosity functions at $z \sim4-7$. 
We confirm that our results are consistent with the previous results in the UV magnitude range fainter than $-22$ mag.
This is because the number density of AGNs are negligibly small compared to that of galaxies in this magnitude range. 
In the brighter magnitude range of $M_{\rm UV} < -23$ mag, our estimates at $z \sim 4-7$ appear to have bright end excesses of number densities compared to the exponential decline, although the uncertainties are large. 
The number densities of bright galaxies at $z\sim6$ are determined more precisely than those at $z\sim5$ thanks to the rich spectroscopic data at $z\sim6$ mainly taken in the SHELLQs project.
Note that the effect of the Eddington bias \citep{1913MNRAS..73..359E} should be small on these bright end excesses, because their magnitude ranges are much brighter than the limiting magnitudes, as discussed in \citet{2018PASJ...70S..10O}.

\subsubsection{Fitting the Galaxy Luminosity Function}\label{ss_galLF_fit}

To characterize the derived galaxy UV luminosity functions, we compare our estimates with the following three functions, a DPL function, the Schechter function, and a lensed Schechter function. 
The forms of the DPL and Schechter functions are already presented in Equations (\ref{eq_dpl}) and (\ref{eq_schechter}).
The lensed Schechter function is a modified Schechter function that considers the effect of gravitational lens magnification by foreground sources (e.g., \citealt{2011Natur.469..181W}; \citealt{2011ApJ...742...15T}; \citealt{2015ApJ...805...79M}; \citealt{2015MNRAS.450.1224B}). 
To take into account the magnification effect on the observed shape of the galaxy UV luminosity functions, we basically follow the method presented by \citet{2011Natur.469..181W} and \citet{2018PASJ...70S..10O}. 
A gravitationally lensed Schechter function can be estimated from the convolution between the intrinsic Schechter function and the magnification distribution of a Singular Isothermal Sphere (SIS), $dP/d\mu$, weighted by the strong lensing optical depth $\tau_{\rm m}$, which is the fraction of strongly lensed random lines of sight.  
The overall magnification distribution can be modeled by using the probability distribution function for magnification of multiply lensed sources over a fraction $\tau_{\rm m}$ of the sky. 
To conserve total flux on the cosmic sphere centered on an observer, we need to consider the de-magnification of unlensed sources:  
\begin{equation}
\mu_{\rm demag}
	= \frac{1 - \left\langle \mu_{\rm mult} \right\rangle \tau_{\rm m}}{1 - \tau_{\rm m}}, 
\end{equation} 
where $\left\langle \mu_{\rm mult} \right\rangle = 4$ is the mean magnification of multiply lensed sources. 
For a given luminosity function $\phi(L)$, a gravitationally lensed luminosity function $\phi_{\rm lensed}(L)$ can be obtained by 
\begin{eqnarray}
\phi_{\rm lensed} (L)
	&=& (1 - \tau_{\rm m}) \frac{1}{\mu_{\rm demag}} \phi \left( \frac{L}{\mu_{\rm demag}} \right) \nonumber \\
	&&\hspace{-10mm} + \tau_{\rm m} \int^\infty_0 d \mu \frac{1}{\mu} 
			\left( \frac{dP_{\rm m,1}}{d\mu} + \frac{dP_{\rm m,2}}{d\mu} \right) \phi\left( \frac{L}{\mu} \right), 
\end{eqnarray}
where 
\begin{equation}
\frac{dP_{\rm m,1}}{d\mu}
= 
\left\{ 
	\begin{array}{ll}
		\frac{2}{(\mu - 1)^3} & ({\rm for} \,\,\, \mu > 2) \\
		0 & ({\rm for} \,\,\, 0< \mu < 2)
	\end{array}
\right. 
\end{equation}
is the magnification distribution as a function of magnification factor $\mu$ for the brighter image in a strongly lensed system given for an SIS and 
\begin{equation}
\frac{dP_{\rm m,2}}{d\mu}
	= \frac{2}{(\mu + 1)^3} \,\,\, ({\rm for} \,\,\, \mu > 0) 
\end{equation}
is the magnification probability distribution of the second image. 
Here we consider two cases of results of optical depth estimates to cover a possible range of systematic uncertainties. 
One is based on the high-resolution ray-tracing simulations of \citet{2011ApJ...742...15T}. 
From their results of the probability distribution function of lensing magnification, the optical depth by foreground sources are estimated to be $\tau_{\rm m}=0.00231$, $0.00315$, $0.00380$, and $0.00446$ at $z=4$, $5$, $6$, and $7$, respectively. 
The other is based on a calibrated Faber-Jackson relation \citep{1976ApJ...204..668F} obtained by \citet{2015MNRAS.450.1224B}: $\tau_{\rm m}=0.0041$, $0.0054$, $0.0065$, and $0.0072$ at $z=4$, $5$, $6$, and $7$, respectively. 
Note that these optical depth estimates would be upper limits, because some fraction of lensed dropout sources might be too close to foreground lensing galaxies to be selected as dropouts in our samples. 
For the Schechter function parameters, we adopt the best-fit values obtained in the Schechter function fitting.

\setlength{\tabcolsep}{0.15cm}
\begin{deluxetable*}{ccccccc}
\tabletypesize{\footnotesize}
\setlength{\tabcolsep}{1pt}
\tablecaption{Fit Parameters for Galaxy Luminosity Functions}
\tablehead{
\colhead{Redshift} & \colhead{Fitted Function} & \colhead{$M_\m{UV}^*$} & \colhead{$\m{log}\phi^*$} & \colhead{$\alpha$} & \colhead{$\beta$} &\colhead{$\chi^2/\m{dof}$} \\
\colhead{}& \colhead{}& \colhead{($\m{Mpc^{-3}}$)}& \colhead{(mag)} &  \colhead{} & \colhead{} & \colhead{}}
\startdata
$z\sim4$ & DPL & $-21.10^{+0.07}_{-0.06}$ & $-3.09^{+0.06}_{-0.05}$ & $-1.87^{+0.04}_{-0.03}$ & $-4.95^{+0.13}_{-0.17}$ & $29.3/40$\\
 & Schechter & $-20.72^{+0.06}_{-0.05}$ & $-2.69^{+0.05}_{-0.05}$ & $-1.68^{+0.04}_{-0.04}$ & \nodata & $38.8/41$\\
 & Lensed Schechter ($\tau_\mathrm{m}$: Takahashi et al. 2011) & $-20.72$ & $-2.69$ & $-1.68$ & \nodata & $37.7/41$\\
 & Lensed Schechter ($\tau_\mathrm{m}$: Barone-Nugent et al. 2015) & $-20.72$ & $-2.69$ & $-1.68$ & \nodata & $36.9/41$\\
$z\sim5$ & DPL & $-21.39^{+0.09}_{-0.07}$ & $-3.48^{+0.07}_{-0.06}$ & $-1.94^{+0.04}_{-0.04}$ & $-4.96^{+0.21}_{-0.18}$ & $7.4/23$\\
 & Schechter & $-21.04^{+0.08}_{-0.07}$ & $-3.10^{+0.06}_{-0.06}$ & $-1.76^{+0.05}_{-0.05}$ & \nodata & $8.6/24$\\
 & Lensed Schechter ($\tau_\mathrm{m}$: Takahashi et al. 2011) & $-21.04$ & $-3.10$ & $-1.76$ & \nodata & $8.2/24$\\
 & Lensed Schechter ($\tau_\mathrm{m}$: Barone-Nugent et al. 2015) & $-21.04$ & $-3.10$ & $-1.76$ & \nodata & $8.0/24$\\
$z\sim6$ & DPL & $-21.23^{+0.18}_{-0.12}$ & $-3.67^{+0.14}_{-0.11}$ & $-2.14^{+0.08}_{-0.06}$ & $-5.03^{+0.26}_{-0.28}$ & $4.8/16$\\
 & Schechter & $-20.90^{+0.13}_{-0.14}$ & $-3.28^{+0.11}_{-0.13}$ & $-1.97^{+0.09}_{-0.08}$ & \nodata & $11.4/17$\\
 & Lensed Schechter ($\tau_\mathrm{m}$: Takahashi et al. 2011) & $-20.90$ & $-3.28$ & $-1.97$ & \nodata & $9.7/17$\\
 & Lensed Schechter ($\tau_\mathrm{m}$: Barone-Nugent et al. 2015) & $-20.90$ & $-3.28$ & $-1.97$ & \nodata & $8.7/17$\\
$z\sim7$ & DPL & $-20.82^{+0.17}_{-0.14}$ & $-3.51^{+0.12}_{-0.11}$ & $-2.05^{+0.06}_{-0.05}$ & $-4.83^{+0.32}_{-0.29}$ & $38.8/26$\\
 & Schechter & $-20.54^{+0.16}_{-0.14}$ & $-3.17^{+0.13}_{-0.11}$ & $-1.89^{+0.08}_{-0.07}$ & \nodata & $39.1/27$\\
 & Lensed Schechter ($\tau_\mathrm{m}$: Takahashi et al. 2011) & $-20.54$ & $-3.17$ & $-1.89$ & \nodata & $38.8/27$\\
 & Lensed Schechter ($\tau_\mathrm{m}$: Barone-Nugent et al. 2015) & $-20.54$ & $-3.17$ & $-1.89$ & \nodata & $38.6/27$
\enddata
\label{tab_LFfitgal}
\end{deluxetable*}

\begin{figure*}
\centering
\includegraphics[width=0.9\hsize, bb=10 17 777 500]{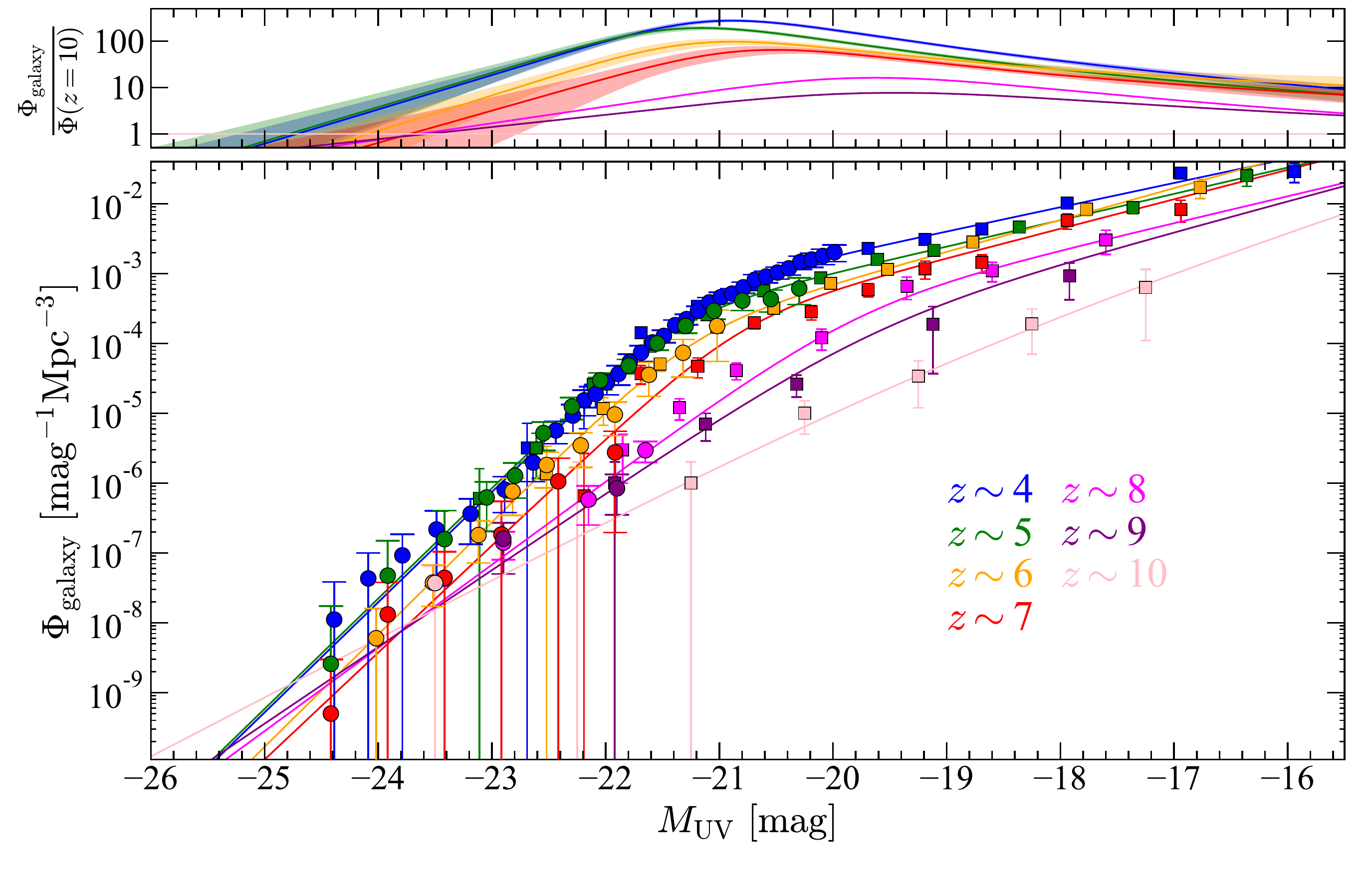}
\caption{
Evolution of the rest-frame UV luminosity functions of galaxies from $z\sim10$ to $z\sim4$.
The bottom panel shows the luminosity functions at $z\sim4-10$, and the blue, green, orange, red, magenta, purple, and pink symbols show results at $z\sim4$, $5$, $6$, $7$, $8$, $9$, and $10$, respectively.
The circles at $z\sim4-7$ show our results based on the HSC-SSP survey data, and those at $z\sim8-10$ are taken from \citet{2020MNRAS.493.2059B}.
The squares show results for galaxies in \citet{2021arXiv210207775B}.
The lines show the best-fit DPL functions in this study at $z\sim4-7$ and \citet{2020MNRAS.493.2059B} at $z\sim8-10$.
The top panel shows ratios of the number densities relative to those at $z\sim10$.
The shaded regions at $z\sim4-7$ correspond to the $2\sigma$ uncertainties of the luminosity function parameters.
The number density of typical galaxies ($M_\m{UV}=M^*_\m{UV}\sim-21$ mag) significantly increases by a factor of $\sim100$ from $\sim10$ to $4$, while that of faint galaxies ($M_\m{UV}\sim-16$ mag) mildly increases by a factor of $\sim10$.
The number density of the bright galaxies at $-25\lesssim M_\m{UV}\lesssim-23$ mag does not significantly change at $z\sim4-10$.
}
\label{fig_LFgal_evol}
\end{figure*}

In Figure \ref{fig_LFgal_wfit}, we show the best-fit functions of these three functional forms with the obtained galaxy UV luminosity function results. 
Table \ref{tab_LFfitgal} summarizes the best-fit parameters and the reduced $\chi^2$ values. 
We find that the DPL and the lensed Schechter functions provide better fits to the observed galaxy UV luminosity functions than the original Schechter functions.
The bright-end shapes of the observed galaxy UV luminosity functions cannot be explained by the Schechter functions.  
The significances of the bright end excess of the number density beyond the Schechter functions are $2.9 \sigma$, $1.9\sigma$, $2.8\sigma$, and $2.0\sigma$ at $z \sim 4$, $5$, $6$, and $7$, respectively. 
Note that the significances are lower than those in \citet{2018PASJ...70S..10O} at $z\sim4$ and $7$ because this time we consider the uncertainties of the spectroscopic galaxy fractions, which are not taken into account in \citet{2018PASJ...70S..10O}.
The physical origin of this bright end excess of the number density beyond the Schechter function will be discussed in Section \ref{ss_dis_bright}. 
The DPL function provides a better fit to the data points than the lensed Schechter function at $z\sim4-6$, although the significance of this difference is low.
The significances of the excess beyond the lensed Schechter functions are $2.5(2.7)\sigma$, $1.4(1.6)\sigma$, $2.1(2.4)\sigma$, and $1.4(1.6)\sigma$ at $z \sim 4$, $5$, $6$, and $7$, respectively, for the optical depth of \citet{2015MNRAS.450.1224B} (\citealt{2011ApJ...742...15T}), slightly smaller than those beyond the Schechter functions.

\subsubsection{Redshift Evolution}

Figure \ref{fig_LFgal_evol} shows our galaxy luminosity functions at $z\sim4-7$ with those of \citet{2021arXiv210207775B} at $z\sim4-10$ and of \citet{2020MNRAS.493.2059B} at $z\sim8-10$.
Although \citet{2020MNRAS.493.2059B} do not subtract AGN contributions from their estimated luminosity functions, the number densities of their bright sources at $z\sim8-10$ are likely dominated by galaxies, not by quasars, given the rapid decrease of the quasar luminosity function from $z\sim3$ to $6$ as discussed in Section \ref{ss_LFall_evol}.
Figure \ref{fig_LFgal_evol} suggests that the number density of typical galaxies ($M_\m{UV}=M^*_\m{UV}\sim-21$ mag) significantly increases by a factor of $\sim100$ from $\sim10$ to $4$, while that of faint galaxies ($M_\m{UV}\sim-16$ mag) mildly increases by a factor of $\sim10$.
Our comparison also shows that the number density of the bright galaxies at $-25\lesssim M_\m{UV}\lesssim-23$ mag does not significantly change at $z\sim4-10$, and is consistent with no evolution within the $2\sigma$ errors.
This is consistent with \citet{2021MNRAS.502..662B}, who report little evolution of the number density of bright galaxies at $z\gtrsim5$, although in their comparison they do not subtract AGN contributions from luminosity functions at $z\sim5-7$.
This agreement is expected because \citet{2021MNRAS.502..662B} compare the number densities of $M_\m{UV}>-24$ mag sources that are not dominated by AGNs.
As shown in Section \ref{ss_LFall_evol}, the number density of $M_\m{UV}\lesssim-24$ mag sources is dominated by AGNs and evolves rapidly, and we need to subtract AGN contributions to fairly compare the galaxy luminosity functions at the bright end.

\section{Clustering Analysis}\label{ss_clustering}

\subsection{Angular Correlation Function} 

We calculate angular correlation functions to evaluate the clustering strength of galaxies at $z\sim2-6$.
We use the galaxy samples at $z\sim1.7$, $2.2$, $3$, $4$, $5$, and $6$ constructed in Sections \ref{ss_selection_z47} and \ref{ss_selection_z23}.
\redc{Note that we do not remove AGNs from our source catalogs.}
To test the dependence of the clustering strength on the luminosity, we divide our galaxy samples into subsamples by UV magnitude thresholds ($m_\m{UV}^\m{th}$). 
The number of dropouts in the subsamples and their magnitude thresholds are summarized in Table \ref{tab_HOD}.
We do not use sources brighter than $m_\m{UV}^\m{cut}=20.0$ mag at each redshift in our analysis.
Changing this cut to a fainter magnitude (e.g., 23.0 mag) to remove AGNs does not change results of the angular correlation functions within the errors, because the number of such bright sources is small \citep[see][]{2018PASJ...70S..11H}.
Note that in the calculations we do not use sources in some part of the fields in the Wide layer whose depths are shallow.

\begin{figure*}
\centering
\includegraphics[width=0.9\hsize, bb=3 15 857 1213]{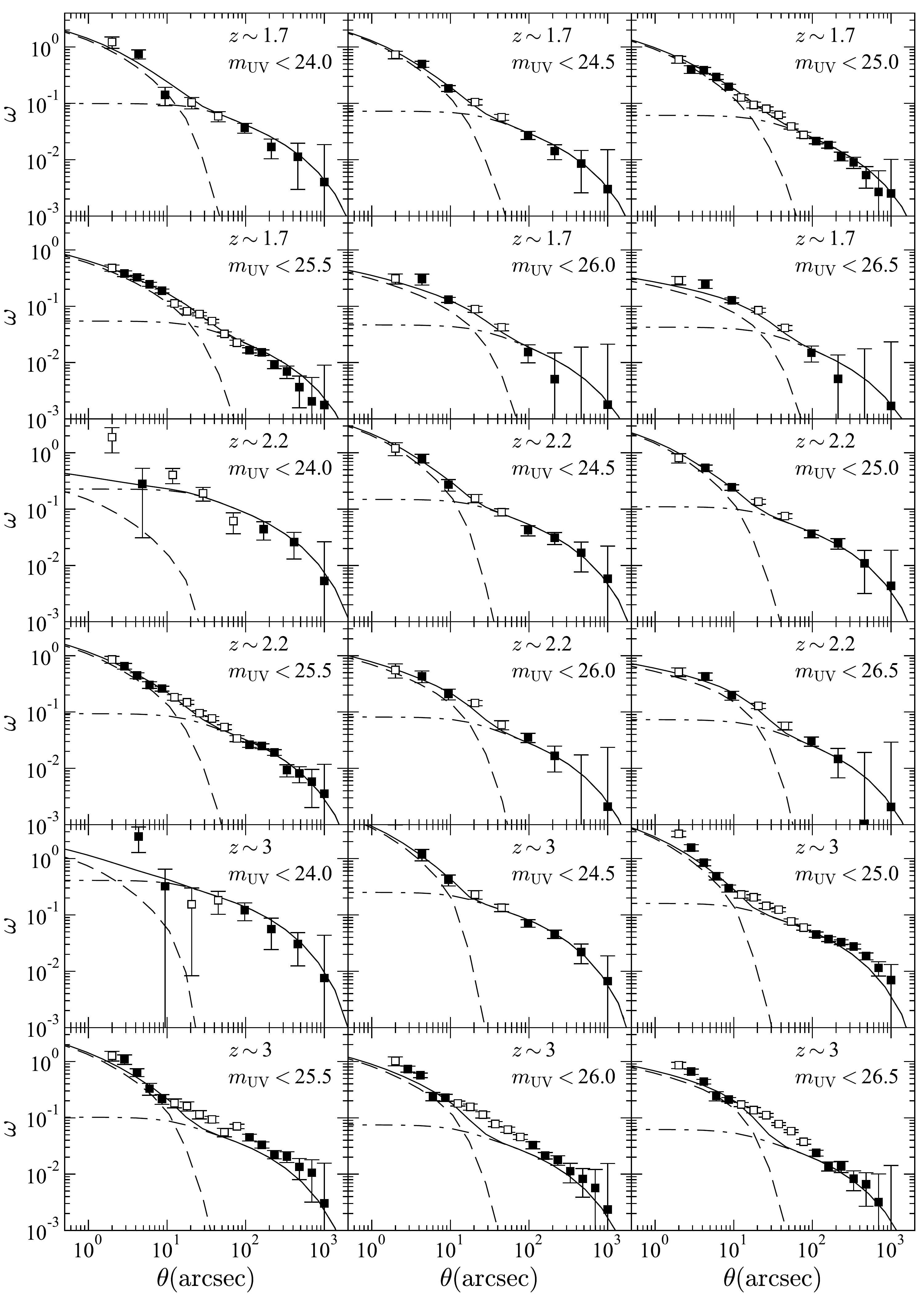}
\caption{Obtained angular correlation functions of the subsamples and their best-fit by the HOD models at $z\sim1.7$, $2.2$, and $3$.
The black squares denote the angular correlation function of the subsample at each redshift.
The data denoted by the open squares are not used in our HOD model fitting because they are at too-small scales or possibly affected by the non-linear halo bias effect.
The dashed and dot-dashed lines represent the one-halo and two-halo terms, respectively, and the solid line is the summations of the one-halo and two-halo terms.}
\label{fig_ACF_lowz}
\end{figure*}

\begin{figure*}
\centering
\includegraphics[width=0.9\hsize, bb=3 205 857 1213]{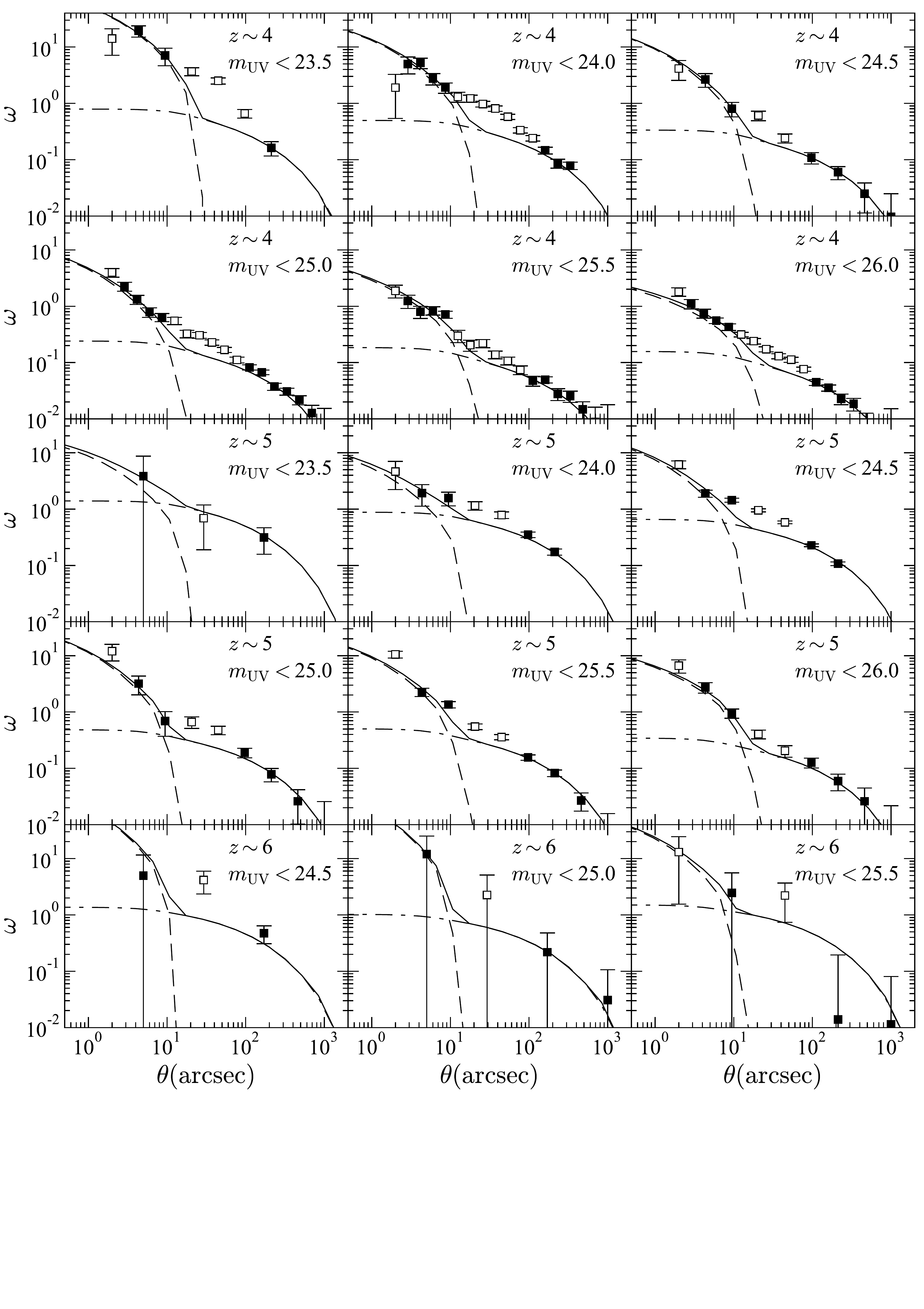}
\caption{Same as Figure \ref{fig_ACF_lowz} but at $z\sim4$, $5$, and $6$.}
\label{fig_ACF_highz}
\end{figure*}

\begin{figure*}
\centering
\includegraphics[width=0.99\hsize, bb=10 14 1106 286]{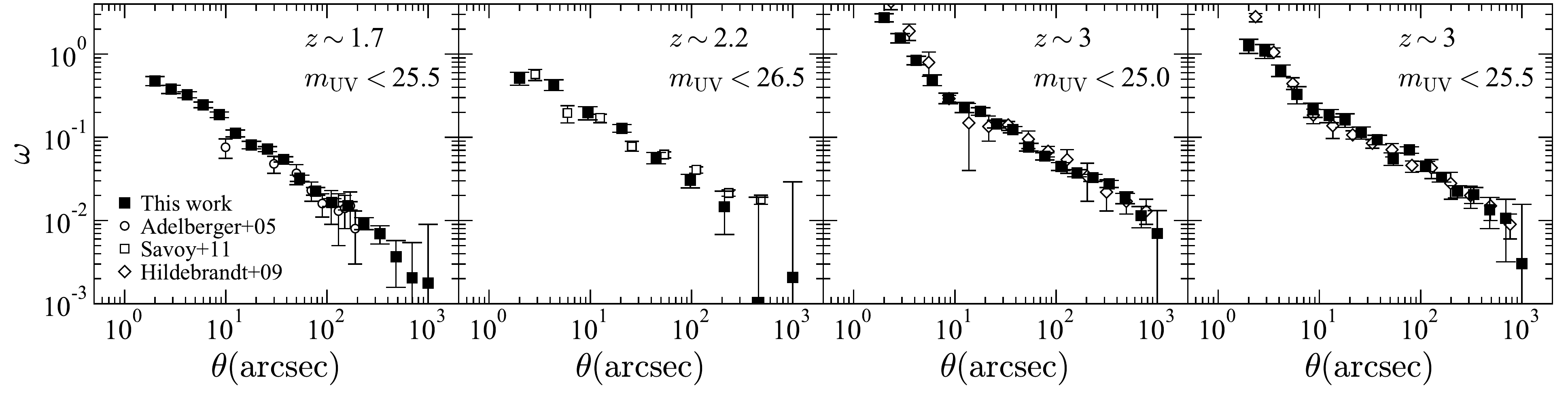}
\caption{Comparisons of the angular correlation functions with the literature.
The black squares represent the correlation functions in this study.
The open circles, squares, and diamonds denote results in \citet{2005ApJ...619..697A}, \cite{2011ApJ...737...92S}, and \citet{2009A&A...498..725H}, respectively.
Our obtained measurements agree well with these previous studies.
}
\label{fig_ACF_compare}
\end{figure*}

We calculate observed angular correlation functions of the subsamples, $\omega_{\m{obs}}(\theta)$, using an estimator proposed by \citet{1993ApJ...412...64L},
\begin{equation}
\omega_{\m{obs}}(\theta)=\frac{DD(\theta)-2DR(\theta)+RR(\theta)}{RR(\theta)},
\end{equation}
where $DD(\theta)$, $DR(\theta)$, and $RR(\theta)$ are the numbers of galaxy-galaxy, galaxy-random, and random-random pairs normalized by the total numbers of pairs.
We use the random catalog whose surface number density is $100\ \m{arcmin^{-2}}$ with the same geometrical shape as the observational data including the mask positions (\citealt{2018PASJ...70S...7C}).
\redc{Corrections for contaminations (e.g., Equation (29) in \cite{2016ApJ...821..123H}) are not applied because the clustering strength of the interlopers is not well-measured.}
We calculate angular correlation functions in individual fields, and obtain the best-estimate that is the mean weighted by the effective area in each field.
Figures \ref{fig_ACF_lowz} and \ref{fig_ACF_highz} show our calculated angular correlation functions of the subsamples at $z\sim2-3$ and $4-6$, respectively.
We compare our obtained correlation functions with the literature in Figure \ref{fig_ACF_compare}.
Our correlations functions are in good agreement with those of \citet{2005ApJ...619..697A}, \cite{2011ApJ...737...92S}, and \citet{2009A&A...498..725H}.


\begin{deluxetable*}{cccccccccccc}
\tabletypesize{\footnotesize}
\setlength{\tabcolsep}{1.5pt}
\tablecaption{Summary of the Clustering Measurements with the HOD Model.}
\tablehead{
\colhead{$\bar z$} & \colhead{$m_\m{UV}^\m{th}$} & \colhead{$N$} & \colhead{$n_\m{obs}$} & \colhead{$M_\m{UV}^\m{th}$} & \colhead{$SFR$} & \colhead{$\m{log}M_\m{min}$} & \colhead{$\m{log}M_\m{sat}$} & \colhead{$\m{log}\left<M_\m{h}\right>$} & \colhead{$b_\m{g}^\m{eff}$} & \colhead{$\m{log}f_\m{sat}$} & \colhead{$\chi^2/\m{dof}$} \\
\colhead{(1)} & \colhead{(2)} & \colhead{(3)} & \colhead{(4)} & \colhead{(5)} & \colhead{(6)} & \colhead{(7)} & \colhead{(8)} & \colhead{(9)} & \colhead{(10)} & \colhead{(11)} & \colhead{(12)} }
\startdata
\multicolumn{12}{c}{$z\sim1.7$}\\
$1.7$ & $24.0$ & $16219$ & $3.6\times10^{-4}$ & $-20.5$ & $31.0$ & $12.46^{+0.03}_{-0.03}$ & $14.18^{+0.19}_{-0.09}$ & $12.82^{+0.02}_{-0.02}$ & $2.42^{+0.03}_{-0.05}$ & $-1.45^{+0.08}_{-0.17}$ & $9.1/5$\\
$1.7$ & $24.5$ & $42719$ & $1.1\times10^{-3}$ & $-20.0$ & $16.3$ & $12.09^{+0.03}_{-0.04}$ & $13.47^{+0.09}_{-0.09}$ & $12.61^{+0.01}_{-0.02}$ & $2.04^{+0.03}_{-0.02}$ & $-1.14^{+0.05}_{-0.08}$ & $2.4/5$\\
$1.7$ & $25.0$ & $87813$ & $2.4\times10^{-3}$ & $-19.5$ & $8.6$ & $11.79^{+0.03}_{-0.04}$ & $12.86^{+0.10}_{-0.06}$ & $12.56^{+0.01}_{-0.02}$ & $1.86^{+0.01}_{-0.02}$ & $-0.93^{+0.03}_{-0.05}$ & $4.1/10$\\
$1.7$ & $25.5$ & $150962$ & $4.5\times10^{-3}$ & $-19.0$ & $4.5$ & $11.55^{+0.04}_{-0.03}$ & $12.48^{+0.06}_{-0.06}$ & $12.55^{+0.02}_{-0.01}$ & $1.76^{+0.01}_{-0.02}$ & $-0.88^{+0.02}_{-0.03}$ & $4.9/10$\\
$1.7$ & $26.0$ & $24950$ & $7.3\times10^{-3}$ & $-18.5$ & $2.4$ & $11.33^{+0.04}_{-0.04}$ & $12.28^{+0.06}_{-0.07}$ & $12.49^{+0.02}_{-0.02}$ & $1.62^{+0.03}_{-0.02}$ & $-0.95^{+0.03}_{-0.03}$ & $2.5/3$\\
$1.7$ & $26.5$ & $36015$ & $1.1\times10^{-2}$ & $-18.0$ & $1.2$ & $11.16^{+0.05}_{-0.04}$ & $12.08^{+0.08}_{-0.07}$ & $12.46^{+0.02}_{-0.03}$ & $1.54^{+0.02}_{-0.03}$ & $-0.99^{+0.02}_{-0.06}$ & $2.9/3$\\
\multicolumn{12}{c}{$z\sim2.2$}\\
$2.2$ & $24.0$ & $4927$ & $8.4\times10^{-5}$ & $-21.0$ & $59.1$ & $12.72^{+0.03}_{-0.02}$ & $(15.91)$ & $12.91^{+0.02}_{-0.01}$ & $3.62^{+0.07}_{-0.05}$ & $-3.03^{+0.06}_{-0.06}$ & $1.1/4$\\
$2.2$ & $24.5$ & $14185$ & $3.6\times10^{-4}$ & $-20.5$ & $31.0$ & $12.30^{+0.03}_{-0.03}$ & $13.92^{+0.11}_{-0.11}$ & $12.61^{+0.02}_{-0.02}$ & $2.86^{+0.04}_{-0.05}$ & $-1.43^{+0.08}_{-0.10}$ & $3.3/5$\\
$2.2$ & $25.0$ & $32241$ & $1.1\times10^{-3}$ & $-20.0$ & $16.3$ & $11.95^{+0.04}_{-0.02}$ & $13.23^{+0.12}_{-0.08}$ & $12.41^{+0.01}_{-0.02}$ & $2.42^{+0.03}_{-0.03}$ & $-1.20^{+0.06}_{-0.07}$ & $0.8/5$\\
$2.2$ & $25.5$ & $59623$ & $2.4\times10^{-3}$ & $-19.5$ & $8.6$ & $11.68^{+0.04}_{-0.03}$ & $12.62^{+0.11}_{-0.05}$ & $12.34^{+0.01}_{-0.02}$ & $2.21^{+0.03}_{-0.02}$ & $-1.02^{+0.02}_{-0.05}$ & $9.8/10$\\
$2.2$ & $26.0$ & $9196$ & $4.5\times10^{-3}$ & $-19.0$ & $4.5$ & $11.45^{+0.02}_{-0.05}$ & $12.23^{+0.15}_{-0.11}$ & $12.32^{+0.04}_{-0.09}$ & $2.06^{+0.04}_{-0.08}$ & $-0.98^{+0.08}_{-0.14}$ & $1.4/3$\\
$2.2$ & $26.5$ & $13949$ & $7.3\times10^{-3}$ & $-18.5$ & $2.4$ & $11.26^{+0.06}_{-0.02}$ & $11.94^{+0.09}_{-0.05}$ & $12.32^{+0.02}_{-0.04}$ & $1.97^{+0.04}_{-0.04}$ & $-0.96^{+0.03}_{-0.05}$ & $3.4/3$\\
\multicolumn{12}{c}{$z\sim3$}\\
$2.9$ & $24.0$ & $4607$ & $5.7\times10^{-5}$ & $-21.5$ & $114.1$ & $12.55^{+0.02}_{-0.02}$ & $(15.39)$ & $12.68^{+0.02}_{-0.01}$ & $4.66^{+0.09}_{-0.04}$ & $-2.75^{+0.02}_{-0.08}$ & $3.0/6$\\
$2.9$ & $24.5$ & $18013$ & $2.5\times10^{-4}$ & $-21.0$ & $59.9$ & $12.19^{+0.02}_{-0.03}$ & $13.80^{+0.19}_{-0.11}$ & $12.42^{+0.01}_{-0.02}$ & $3.71^{+0.05}_{-0.05}$ & $-1.57^{+0.10}_{-0.15}$ & $1.1/5$\\
$2.9$ & $25.0$ & $57199$ & $6.7\times10^{-4}$ & $-20.5$ & $31.5$ & $11.92^{+0.03}_{-0.02}$ & $13.12^{+0.12}_{-0.07}$ & $12.26^{+0.01}_{-0.02}$ & $3.23^{+0.04}_{-0.03}$ & $-1.30^{+0.04}_{-0.06}$ & $9.7/10$\\
$2.9$ & $25.5$ & $11257$ & $1.5\times10^{-3}$ & $-20.0$ & $16.5$ & $11.71^{+0.04}_{-0.02}$ & $12.55^{+0.14}_{-0.04}$ & $12.21^{+0.01}_{-0.02}$ & $2.90^{+0.04}_{-0.03}$ & $-1.11^{+0.05}_{-0.06}$ & $10.5/10$\\
$2.9$ & $26.0$ & $23231$ & $2.6\times10^{-3}$ & $-19.5$ & $8.7$ & $11.55^{+0.03}_{-0.04}$ & $12.20^{+0.06}_{-0.07}$ & $12.20^{+0.01}_{-0.02}$ & $2.72^{+0.02}_{-0.05}$ & $-1.03^{+0.03}_{-0.03}$ & $15.4/10$\\
$2.9$ & $26.5$ & $43111$ & $4.4\times10^{-3}$ & $-19.0$ & $4.6$ & $11.36^{+0.05}_{-0.03}$ & $11.84^{+0.10}_{-0.04}$ & $12.21^{+0.02}_{-0.01}$ & $2.55^{+0.04}_{-0.04}$ & $-0.97^{+0.02}_{-0.03}$ & $15.7/10$\\
\multicolumn{12}{c}{$z\sim4$}\\
$3.8$ & $23.5$ & $4971$ & $1.1\times10^{-6}$ & $-22.5$ & $186.0$ & $13.08^{+0.03}_{-0.01}$ & $15.25^{+0.12}_{-0.09}$ & $13.01^{+0.01}_{-0.01}$ & $8.55^{+0.08}_{-0.09}$ & $-2.25^{+0.09}_{-0.11}$ & $0.4/2$\\
$3.8$ & $24.0$ & $19125$ & $1.0\times10^{-5}$ & $-22.0$ & $106.1$ & $12.71^{+0.02}_{-0.02}$ & $14.80^{+0.09}_{-0.08}$ & $12.73^{+0.01}_{-0.02}$ & $6.70^{+0.05}_{-0.09}$ & $-2.12^{+0.07}_{-0.07}$ & $4.9/6$\\
$3.8$ & $24.5$ & $8059$ & $6.1\times10^{-5}$ & $-21.5$ & $60.5$ & $12.32^{+0.03}_{-0.02}$ & $13.96^{+0.34}_{-0.09}$ & $12.43^{+0.01}_{-0.02}$ & $5.35^{+0.07}_{-0.07}$ & $-1.68^{+0.08}_{-0.31}$ & $0.2/5$\\
$3.8$ & $25.0$ & $27735$ & $2.6\times10^{-4}$ & $-21.0$ & $34.5$ & $11.98^{+0.03}_{-0.02}$ & $13.23^{+0.13}_{-0.08}$ & $12.18^{+0.02}_{-0.01}$ & $4.41^{+0.06}_{-0.06}$ & $-1.50^{+0.05}_{-0.10}$ & $8.2/10$\\
$3.8$ & $25.5$ & $8059$ & $8.1\times10^{-4}$ & $-20.5$ & $19.7$ & $11.66^{+0.02}_{-0.03}$ & $12.24^{+0.07}_{-0.10}$ & $12.02^{+0.01}_{-0.02}$ & $3.85^{+0.04}_{-0.05}$ & $-1.23^{+0.05}_{-0.04}$ & $14.1/10$\\
$3.8$ & $26.0$ & $16494$ & $1.6\times10^{-3}$ & $-20.0$ & $11.2$ & $11.48^{+0.04}_{-0.02}$ & $11.94^{+0.10}_{-0.04}$ & $11.94^{+0.02}_{-0.01}$ & $3.44^{+0.05}_{-0.04}$ & $-1.27^{+0.03}_{-0.05}$ & $1.4/10$\\
\multicolumn{12}{c}{$z\sim5$}\\
$4.9$ & $23.5$ & $1650$ & $2.2\times10^{-7}$ & $-22.9$ & $325.1$ & $12.95^{+0.02}_{-0.01}$ & $16.65^{+1.57}_{-0.12}$ & $12.81^{+0.01}_{-0.01}$ & $11.02^{+0.08}_{-0.12}$ & $-3.86^{+0.12}_{-1.57}$ & $0.1/1$\\
$4.9$ & $24.0$ & $21453$ & $2.9\times10^{-6}$ & $-22.4$ & $180.4$ & $12.60^{+0.01}_{-0.02}$ & $15.70^{+0.82}_{-0.07}$ & $12.55^{+0.01}_{-0.01}$ & $8.64^{+0.08}_{-0.10}$ & $-3.21^{+0.14}_{-0.69}$ & $2.6/3$\\
$4.9$ & $24.5$ & $63308$ & $1.7\times10^{-5}$ & $-21.9$ & $100.1$ & $12.29^{+0.03}_{-0.00}$ & $14.63^{+0.11}_{-0.06}$ & $12.32^{+0.02}_{-0.01}$ & $7.11^{+0.10}_{-0.06}$ & $-2.49^{+0.06}_{-0.08}$ & $27.2/3$\\
$4.9$ & $25.0$ & $5305$ & $6.4\times10^{-5}$ & $-21.4$ & $55.6$ & $12.00^{+0.02}_{-0.01}$ & $13.45^{+0.46}_{-0.08}$ & $12.09^{+0.01}_{-0.02}$ & $6.17^{+0.06}_{-0.08}$ & $-1.89^{+0.09}_{-0.40}$ & $1.7/5$\\
$4.9$ & $25.5$ & $13688$ & $1.9\times10^{-4}$ & $-20.9$ & $30.8$ & $11.76^{+0.02}_{-0.02}$ & $12.57^{+0.13}_{-0.09}$ & $11.93^{+0.02}_{-0.01}$ & $5.37^{+0.06}_{-0.06}$ & $-1.62^{+0.06}_{-0.07}$ & $9.7/5$\\
$4.9$ & $26.0$ & $3349$ & $4.2\times10^{-4}$ & $-20.4$ & $17.1$ & $11.57^{+0.03}_{-0.02}$ & $11.86^{+0.11}_{-0.06}$ & $11.83^{+0.01}_{-0.02}$ & $4.99^{+0.07}_{-0.06}$ & $-1.44^{+0.04}_{-0.07}$ & $1.4/5$\\
\multicolumn{12}{c}{$z\sim6$}\\
$5.9$ & $24.5$ & $2026$ & $1.6\times10^{-6}$ & $-22.2$ & $163.3$ & $12.33^{+0.02}_{-0.02}$ & $(14.67)$ & $12.24^{+0.01}_{-0.01}$ & $10.09^{+0.09}_{-0.09}$ & $-2.63^{+0.03}_{-0.02}$ & $2.9/2$\\
$5.9$ & $25.0$ & $328$ & $7.7\times10^{-6}$ & $-21.7$ & $85.8$ & $12.09^{+0.01}_{-0.02}$ & $13.73^{+5.67}_{-0.28}$ & $12.07^{+0.01}_{-0.02}$ & $8.77^{+0.04}_{-0.14}$ & $-2.19^{+0.40}_{-4.01}$ & $1.0/2$\\
$5.9$ & $25.5$ & $480$ & $3.2\times10^{-5}$ & $-21.2$ & $45.1$ & $11.78^{+0.02}_{-0.02}$ & $(12.93)$ & $11.81^{+0.02}_{-0.01}$ & $7.68^{+0.05}_{-0.11}$ & $-2.41^{+0.01}_{-0.00}$ & $1.9/3$
\enddata
\tablecomments{(1) Mean redshift.
(2) Threshold apparent magnitude in the rest-frame UV band. 
(3) Number of galaxies in the subsample. 
(4) Number density of galaxies in the subsample in units of $\m{Mpc^{-3}}$. 
(5) Threshold absolute magnitude in the rest-frame UV band. 
(6) SFR corresponding to $M_\m{UV}^\m{th}$ after the extinction correction in units of $M_\odot\ \m{yr^{-1}}$.  
(7) Best-fit value of $M_\m{min}$ in units of $M_\odot$. 
(8) Best-fit value of $M_\m{sat}$ in units of $M_\odot$. The value in parenthesis is derived from $M_\m{min}$ via Equation (\ref{eq_Msat}). 
(9) Mean halo mass in units of $M_\odot$. 
(10) Effective bias. 
(11) Satellite fraction. 
(12) Reduced $\chi^{2}$ value.
}
\label{tab_HOD}
\end{deluxetable*}

Due to the finite size of our survey fields, the observed correlation functions is underestimated by a constant value known as the integral constraint, $IC$ \citep{1977ApJ...217..385G}. 
Including a correction for the number of objects in the sample, $N$ \citep{1980lssu.book.....P}, the true angular correlation function is given by
\begin{equation}
\omega(\theta)=\omega_{\m{obs}}(\theta)+IC+\frac{1}{N}.
\label{eq_acf}
\end{equation}
We estimate the integral constraint with 
\begin{equation}
IC=\frac{\Sigma_{i}RR(\theta_i)\omega_{\m{model}}(\theta_i)}{\Sigma_{i}RR(\theta_i)},
\label{eq_ic}
\end{equation}
where $\omega_{\m{model}}(\theta)$ is the best-fit model of the correlation function, and $i$ refers the angular bin.
$IC$ and $\omega_{\m{model}}(\theta)$ are simultaneously determined in the model fitting in Section \ref{ss_HODmodel}.

We estimate statistical errors of the angular correlation functions using the Jackknife estimator.
We divide each subsample into Jackknife samples of about $1000^2\ \m{arcsec}^2$, whose size is larger than the largest angular scale in the correlation function.
Removing one Jackknife sample at a time for each realization, we compute the covariance matrix as
\begin{equation}
C_{ij}=\frac{N_\m{Jack}-1}{N_\m{Jack}}\sum^{N_\m{Jack}}_{l=1}\left[\omega^l(\theta_i)-\bar{\omega}(\theta_i)\right]\left[\omega^l(\theta_j)-\bar{\omega}(\theta_j)\right].
\end{equation}
where $N_\m{Jack}$ is the total number of the Jackknife samples, $\omega^l$ is the estimated correlation function from the $l$th realization, and $\bar{\omega}$ is the mean correlation function.
We apply a correction factor given by \citet{2007A&A...464..399H} to an inverse covariance matrix in order to compensate for the bias introduced by the noise.
The inverse of the square root of the inverse covariance matrix is plotted in Figures \ref{fig_ACF_lowz}, \ref{fig_ACF_highz}, \ref{fig_ACF_compare} as uncertainties.

\subsection{Halo Occupation Distribution (HOD) Model Fitting}\label{ss_HODmodel}

We use an HOD model to investigate the relationship between galaxies and their dark matter halos.
The HOD model is an analytic framework quantifying a probability distribution of the number of galaxies in dark matter halos \citep[e.g.,][]{2000MNRAS.318..203S,2000MNRAS.318.1144P,2000ApJ...543..503M}.
The key assumption in the HOD model is that the probability depends only on the halo mass, $M_\m{h}$.
We can analytically calculate correlation functions and number densities from the HOD model.
Details of the calculations are presented in \citet{2016ApJ...821..123H}.

We fit our HOD model to the observed angular correlation functions and number densities. 
In the fitting procedures, the best-fit parameters are determined by minimizing the $\chi^2$ value,
\begin{eqnarray}\label{eq_chi2}
\chi^2=&\sum_{i,j}& \left[\omega(\theta_i)-\omega_\m{model}(\theta_i)\right]C^{-1}_{i,j}\left[\omega(\theta_j)-\omega_\m{model}(\theta_j)\right]\nonumber\\
&+&\frac{\left[\m{log}\,n_\m{g}^\m{obs}-\m{log}\,n_\m{g}^\m{model}\right]^2}{\sigma^2_{\m{log}n_\m{g}}},
\end{eqnarray}
where $C^{-1}_{i,j}$ is the inverse covariance matrix, $n_\m{g}$ is a space number density of galaxies in the subsample, and $\sigma_{\m{log}n_\m{g}}$ is its error.
We calculate the number density of galaxies corrected for incompleteness using the galaxy UV luminosity functions derived in this work (Section \ref{ss_galLF}) and \citet{2021arXiv210207775B}.
The galaxy number density of each subsample is presented in Table \ref{tab_HOD}.
We assume $10\%$ fractional uncertainties in the number densities as \citet{2007ApJ...667..760Z}.
This $10\%$ uncertainty is a conservative assumption, because the actual statistical uncertainty is typically less than $5\%$.
We constrain the parameters of our HOD model using the Markov Chain Monte Carlo (MCMC) parameter estimation technique.

In our HOD model, an occupation function for central galaxies follows a step function with a smooth transition,
\begin{eqnarray}
\left< N_\m{cen}(M_\m{h})\right>=\frac{1}{2}\left[1+\m{erf}\left(\frac{\m{log}M_\m{h}-\m{log}M_\m{min}}{\sqrt{2}\sigma_{\m{log}M_\m{h}}}\right)\right].
\end{eqnarray}
An occupation function for satellite galaxies is expressed by a power law with a mass cut,
\begin{equation}
\left< N_\m{sat}(M_\m{h})\right>=\left< N_\m{cen}(M_\m{h})\right>\left(\frac{M_\m{h}-M_\m{cut}}{M_\m{sat}}\right)^\alpha.
\end{equation}
The total occupation function is 
\begin{eqnarray}
\left< N_\m{tot}(M_\m{h})\right>=\left< N_\m{cen}(M_\m{h})\right>+\left< N_\m{sat}(M_\m{h})\right>.
\end{eqnarray}
These functional forms are motivated by N-body simulations, smoothed particle hydrodynamic simulations, and semi-analytic models for both low and high redshift galaxies \citep[e.g.,][]{2004ApJ...609...35K,2005ApJ...633..791Z,2015MNRAS.450.1279G}.
Indeed, previous studies demonstrate that this HOD model can explain observed angular correlation functions of high redshift galaxies \citep{2016ApJ...821..123H,2018PASJ...70S..11H,2017ApJ...841....8I}.

We calculate the mean dark matter halo mass of galaxies including both the central and satellite galaxies, $\left<M_\m{h}\right>$, effective galaxy bias, $b_\m{g}^\m{eff}$, and the satellite fraction, $f_\m{sat}$, as follows:
\begin{eqnarray}
\left<M_\m{h}\right> &=& \frac{1}{n_\m{g}}\int dM_\m{h}\frac{dn}{dM_\m{h}}(M_\m{h},z)N_\m{tot}(M_\m{h})M_\m{h},\\
b_\m{g}^\m{eff} &=& \frac{1}{n_\m{g}}\int dM_\m{h}\frac{dn}{dM_\m{h}}(M_\m{h},z)N_\m{tot}(M_\m{h})b_\mathrm{h}(M_\m{h},z),\\
f_\m{sat} &=& \frac{1}{n_\m{g}}\int dM_\m{h}\frac{dn}{dM_\m{h}}(M_\m{h},z)N_\m{sat}(M_\m{h}),
\end{eqnarray}
where $\frac{dn}{dM_\m{h}}(M_\m{h},z)$, $b_\mathrm{h}(M_\m{h},z)$, and $n_\m{g}$ are the halo mass function, halo bias, and the galaxy number density in the model \citep[Equation (51) in][]{2016ApJ...821..123H}, respectively.
We use the \citet{2013ApJ...770...57B} halo mass function, which is a modification of the \citet{2008ApJ...688..709T} mass function, and is calibrated at $z>3$, the NFW dark matter halo profile \citep{1996ApJ...462..563N,1997ApJ...490..493N}, the \citet{2008MNRAS.390L..64D} concentration parameter, and the \citet{2003MNRAS.341.1311S} non-linear matter power spectrum.

Some theoretical studies claim that the halo bias is scale-dependent in the quasi-linear scale of $r\sim50\ \m{Mpc}$ \citep[the non-linear halo bias effect;][]{2009MNRAS.394..624R,2013MNRAS.429.2333J,2016MNRAS.463..270J,2017MNRAS.469.4428J}.
However, in this study, we assume the scale-independent linear halo bias of \citet{2010ApJ...724..878T}, $b(M_\m{h},z)$.
Instead, we do not use the angular correlation functions at $10\arcsec<\theta<90\arcsec$ ($10\arcsec<\theta<120\arcsec$) in the UltraDeep and Deep (Wide) layers, because they could be affected by the non-linear halo bias effect.
We also do not use the measurements at $\theta\leq2\arcsec$ that are possibly affected by the source confusion.

The HOD model has five parameters, $M_\m{min}$, $\sigma_{\m{log}M_\m{h}}$, $M_\m{cut}$, $M_\m{sat}$, and $\alpha$. 
We take $M_\m{min}$ and $M_\m{sat}$ as free parameters, which control typical masses of halos having one central and satellite galaxies, respectively, as previous studies \citep{2016ApJ...821..123H,2018PASJ...70S..11H}.
We fix $\sigma_{\m{log}M_\m{h}}=0.2$ and $\alpha=1.0$, following results of previous studies \citep[e.g.,][]{2004ApJ...609...35K,2005ApJ...633..791Z,2006ApJ...647..201C,2017ApJ...841....8I}.
To derive $M_\m{cut}$ from $M_\m{min}$, we use the relation
\begin{equation}
M_\m{cut}=M_\m{min}^{-0.5},
\end{equation}
which is given by \citet{2015MNRAS.449.1352C}.
Because the exact value of $M_\m{cut}$ has very little importance compared to the other parameters, this assumption does not change any of our conclusions.
For subsamples whose correlation functions are not accurately determined due to the small number of galaxies (subsamples of $z\sim2.2$ $m_\m{UV}^\m{th}=24.0$ mag, $z\sim3$ $m_\m{UV}^\m{th}=24.0$ mag, and $z\sim6$ $m_\m{UV}^\m{th}=24.5,\ 25.5$), we also use the following relation calibrated with results in \citet{2018PASJ...70S..11H}.
\begin{equation}
\m{log}M_\m{sat}=3.16\ \m{log}M_\m{min}-24.33. \label{eq_Msat}
\end{equation}

\begin{figure}
\centering
\includegraphics[width=0.99\hsize, bb=15 5 566 349]{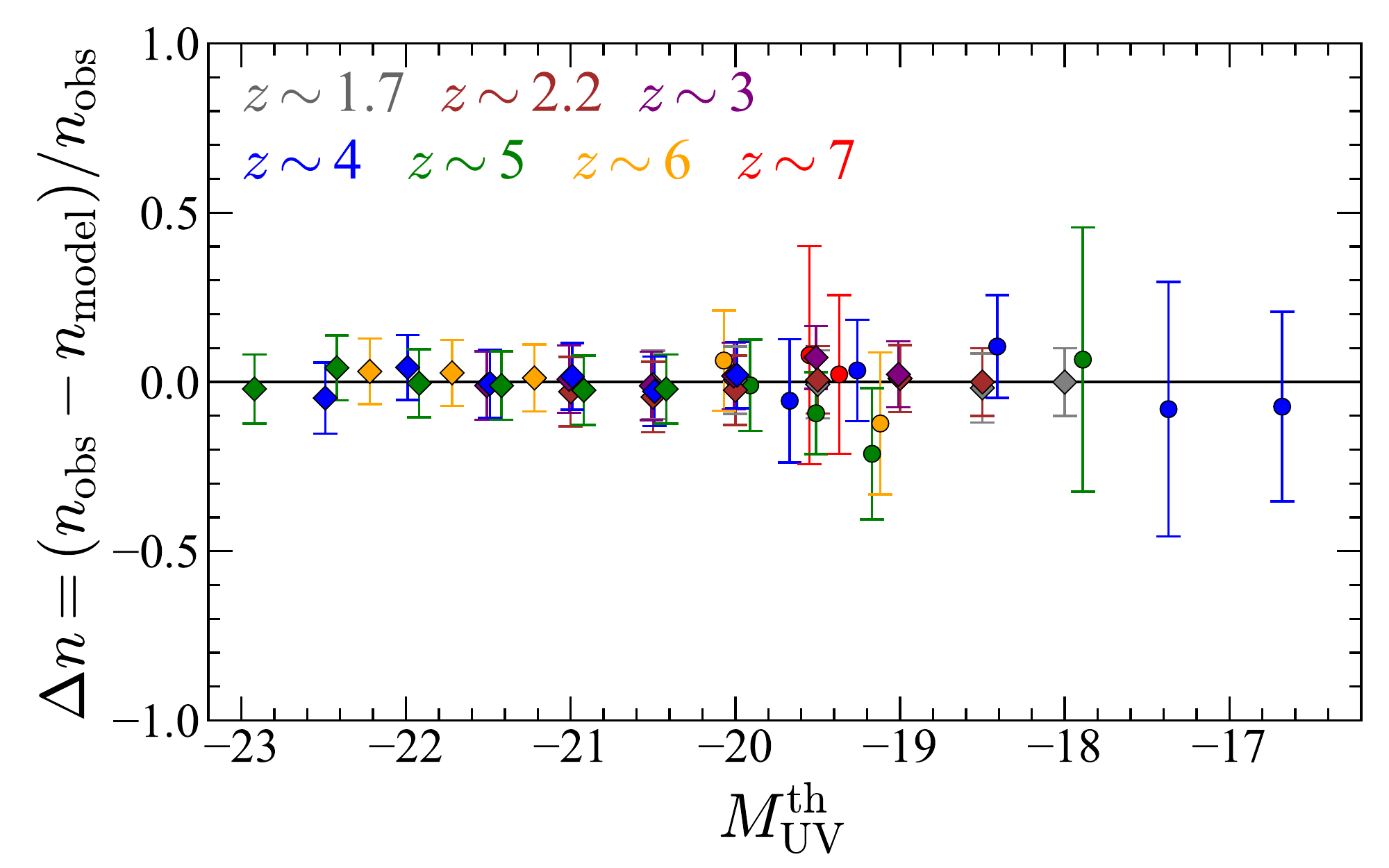}
\caption{Comparison of the number densities between the HOD models and observations.
The gray, brown, purple, blue, green, orange, and red squares (circles) represent the relative differences of the number densities between the HOD models and observations for the subsamples in this work (in \citealt{2016ApJ...821..123H}), at $z\sim1.7$, $2.2$, $3$, $4$, $5$, $6$, and $7$, respectively, as a function of the threshold absolute UV magnitudes, $M_\m{UV}^\m{th}$.}
\label{fig_dnum}
\end{figure}

We plot the observed angular correlation functions and predictions from their best-fit HOD models in Figures \ref{fig_ACF_lowz} and \ref{fig_ACF_highz}.
The best-fit parameters and their $1\sigma$ errors are presented in Table \ref{tab_HOD}.
The HOD models can reproduce the observed correlation functions at small ($2\lesssim\theta\lesssim10\arcsec$) and large ($\theta\gtrsim100\arcsec$) scales.
However, the models underpredict the correlation functions by a factor of $1.5-6$ in $10\arcsec\lesssim\theta\lesssim100\arcsec$, the transition scale between 1- and 2-halo terms (the quasi-linear scale), especially in the subsamples at $z\gtrsim3$.
These results indicate that the correlation functions at $10\arcsec\lesssim\theta\lesssim100\arcsec$ can not be explained by the scale-independent halo bias due to the non-linear halo bias effect in this quasi-linear scale \citep{2013MNRAS.429.2333J,2016MNRAS.463..270J,2017MNRAS.469.4428J}.
We also find that the best-fit HOD models slightly underpredict the correlation functions of the subsamples of $z\sim3$ and $m_\m{UV}<25.5$ and $26.0$ at $\theta\geq100\arcsec$, although the reduced $\chi^2$ values are not bad ($\chi^2/\m{dof}=1.05,1.54$).
We have tried to fit the correlation functions of these subsamples with taking $M_\m{min}$, $\sigma_{\m{log}M_\m{h}}$, $M_\m{sat}$, and $\alpha$ as free parameters, but the results does not significantly change. 
In Figure \ref{fig_dnum}, we compare the observed number densities and predictions from the best-fit HOD models, showing good agreement.

\subsection{$M_\mathrm{UV}-M_\m{h}$ Relation}

\begin{figure}
\centering
\includegraphics[width=0.99\hsize, bb=5 7 330 357]{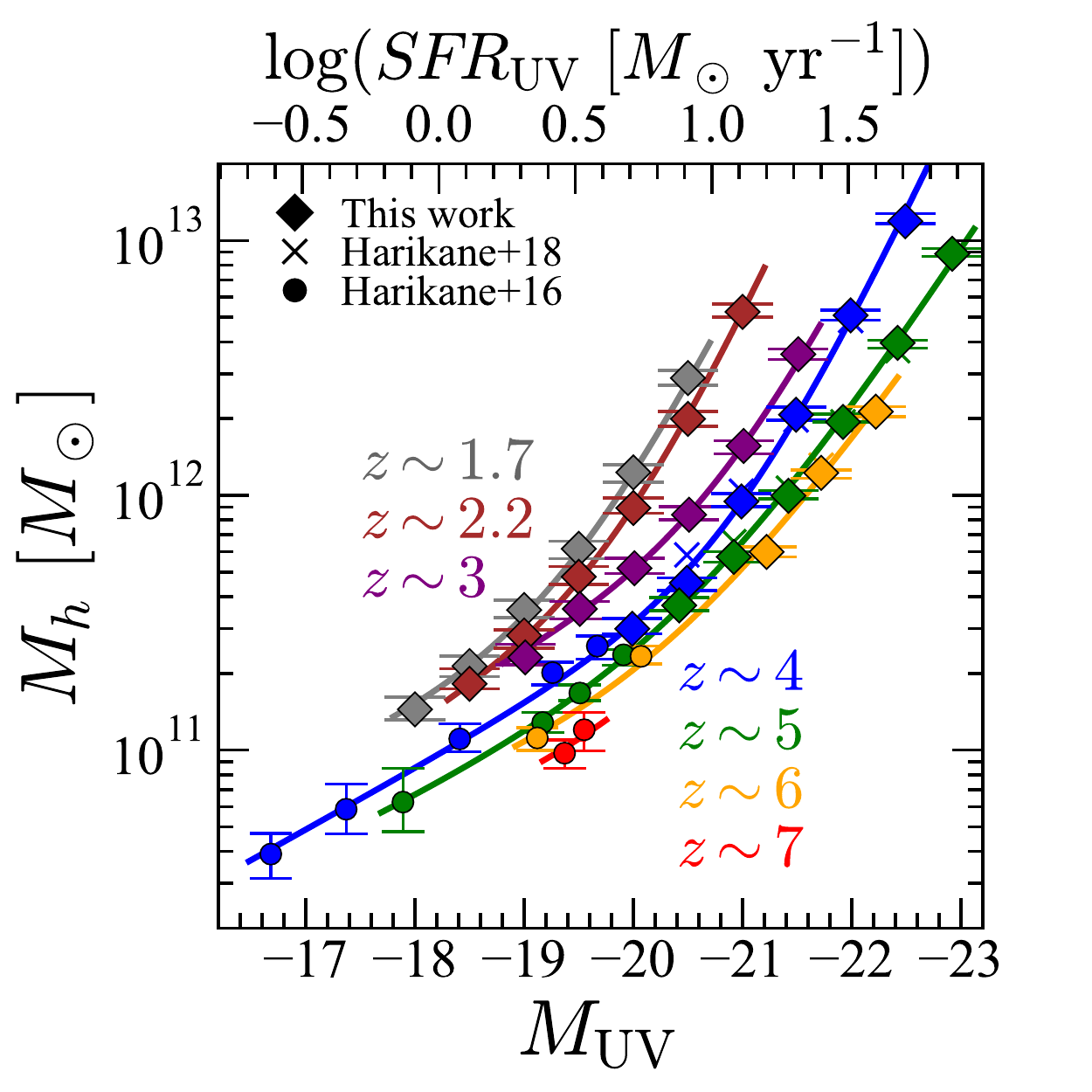}
\caption{$M_\m{UV}-M_\m{h}$ relation.
The gray, brown, purple, blue, green, orange, and red filled diamonds (circles) denote the halo masses as a function of the UV magnitude at $z\sim1.7$, $2.2$, $3$, $4$, $5$, $6$, and $7$, respectively, for the subsamples in this work (in \citealt{2016ApJ...821..123H}).
We plot $M_\m{min}$ and $M_\m{UV}^\m{th}$ as $M_\m{h}$ and $M_\m{UV}$, respectively.
The crosses are results of the previous work based on the early HSC-SSP data \citep{2018ApJ...859...84H}.
The solid curves show the best-fit relations of Equation (\ref{eq_muv_mh}).
}
\label{fig_MUV_Mh}
\end{figure}

Figure \ref{fig_MUV_Mh} shows our results of the halo mass, $M_\m{h}$, at $z\sim2-6$ as a function of the UV magnitude, $M_\m{UV}$, with previous studies \citep{2016ApJ...821..123H,2018PASJ...70S..11H} at $z\sim4-7$.
We plot $M_\m{min}$ and $M_\m{UV}^\m{th}$ as $M_\m{h}$ and $M_\m{UV}$, respectively.
Table \ref{tab_BCE} summarizes the results of this work at $z\sim2-6$ and of \citet{2016ApJ...821..123H}.
We find that the new results obtained in this work are consistent with our previous measurements in \citet{2018PASJ...70S..11H}, which are indicated by the crosses in Figure \ref{fig_MUV_Mh}.
The halo mass of $z\sim4-6$ galaxies identified in the HSC data ranges from $3\times10^{11}\ M_\odot$ to $1\times10^{13}\ M_\odot$, which is more massive than those of galaxies identified in the {\it Hubble} data \citep{2016ApJ...821..123H}.
The combination of the {\it Hubble} and HSC data allows us to investigate the $M_\m{UV}-M_\m{h}$ relation over two orders of magnitude in the halo mass at $z\sim4$ and $5$.
Thus in the following discussion, we will mainly use the results of this work and of \citet{2016ApJ...821..123H}.
There is a positive correlation between the UV luminosities and the halo masses at all redshifts, indicating that more UV-luminous galaxies reside in more massive halos, as suggested by previous studies.
The slope of the $M_\m{UV}-M_\m{h}$ relation becomes steeper at the brighter magnitude, which is similar to the local $M_\m{*}-M_\m{h}$ relation \citep[e.g.,][]{2012ApJ...744..159L,2013ApJ...770...57B,2019MNRAS.488.3143B,2013MNRAS.428.3121M,2018MNRAS.477.1822M,2015MNRAS.449.1352C}.
Note that the uncertainty of the halo mass at $z\sim6$ is as small as those at $z\sim4-5$ albeit with the large errors in the correlation function measurement at $z\sim6$, because the halo mass is mainly determined by the number density whose uncertainty is $10\%$, and the slope of the halo mass function is very steep at high redshift.

\begin{deluxetable}{ccccc}
\tabletypesize{\footnotesize}
\setlength{\tabcolsep}{1.5pt}
\tablecaption{Halo mass and $SFR/\dot{M}_\m{h}$ in This Study.}
\tablehead{
\colhead{$\bar{z}$} & \colhead{$M_\m{UV}^\m{th}$} & $SFR$ & \colhead{$\m{log}M_\m{h}$} & \colhead{$SFR/\dot{M}_\m{h}\ (10^{-2})$} \\
\colhead{(1)} & \colhead{(2)} & \colhead{(3)} & \colhead{(4)} & \colhead{(5)} }
\startdata
1.7 & -20.5 & $31.0$ & $12.46_{-0.03}^{+0.03}$ & $3.12_{-0.26}^{+0.23}$\\
1.7 & -20.0 & $16.3$ & $12.09_{-0.04}^{+0.03}$ & $4.43_{-0.36}^{+0.44}$\\
1.7 & -19.5 & $8.6$ & $11.79_{-0.04}^{+0.03}$ & $5.20_{-0.44}^{+0.51}$\\
1.7 & -19.0 & $4.5$ & $11.55_{-0.03}^{+0.04}$ & $5.19_{-0.57}^{+0.40}$\\
1.7 & -18.5 & $2.4$ & $11.33_{-0.04}^{+0.04}$ & $4.91_{-0.53}^{+0.52}$\\
1.7 & -18.0 & $1.2$ & $11.16_{-0.04}^{+0.05}$ & $4.06_{-0.56}^{+0.42}$\\
2.2 & -21.0 & $59.1$ & $12.72_{-0.02}^{+0.03}$ & $1.91_{-0.16}^{+0.10}$\\
2.2 & -20.5 & $31.0$ & $12.30_{-0.03}^{+0.03}$ & $3.04_{-0.24}^{+0.23}$\\
2.2 & -20.0 & $16.3$ & $11.95_{-0.02}^{+0.04}$ & $4.02_{-0.45}^{+0.21}$\\
2.2 & -19.5 & $8.6$ & $11.68_{-0.03}^{+0.04}$ & $4.31_{-0.49}^{+0.32}$\\
2.2 & -19.0 & $4.5$ & $11.45_{-0.05}^{+0.02}$ & $4.15_{-0.25}^{+0.52}$\\
2.2 & -18.5 & $2.4$ & $11.26_{-0.02}^{+0.06}$ & $3.59_{-0.60}^{+0.19}$\\
2.9 & -21.5 & $114.1$ & $12.55_{-0.02}^{+0.02}$ & $3.45_{-0.18}^{+0.18}$\\
2.9 & -21.0 & $59.9$ & $12.19_{-0.03}^{+0.02}$ & $4.59_{-0.24}^{+0.35}$\\
2.9 & -20.5 & $31.5$ & $11.92_{-0.02}^{+0.03}$ & $4.85_{-0.40}^{+0.25}$\\
2.9 & -20.0 & $16.5$ & $11.71_{-0.02}^{+0.04}$ & $4.39_{-0.47}^{+0.23}$\\
2.9 & -19.5 & $8.7$ & $11.55_{-0.04}^{+0.03}$ & $3.49_{-0.27}^{+0.34}$\\
2.9 & -19.0 & $4.6$ & $11.36_{-0.03}^{+0.05}$ & $3.00_{-0.42}^{+0.22}$\\
3.8 & -22.5 & $186.0$ & $13.08_{-0.01}^{+0.03}$ & $0.90_{-0.07}^{+0.02}$\\
3.8 & -22.0 & $106.1$ & $12.71_{-0.02}^{+0.02}$ & $1.31_{-0.07}^{+0.06}$\\
3.8 & -21.5 & $60.5$ & $12.32_{-0.02}^{+0.03}$ & $2.00_{-0.16}^{+0.10}$\\
3.8 & -21.0 & $34.5$ & $11.98_{-0.02}^{+0.03}$ & $2.71_{-0.22}^{+0.14}$\\
3.8 & -20.5 & $19.7$ & $11.66_{-0.03}^{+0.02}$ & $3.49_{-0.19}^{+0.26}$\\
3.8 & -20.0 & $11.2$ & $11.48_{-0.02}^{+0.04}$ & $3.14_{-0.34}^{+0.16}$\\
3.8 & -19.7 & $7.8$ & $11.41_{-0.05}^{+0.04}$ & $2.34_{-0.25}^{+0.28}$\\
3.8 & -19.3 & $5.0$ & $11.30_{-0.05}^{+0.04}$ & $1.93_{-0.21}^{+0.23}$\\
3.8 & -18.4 & $1.9$ & $11.04_{-0.05}^{+0.06}$ & $1.44_{-0.24}^{+0.17}$\\
3.8 & -17.4 & $0.6$ & $10.77_{-0.10}^{+0.10}$ & $0.91_{-0.26}^{+0.20}$\\
3.8 & -16.7 & $0.3$ & $10.59_{-0.10}^{+0.08}$ & $0.66_{-0.15}^{+0.14}$\\
4.9 & -22.9 & $325.1$ & $12.95_{-0.01}^{+0.02}$ & $1.36_{-0.07}^{+0.03}$\\
4.9 & -22.4 & $180.4$ & $12.60_{-0.02}^{+0.01}$ & $1.80_{-0.05}^{+0.09}$\\
4.9 & -21.9 & $100.1$ & $12.29_{-0.00}^{+0.03}$ & $2.16_{-0.17}^{+0.00}$\\
4.9 & -21.4 & $55.6$ & $12.00_{-0.01}^{+0.02}$ & $2.46_{-0.13}^{+0.07}$\\
4.9 & -20.9 & $30.8$ & $11.76_{-0.02}^{+0.02}$ & $2.49_{-0.13}^{+0.12}$\\
4.9 & -20.4 & $17.1$ & $11.57_{-0.02}^{+0.03}$ & $2.22_{-0.18}^{+0.11}$\\
4.9 & -19.9 & $9.4$ & $11.37_{-0.04}^{+0.02}$ & $1.90_{-0.10}^{+0.18}$\\
4.9 & -19.5 & $5.9$ & $11.22_{-0.03}^{+0.03}$ & $1.73_{-0.14}^{+0.13}$\\
4.9 & -19.2 & $3.9$ & $11.11_{-0.04}^{+0.04}$ & $1.55_{-0.16}^{+0.15}$\\
4.9 & -17.9 & $0.9$ & $10.80_{-0.12}^{+0.13}$ & $0.76_{-0.30}^{+0.19}$\\
5.9 & -22.2 & $163.3$ & $12.33_{-0.02}^{+0.02}$ & $2.21_{-0.11}^{+0.10}$\\
5.9 & -21.7 & $85.8$ & $12.09_{-0.02}^{+0.01}$ & $2.09_{-0.06}^{+0.11}$\\
5.9 & -21.2 & $45.1$ & $11.78_{-0.02}^{+0.02}$ & $2.36_{-0.12}^{+0.11}$\\
5.9 & -20.1 & $10.2$ & $11.37_{-0.03}^{+0.04}$ & $1.43_{-0.15}^{+0.10}$\\
5.9 & -19.1 & $3.0$ & $11.05_{-0.05}^{+0.04}$ & $0.93_{-0.10}^{+0.11}$\\
6.8 & -19.5 & $4.8$ & $11.08_{-0.08}^{+0.07}$ & $0.98_{-0.18}^{+0.18}$\\
6.8 & -19.3 & $3.8$ & $10.99_{-0.06}^{+0.05}$ & $0.98_{-0.13}^{+0.13}$
\enddata
\tablecomments{(1) Mean redshift.
(2) Threshold absolute magnitude in the rest-frame UV band. 
(3) SFR corresponding to $M_\m{UV}^\m{th}$ after the extinction correction in units of $M_\odot\ \m{yr^{-1}}$.  
(4) Dark matter halo mass ($M_\m{min}$) in units of $M_\odot$.
(5) Ratio of the SFR to the dark matter accretion rate in units of $10^{-2}$. 
}
\label{tab_BCE}
\end{deluxetable}

\begin{figure*}
\centering
\includegraphics[width=0.99\hsize, bb=5 8 1213 354]{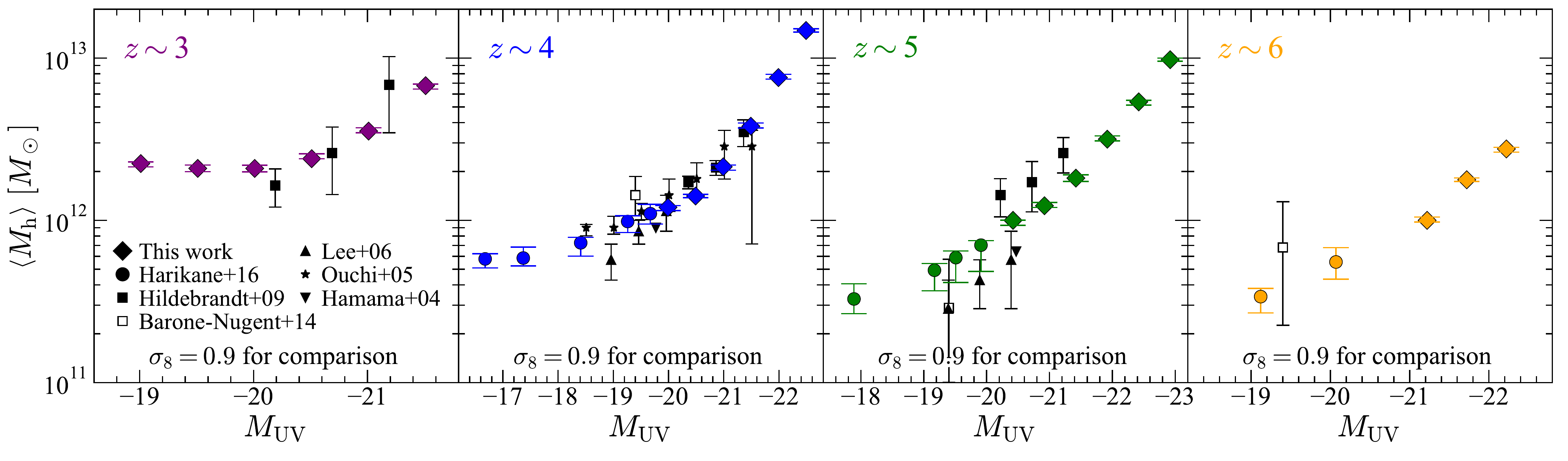}
\caption{Comparison of the mean dark matter halo masses, $\left<M_\m{h}\right>$, with the literature under the same cosmology.
We plot $M_\m{UV}^\m{th}$ as $M_\m{UV}$.
The diamonds represent the mean dark matter halo masses in this work with the cosmological parameters of $(h,\Omega_\m{m},\Omega_\m{\Lambda},\sigma_8) = (0.7,0.3,0.7,0.9)$.
The circles are the results of \citet[][]{2016ApJ...821..123H} with the same cosmology.
The black symbols denote results of the previous studies.
We plot the results of \citet[][squares]{2009A&A...498..725H}, \citet[][upward triangles]{2006ApJ...642...63L}, \citet[][stars]{2005ApJ...635L.117O}, and \citet[][downward triangle]{2004MNRAS.347..813H}.
The downward triangles have no error bars, because \citet{2004MNRAS.347..813H} do not provide errors of the mean dark matter halo mass.
We also show the results of \citet[][]{2014ApJ...793...17B} as black open squares, which are re-calcutaed with the cosmological parameters of $(h,\Omega_\m{m},\Omega_\m{L},\sigma_8) = (0.7,0.3,0.7,0.9)$.
The mean dark matter halo mass of the faintest subsample at $z\sim3$ is slightly more massive than that of the next faintest subsample, because of the higher fraction of satellite galaxies that are typically reside in massive halos (see Table \ref{tab_HOD}).
}
\label{fig_MUVMhave_com}
\end{figure*}

We find a redshift evolution of the $M_\m{UV}-M_\m{h}$ relation from $z\sim1.7$ to $7$.
For example, $M_\m{h}$ monotonically decreases from $z\sim7$ to $1.7$ (from $z\sim6$ to $2.2$) by a factor of 5 (9) at $M_\m{UV}=-19.5\ (-20.5)$.
This redshift evolution indicates that the dust-uncorrected SFR increases with increasing redshift at fixed dark matter halo mass.
We also plot the best-fit $M_\m{UV}-M_\m{h}$ relations at $z\sim1.7$, $2.2$, $3$, $4$, $5$, $6$, and $7$ in Figure \ref{fig_MUV_Mh}.
These relations are expressed with the following double power law function:
\begin{equation}\label{eq_muv_mh}
M_\m{h}=\frac{M_\m{h,0}}{2}\left[10^{-0.4(M_\m{UV}-M_\m{UV,0})\alpha}+10^{-0.4(M_\m{UV}-M_\m{UV,0})\beta}\right]
\end{equation}
where $M_\m{h,0}$ and $M_\m{UV,0}$ are characteristic halo mass and UV magnitude, respectively, and $\alpha$ and $\beta$ are faint and bright end power-law slopes, respectively.
In Figure \ref{fig_MUV_Mh}, we use parameter sets of $(\m{log}M_\m{h,0},M_\m{UV,0},\alpha,\beta)=(11.67,-19.30,0.47,2.11)$, $(11.82,-19.77,0.61,2.34)$, $(12.05,-20.75,0.56,2.15)$, $(11.92,-20.90,0.60,2.29)$, $(11.62,-20.58,0.51,1.67)$, $(11.44,-20.35,0.36,1.54)$, and $(11.32,-20.35,0.36,1.54)$ for $z\sim1.7$, $2.2$, $3$, $4$, $5$, $6$, and $7$, respectively.
At $z\sim7$, $M_\m{UV,0}$, $\alpha$, and $\beta$ are fixed to the values at $z\sim6$.

In Figure \ref{fig_MUVMhave_com}, we compare the mean halo masses, $\left<M_\m{h}\right>$, of our subsamples with the literature.
Because most of the previous studies assume the cosmological parameter set of $(\Omega_\m{m},\Omega_\m{\Lambda},h,\sigma_8) = (0.3,0.7,0.7,0.9)$ that is different from our assumption, we obtain HOD model fitting results for our data with $(\Omega_\m{m},\Omega_\m{\Lambda},h,\sigma_8) = (0.3,0.7,0.7,0.9)$ for comparison.
Similarly, the results of the previous studies are re-calculated with the same cosmological parameter sets if different cosmological parameter set is assumed.
In this way, we conduct our comparisons using an equivalent set of cosmological parameters across all datasets.
In Figure \ref{fig_MUVMhave_com}, we find that our results at $z\sim3$ and $4$ are consistent with those of the previous studies within the uncertainties.
While the previous results at $z\sim 5$ are largely scattered, our $z\sim 5$ results are placed near the center of the distribution of the previous studies. 
At $z\sim 6$, our results agree with that of \citet{2014ApJ...793...17B}.
In summary, our results are consistent with most of the previous studies.
Furthermore, our results improve on both the statistics and the dynamic range covered in $M_\mathrm{UV}$.

\subsection{$SFR/\dot{M}_\m{h}-M_\m{h}$ Relation}\label{ss_BCE}

We estimate a ratio of the SFR to the dark matter accretion rate, $SFR/\dot{M}_\m{h}$, or the baryon conversion efficiency, $SFR/(f_\m{b} \dot{M}_\m{h})$, where $f_\m{b}=\Omega_\m{b}/\Omega_\m{m}$ is the cosmic baryon fraction.
Since the baryon gas accretes into the halo together with dark matter, this ratio indicates the star formation efficiency.
In this paper, the star formation efficiency indicates $SFR/\dot{M}_\m{h}$ or $SFR/(f_\m{b} \dot{M}_\m{h})$, not the ratio of the SFR to the gas mass ($SFR/M_\m{gas}$), which is usually used in radio astronomy. 
We derive the dust-uncorrected SFRs ($SFR_\m{UV}$) from UV luminosities using the calibration used in \citet{2014ARA&A..52..415M} with the \citet{1955ApJ...121..161S} IMF:
\begin{equation}
SFR_\m{UV}\ (M_\odot\ \m{yr}^{-1})=1.15\times10^{-28} L_\m{UV}\ (\m{erg\ s^{-1}\ Hz^{-1}}).\label{eq_LUVSFR}
\end{equation}
We correct the SFR for the dust extinction using an attenuation-UV slope ($\beta_\m{UV}$) relation \citep{1999ApJ...521...64M} and $\beta_\m{UV}-M_\m{UV}$ relation at each redshift.
We use the $\beta_\m{UV}-M_\m{UV}$ relation in \citet{2014ApJ...793..115B} at $z\gtrsim4$ and linearly extrapolate the relation with fixing the slope at $z\lesssim4$.
The estimated SFRs are presented in Table \ref{tab_HOD}.
We calculate $\dot{M}_\m{h}$ as a function of halo mass and redshift using an analytic formula obtained from N-body simulation results in \citet[][Equation (B8)]{2015ApJ...799...32B}.
Note that the accretion rates in \citet{2015ApJ...799...32B} are typically $\sim2$ times lower than those calculated based on Equations (E2)-(E6) in \citet{2013ApJ...770...57B}, which are used in our previous work \citep{2018PASJ...70S..11H}, because the \citet{2013ApJ...770...57B} accretion rates only trace the progenitors of $z=0$ halos.

\begin{figure}
\centering
\includegraphics[width=0.99\hsize, bb=7 6 428 286]{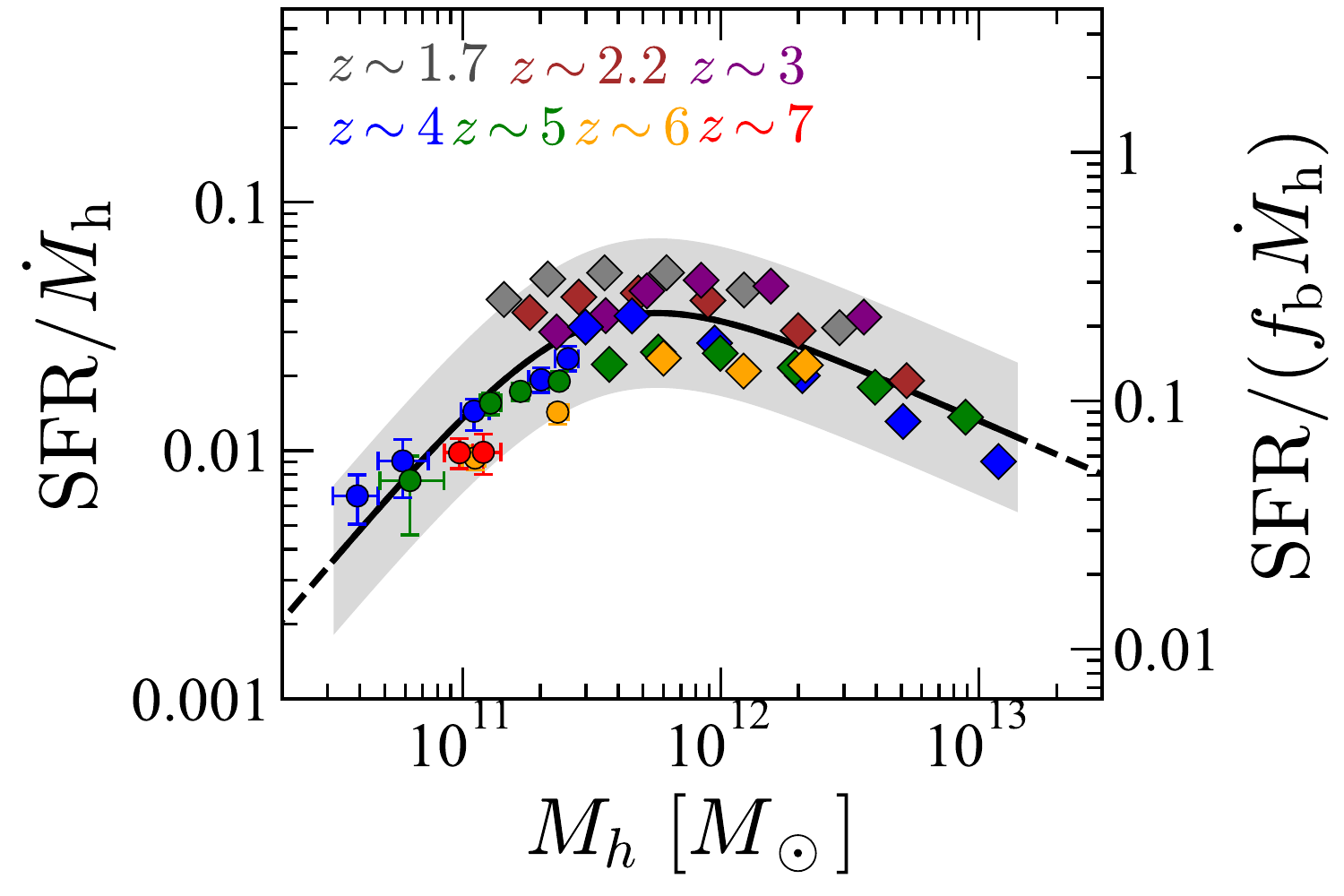}
\caption{$\m{SFR}/\dot{M}_\m{h}$ and baryon conversion efficiency ($\m{SFR}/(f_\m{b}\dot{M}_\m{h})$) as a function of the halo mass.
The gray, brown, purple, blue, green, orange, and red filled diamonds (circles) denote the ratios as a function of the halo mass, at $z\sim1.7$, $2.2$, $3$, $4$, $5$, $6$, and $7$, respectively, for the subsamples in this work (in \citealt{2016ApJ...821..123H}).
The statistical errors for our data are smaller than the symbols (diamonds).
The black solid curve is the fitting formulae of Equation (\ref{eq_SFRdMdt_1}) for the $SFR/\dot{M}_\m{h}-M_\m{h}$ relation at $z\sim2-7$, and the gray shaded region represents the $2\sigma$ typical scatter ($0.3\ \m{dex}$) of the data points compared to the relation.}
\label{fig_SFR_dMdt}
\end{figure}

We plot the dust-corrected $SFR/\dot{M}_\m{h}$ ratios at $z\sim2-7$ as a function of the halo mass in Figure \ref{fig_SFR_dMdt}.
The results are also summarized in Table \ref{tab_BCE}.
The black solid curve in Figure \ref{fig_SFR_dMdt} represents the following $SFR/\dot{M}_\m{h}-M_\m{h}$ relation:
\begin{eqnarray}
\frac{SFR}{\dot{M}_\m{h}}=\frac{2\times3.2\times10^{-2}}{(M_\m{h}/10^{11.5})^{-1.2}+(M_\m{h}/10^{11.5})^{0.5}}\label{eq_SFRdMdt_1}
\end{eqnarray}
This relation agrees with the measured $SFR/\dot{M}_\m{h}$ ratios at $z\sim2-7$ within 0.3 dex that is a typical $2\sigma$ scatter.
This good agreement indicates that the star formation efficiency does not significantly change beyond 0.3 dex in the wide redshift range of $z\sim2-7$, suggesting the existence of the fundamental relation between the star formation and the mass accretion (the growth of the galaxy and its dark matter halo assembly), as discussed in \citet{2018PASJ...70S..11H} at $z\sim4-7$ (see also \citealt{2013ApJ...774...28B}).

\begin{figure*}
\centering
\includegraphics[width=0.9\hsize, bb=9 2 1215 354]{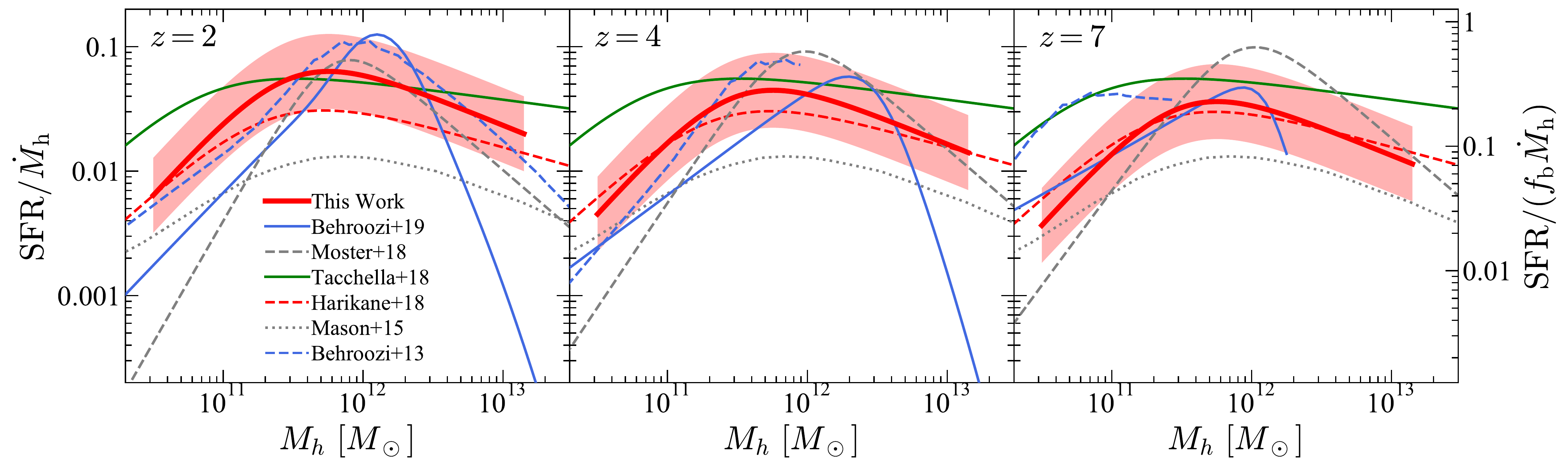}
\caption{
Comparison of the $\m{SFR}/\dot{M}_\m{h}$ ratio with the literature.
The red solid curve shows the redshift-dependent relation constrained in this work (Equation (\ref{eq_SFRdMdt_2})), and the shaded region represents the $2\sigma$ typical uncertainties.
Results compared include those from our previous work (Equation (25) in \citealt{2018PASJ...70S..11H}, red dashed curve), \citet[][blue solid curve]{2019MNRAS.488.3143B}, \citet[][gray dashed curve]{2018MNRAS.477.1822M}, \citet[][green solid curve]{2018ApJ...868...92T}, \citet[][gray dotted curve]{2015ApJ...813...21M}, and \citet[][blue dashed curve]{2013ApJ...770...57B}.
All results are converted to use the \citet{1955ApJ...121..161S} IMF (equation (\ref{eq_LUVSFR})).
The result of \citet[][]{2018PASJ...70S..11H} is re-calculated based on the accretion rate of \citet{2015ApJ...799...32B}, and thus is $\sim2$ times higher than the original relation in \citet[][]{2018PASJ...70S..11H}.
}
\label{fig_SFR_dMdt_com}
\end{figure*}

On the other hand, the $SFR/\dot{M}_\m{h}$ ratio gradually increases with increasing redshift within 0.3 dex from $z\sim5$ to $1.7$.
If we take this possible evolution into account, the ratio can be expressed as, 
\begin{eqnarray}
\frac{SFR}{\dot{M}_\m{h}}
&=&\frac{2\times3.2\times10^{-2}}{(M_\m{h}/10^{11.5})^{-1.2}+(M_\m{h}/10^{11.5})^{0.5}}\notag\\
&&\times(0.53\,\m{tanh}[0.54(2.9-z)]+1.53)\ \ \ \ \ 
\label{eq_SFRdMdt_2}
\end{eqnarray}
This relation indicates that the star formation efficiency does not significantly change from $z\sim7$ to $5$, and then gradually increases within a factor of $\sim2$ from $z\sim5$ to $1.7$, still consistent with the results of \citet{2018PASJ...70S..11H}, who have identified redshift-independent relation at $z\sim4-7$ within 0.15 dex.
The reason for this elevated efficiency at $z<5$ is not clear.
One possibility is an increase of the metallicity in galaxies toward lower redshift, resulting in more efficient gas cooling.

In Figure \ref{fig_SFR_dMdt_com}, we compare our $\m{SFR}/\dot{M}_\m{h}-M_\m{h}$ relation with results in the literature \citep{2013ApJ...770...57B,2015ApJ...813...21M,2018PASJ...70S..11H,2018ApJ...868...92T,2018MNRAS.477.1822M,2019MNRAS.488.3143B}.
The relations in the literature show similar trends to our result; the $\m{SFR}/\dot{M}_\m{h}$ ratio has a peak of $\m{SFR}/\dot{M}_\m{h}\sim0.1-0.01$ around the halo mass of $10^{11}-10^{12}\ M_\odot$.
However, the slopes of the relation at the high-mass and low-mass ends are different between these studies.
These differences are possibly due to differences in used observational datasets, the halo mass functions, and details of modeling.
\redc{We also note that there are several systematic uncertainties on the $\m{SFR}/\dot{M}_\m{h}$ measurements.
Assumptions on the IMF and dust attenuation have impacts on the SFR.
The conversion factor between the SFR and UV luminosity (Equation (\ref{eq_LUVSFR})) depends on the stellar age and metallicity \citep[e.g.,][]{2014ARA&A..52..415M}.}

\section{Discussion}\label{ss_discussion}

\subsection{Physical Origin of the Cosmic SFR Density Evolution}

In Section \ref{ss_BCE}, we find the fundamental $SFR/\dot{M}_\m{h}-M_\m{h}$ relation; the value of $SFR/\dot{M}_\m{h}$ at fixed $M_\m{h}$ does not significantly change beyond $0.3\ \mathrm{dex}$ at $z\sim2-7$.
We examine whether this fundamental $SFR/\dot{M}_\m{h}-M_\m{h}$ relation is consistent with the observational results, i.e., cosmic SFR densities and the UV luminosity functions.
We calculate the cosmic SFR density as follows:
\begin{equation}
\rho_\m{SFR}=\int dM_\m{h} \frac{dn}{dM_\m{h}}SFR=\int dM_\m{h} \frac{dn}{dM_\m{h}} \dot{M}_\m{h} \frac{SFR}{\dot{M}_\m{h}},\label{eq_cSFR}
\end{equation}
where $SFR/\dot{M}_\m{h}$ at $z\sim2-7$ is obtained as a function of $M_\m{h}$ in Section \ref{ss_BCE}  (Equations (\ref{eq_SFRdMdt_1}) or (\ref{eq_SFRdMdt_2})).
We integrate down to the halo mass corresponding to the SFR of $0.3\ M_\odot\ \m{yr^{-1}}$ ($M_\m{UV}=-17$ mag with the \citealt{2014ARA&A..52..415M} calibration), as previous studies \citep{2015ApJ...803...34B,2020ApJ...902..112B,2015ApJ...810...71F,2018ApJ...855..105O}.

\begin{figure}
\centering
\includegraphics[width=0.99\hsize, bb=7 9 358 356]{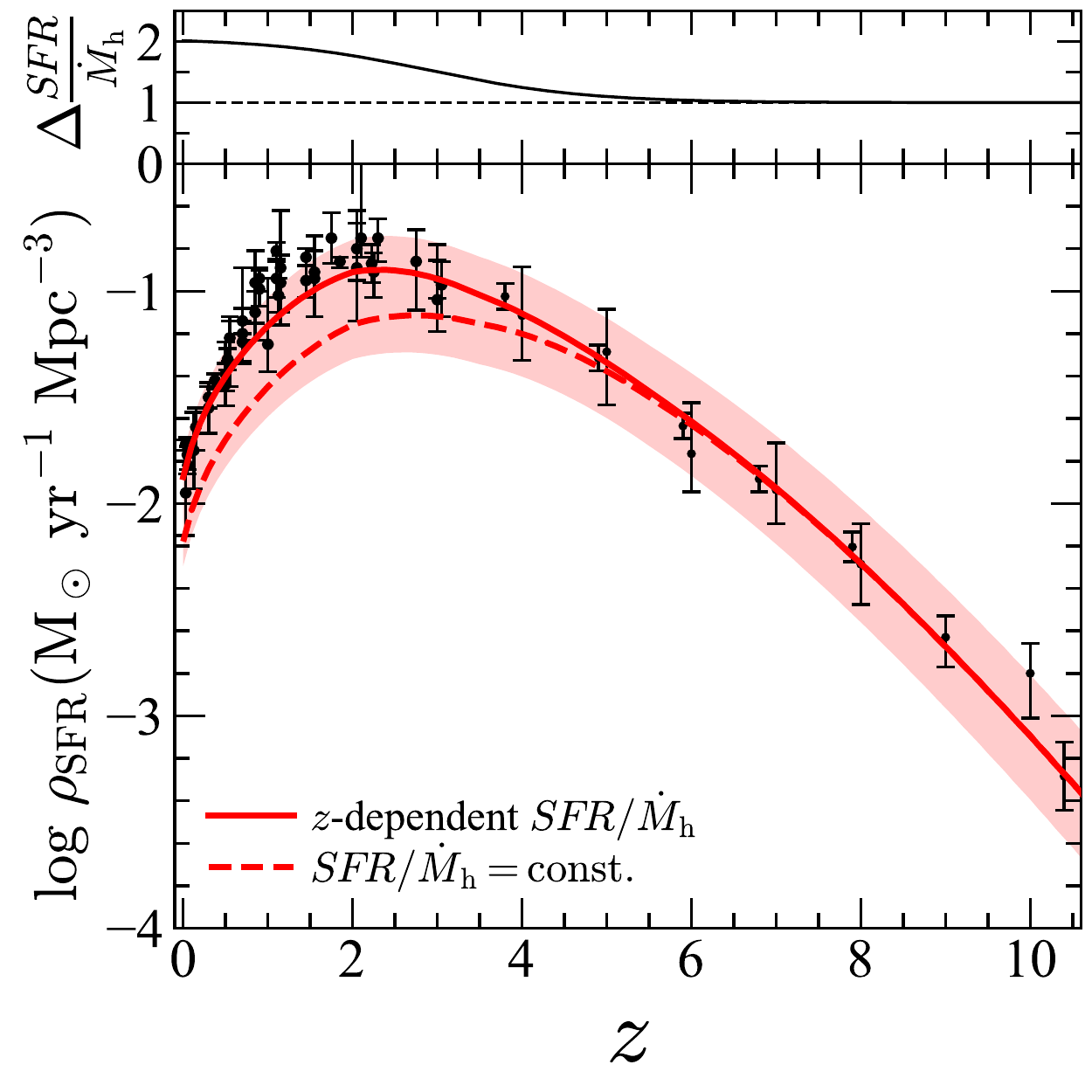}
\caption{
Cosmic SFR density.
The bottom panel shows the comparison of the cosmic SFR densities. 
The red curves with the shade represent cosmic SFR densities derived from Equation (\ref{eq_cSFR}). 
For the solid and dashed lines, we assume the redshift-dependent and independent $SFR/\dot{M}_\m{h}-M_\m{h}$ relations (Equations (\ref{eq_SFRdMdt_2}) and (\ref{eq_SFRdMdt_1})), respectively, which are constrained in this paper at $2\lesssim z\lesssim7$.
The shaded regions correspond to the $1\sigma$ (0.15 dex) scatter in the $SFR/\dot{M}_\m{h}-M_\m{h}$ relation.
We integrate down to the halo mass corresponding to the SFR of $0.3\ M_\odot\ \m{yr^{-1}}$ ($M_\m{UV}=-17$ mag with the \citealt{2014ARA&A..52..415M} calibration), as previous studies \citep{2015ApJ...803...34B,2020ApJ...902..112B,2015ApJ...810...71F,2018ApJ...855..105O}.
The black circles show observed cosmic SFR densities taken from \citet{2014ARA&A..52..415M}, \citet{2015ApJ...810...71F}, \citet{2016MNRAS.459.3812M}, and \citet{2020ApJ...902..112B}.
All results are converted to use the calibration of \citet{2014ARA&A..52..415M} with the \citet{1955ApJ...121..161S} IMF (Equation (\ref{eq_LUVSFR})).
The top panel shows $\Delta\frac{SFR}{\dot{M}_\m{h}}=\frac{SFR}{\dot{M}_\m{h}}(z)/\frac{SFR}{\dot{M}_\m{h}}(z=5)$ at a given halo mass.
The sold and dashed curves are calculated with Equations (\ref{eq_SFRdMdt_2}) and (\ref{eq_SFRdMdt_1}), respectively.
}
\label{fig_cSFR}
\end{figure}

Firstly we assume the redshift-independent $SFR/\dot{M}_\m{h}-M_\m{h}$ relation (Equation (\ref{eq_SFRdMdt_1})).
Figure \ref{fig_cSFR} compares our calculated SFR densities (the dashed curve) with observational results in the literature \citep{2014ARA&A..52..415M,2015ApJ...810...71F, 2016MNRAS.459.3812M,2018ApJ...855..105O,2020ApJ...902..112B}.
The results in the literature are all converted to use the calibration of \citet{2014ARA&A..52..415M} with the \citet{1955ApJ...121..161S} IMF (Equation (\ref{eq_LUVSFR})).
We find that our calculation well reproduces the overall trend of the cosmic SFR density evolution; the calculated density increases from $z\sim10$ to $4-2$, and decreases from $z\sim4-2$ to $0$.
However, the SFR densities are underpredicted compared to the observations at $z\sim1-2$ by $\sim0.3$ dex.

Then we use the gradually evolving $SFR/\dot{M}_\m{h}-M_\m{h}$ relation (Equation (\ref{eq_SFRdMdt_2})), instead of Equation (\ref{eq_SFRdMdt_1}).
As shown in Figure \ref{fig_cSFR}, our calculated cosmic SFR densities (the solid curve) based on Equation (\ref{eq_SFRdMdt_2}) agree well with the observations especially at $z\sim1-2$, compared to the calculation based on the redshift-independent relation (the dashed curve).
Quantitatively, the reduced $\chi^2$ value improves significantly from $\chi^2/\m{dof}=25.4$ to $3.0$.
These analyses indicate that the overall trend of the redshift evolution can be reproduced by the redshift-independent $SFR/\dot{M}_\m{h}-M_\m{h}$ relation (Equation (\ref{eq_SFRdMdt_1})), but the gradual increase of the star formation efficiency at $z<5$ (Equation (\ref{eq_SFRdMdt_2})) is needed to quantitatively reproduce the observed SFR densities.
Note that we have constrained the $SFR/\dot{M}_\m{h}-M_\m{h}$ relations by using normal star forming galaxies (dropout, BX, and BM galaxies), and not considered quiescent or dusty starburst galaxies.
Quiescent and dusty starburst galaxies are expected to have lower and higher $SFR/\dot{M}_\m{h}$ ratios than the normal star forming galaxies, and \redc{effects of these galaxies are thought to be less dominant at $z\gtrsim4$ \citep[e.g.,][]{2013ApJ...777...18M,2020ApJ...902..112B,2021arXiv210910378F}}.
Although quiescent and dusty starburst galaxies would be non-negligible at $z\lesssim3$, the good agreement at $z\sim1-2$ still indicates that the star formation efficiency averaged over all galaxy populations gradually increases at $z<5$, as long as the observed SFR densities (including the treatment of the dust extinction correction) are correct.

\begin{figure}
\centering
\includegraphics[width=0.99\hsize, bb=36 29 548 860]{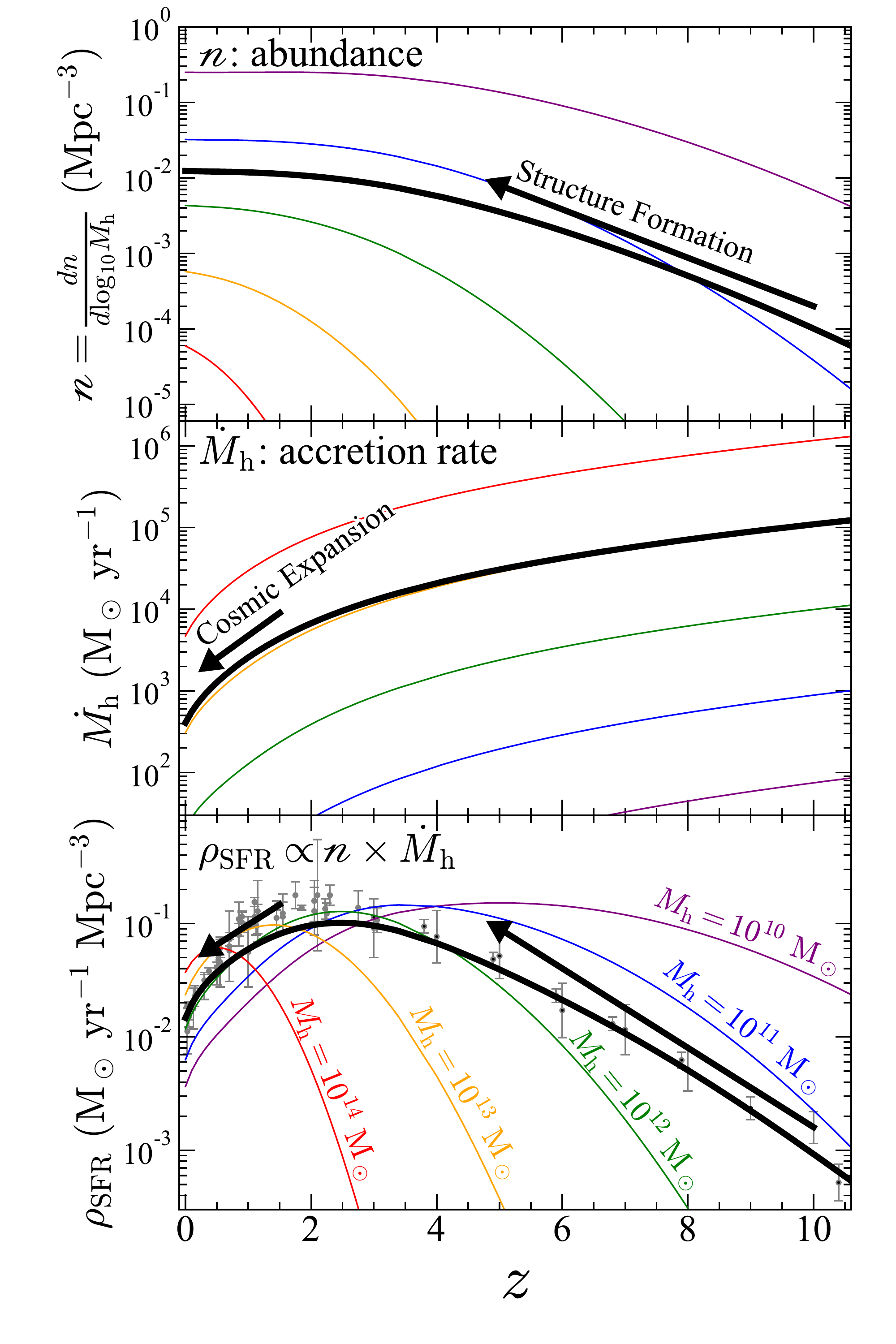}
\caption{Mechanism of the cosmic SFR density evolution. 
{\it Top panel:} the purple, blue, green, orange, and red curves indicate the number density of halos whose masses are $M_\m{h}=10^{10}$, $10^{11}$, $10^{12}$, $10^{13}$, and $10^{14}\ M_\odot$, respectively. 
The number density is calculated by using the \citet{2013ApJ...770...57B} halo mass function.
The black curve represents a weighted number density based on Equation (\ref{eq_SFRdMdt_2}).
{\it Middle panel:} same as the top panel but for the dark matter accretion rate calculated by using the formula in \citet{2015ApJ...799...32B}.
{\it Bottom panel:} same as the top panel but for the cosmic SFR density.
The gray circles are observed cosmic SFR densities taken from \citet{2014ARA&A..52..415M}, \citet{2015ApJ...810...71F}, \citet{2016MNRAS.459.3812M}, and \citet{2020ApJ...902..112B}.
Since the cosmic SFR density at a given halo mass is proportional to the galaxy (halo) number densities and SFRs (accretion rates) as shown in Equation (\ref{eq_cSFR}), the redshift evolution of the cosmic SFR density is made by the monotonic steep increase of the halo number density from $z\sim10$ to $z\sim4$ and the monotonic decrease of the accretion rate from $z\sim2$ to $z\sim0$, resulting a peak of the SFR density around $z\sim2-3$.
}
\label{fig_HMF_dMdt_panel}
\end{figure}

The good agreement with the overall trend indicates that the evolution of the cosmic SFR densities is primarily driven by the monotonic increase of the halo number density and the monotonic decrease of the accretion rate, given the weak redshift evolution of the $SFR/\dot{M}_\m{h}-M_\m{h}$ relation, as discussed in \citet{2018PASJ...70S..11H}.
The number density of halos at a given halo mass increases due to structure formation from $z\sim10$ 
to a certain redshift at $z\lesssim4$ depending on the mass and then becomes almost constant after that (the top panel in Figure \ref{fig_HMF_dMdt_panel}), resulting in the increase of the galaxy number density from $z\sim10$ to $z\lesssim4$.
The dark matter (and gas) accretion rate monotonically decreases over the whole redshift range due to the cosmic expansion, with a steep drop from $z\sim2$ to $0$ (the middle panel in Figure \ref{fig_HMF_dMdt_panel}), resulting in the monotonic decrease SFR of each galaxy at a given halo mass.
Because the cosmic SFR density at a given halo mass is proportional to the number density and mass accretion rate (or SFR) as shown in Equation (\ref{eq_cSFR}), the calculated cosmic SFR density has a peak at $z\sim2-3$ (the bottom panel in Figure \ref{fig_HMF_dMdt_panel}).
More specifically, the product of the number density and mass accretion rate for each halo mass has a peak at a certain redshift due to the increase of the number density and decrease of the accretion rate, with the peak redshift depending on the halo mass, and the $SFR/\dot{M}_\odot-M_\m{h}$ relation determines the peak redshift of the cosmic SFR density integrated over the halo mass, as shown in the bottom panel in Figure \ref{fig_HMF_dMdt_panel}).

In our calculation we integrate Equation (\ref{eq_cSFR}) down to the SFR of $0.3\ M_\odot\ \m{yr^{-1}}$, corresponding to the halo mass of $\sim3\times10^{10}\ M_\odot$ at $z\sim7$.
This integration limit is chosen to match the calculations at $z\gtrsim2$ in previous studies \citep{2015ApJ...803...34B,2020ApJ...902..112B,2015ApJ...810...71F,2018ApJ...855..105O}, and slightly different from the calculations in \citet{2014ARA&A..52..415M}, who integrate down to $0.03L^*$.
This difference does not affect our discussions above.
\citet{2021arXiv210207775B} show that the $M_\m{UV}^*$ parameter of the luminosity function is almost constant ($M_\m{UV}^*\sim-21$ mag) from $z\sim8$ to $2$, and then decreases to $M_\m{UV}^*\sim-18$ mag toward $z\sim0$.
At $z\sim2-8$, the SFR corresponding to $0.03L^*$ with $M_\m{UV}^*=-21$ mag is roughly $0.5\ M_\odot\ \m{yr^{-1}}$, comparable to our integration limit.
Below $z\sim2$, the corresponding SFR is smaller than our limit, e.g., $0.03\ M_\odot\ \m{yr^{-1}}$ with $M_\m{UV}^*\sim-18$ mag at $z\sim0$. 
However, even if we integrate Equation (\ref{eq_cSFR}) down to this SFR limit, the calculated cosmic SFR density increases only by 0.1 dex at $z<3$, and cannot explain the 0.3 dex difference between the observed SFR densities and the dashed curve in Figure \ref{fig_cSFR}.

\begin{figure}
\centering
\includegraphics[width=0.99\hsize, bb=3 11 394 283]{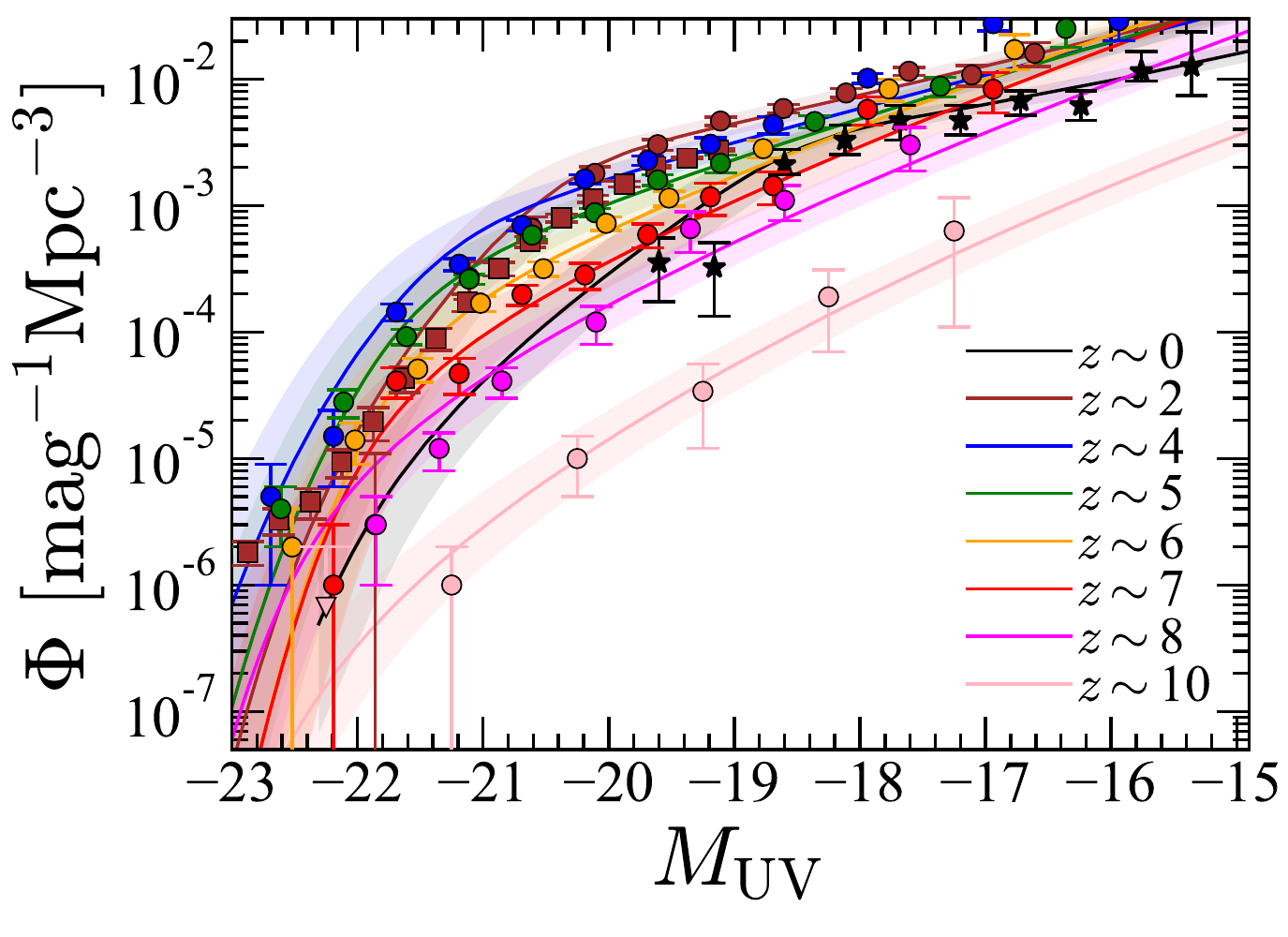}
\caption{Comparison of the rest-frame UV luminosity functions.
Solid curves are the calculated luminosity functions from Equation (\ref{eq_phi_acc}) with the $\m{SFR}/\dot{M}_\m{h}-M_\m{h}$ relation of Equation (\ref{eq_SFRdMdt_2}).
The points show the observed galaxy luminosity functions of \citet[][circles]{2021arXiv210207775B}, \citet[][squares]{2020MNRAS.494.1894M}, and \citet[][stars]{2005ApJ...619L..43A}.
The shaded regions correspond to the $1\sigma$ (0.15 dex) scatter in the $SFR/\dot{M}_\m{h}-M_\m{h}$ relation.
}
\label{fig_LF_SFRdMdt}
\end{figure}

We also calculate the UV luminosity function at each redshift as follows:
\begin{equation}\label{eq_phi_acc}
\Phi(M_\m{UV})=\frac{dn}{dM_\m{h}}\frac{dM_\m{h}}{dM_\m{UV}}.
\end{equation}
From Equation (\ref{eq_SFRdMdt_2}), we can obtain the $M_\m{UV}-M_\m{h}$ relation and $\frac{dM_\m{h}}{dM_\m{UV}}$ at each redshift, since $\dot{M}_\m{h}$ can be expressed as a function of $M_\m{h}$ and $z$ \citep{2015ApJ...799...32B}.
Note that $M_\m{UV}$ is the observed absolute magnitude after dust extinction assuming the attenuation-UV slope ($\beta_\m{UV}$) relation \citep{1999ApJ...521...64M} and $\beta_\m{UV}-M_\m{UV}$ relations \citep{2014ApJ...793..115B}.
We correct for satellite galaxies using satellite fractions measured in previous studies \citep{2011ApJ...728...46W,2015MNRAS.446..169M,2015MNRAS.449..901M,2016ApJ...821..123H,2018PASJ...70S..11H}, although the correction is not large at high redshift.
The calculated UV luminosity functions at $z\sim0-10$ are plotted in Figure \ref{fig_LF_SFRdMdt}.
We find that the calculated luminosity functions are in rough agreement with observed results given the $0.15\ \m{dex}$ ($1\sigma$) uncertainty in $SFR/\dot{M}_\m{h}$, indicating that our $SFR/\dot{M}_\m{h}-M_\m{h}$ relation is consistent with the observed redshift evolution of the UV luminosity function.

\citet{2015ApJ...813...21M} and \citet{2018ApJ...868...92T} also report that the constant star formation efficiency model can reproduce the UV luminosity functions at $z\gtrsim4$.
This is consistent with our results of the $SFR/\dot{M}_\m{h}-M_\m{h}$ relation in this study and our previous work \citep{2018PASJ...70S..11H}.
\citet{2021arXiv210207775B} claim that the evolution of the luminosity function at $z\sim2.5-10$ can be explained by the halo mass function and the constant star formation efficiency model.
This is qualitatively consistent with our results, but the gradual increase of the star formation efficiency at $z<5$ is needed to quantitatively reproduce the redshift evolution of the cosmic SFR density, as discussed above.

\subsection{Future Prospects for Star Formation at $z>10$}

\begin{figure*}
\centering
\includegraphics[width=0.9\hsize, bb=15 6 707 355]{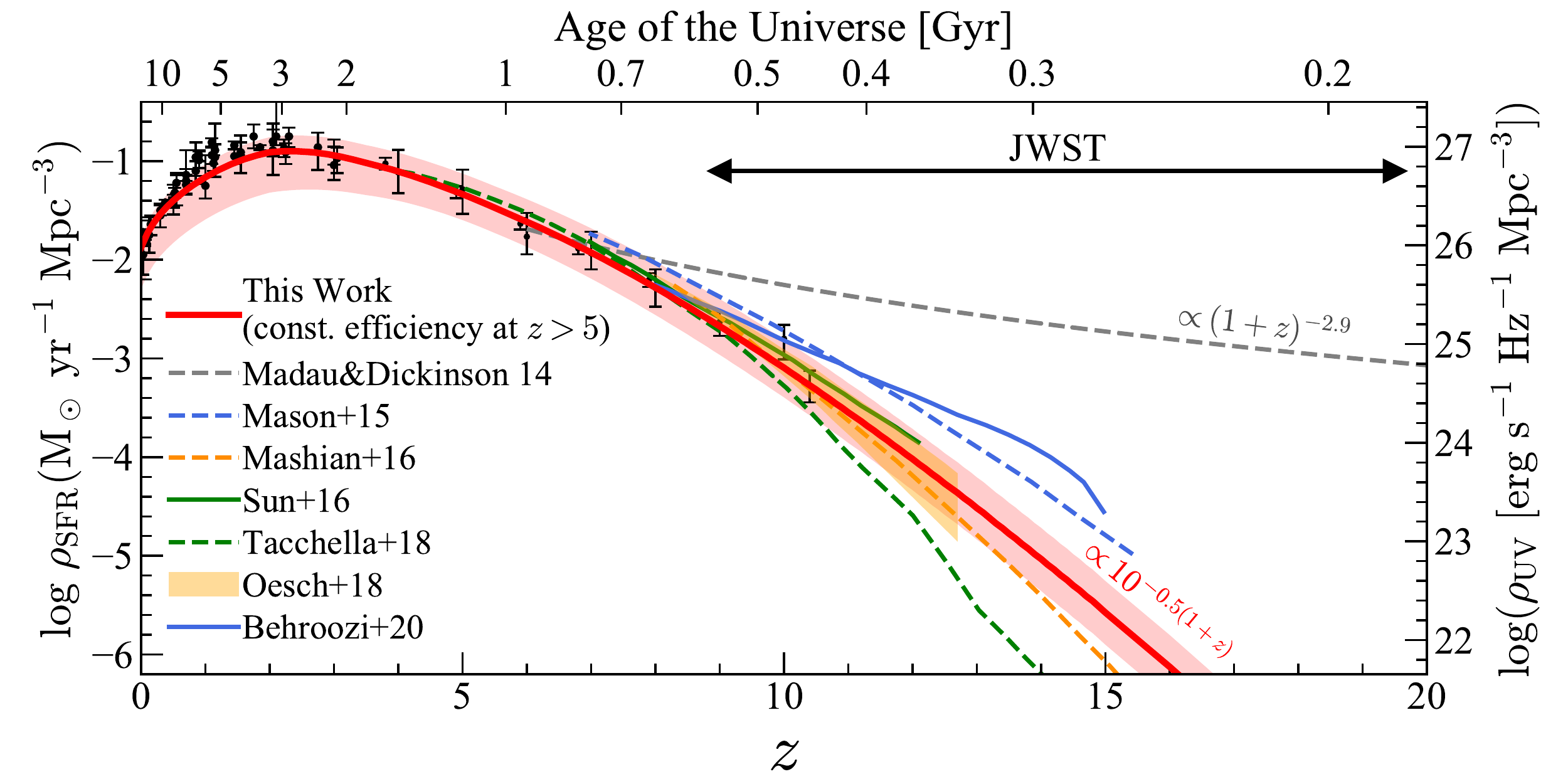}
\caption{
Comparison of the cosmic SFR density at $z>7$.
The red curve with the shade represents the cosmic SFR density calculated in this work based on the constant star formation efficiency at $z>5$ (Equation (\ref{eq_SFRdMdt_2})), integrated down to the SFR of $0.3\ M_\odot\ \m{yr^{-1}}$ ($M_\m{UV}=-17$ mag), as previous studies \citep{2015ApJ...803...34B,2020ApJ...902..112B,2015ApJ...810...71F,2018ApJ...855..105O}.
The gray dashed curve shows the extrapolation of the relation of \citet{2014ARA&A..52..415M} at $z>6$.
The other curves show predictions from models of \citet[][blue dashed curve]{2015ApJ...813...21M}, \citet[][orange dashed curve]{2016MNRAS.455.2101M}, \citet[][green solid curve]{2016MNRAS.460..417S}, \citet[][green dashed curve]{2018ApJ...868...92T}, and \citet[][blue solid curve]{2020MNRAS.499.5702B}.
The orange shaded region indicates a prediction of the halo evolution model in \citet{2018ApJ...855..105O}.
All results are converted to use the \citet{1955ApJ...121..161S} IMF (equation (\ref{eq_LUVSFR})).
}
\label{fig_cSFR_highz}
\end{figure*}

In the previous section, we show that the constant star formation efficiency (Equation (\ref{eq_SFRdMdt_1})) can reproduce the evolution of the cosmic SFR density at $5\lesssim z\lesssim10$.
By assuming that this $SFR/\dot{M}_\m{h}-M_\m{h}$ relation does not evolve to higher redshifts, we can predict the cosmic SFR density at $z>10$ based on the evolution of the halo mass function and dark matter accretion rate.

\begin{figure}
\centering
\includegraphics[width=0.9\hsize, bb=11 7 354 283]{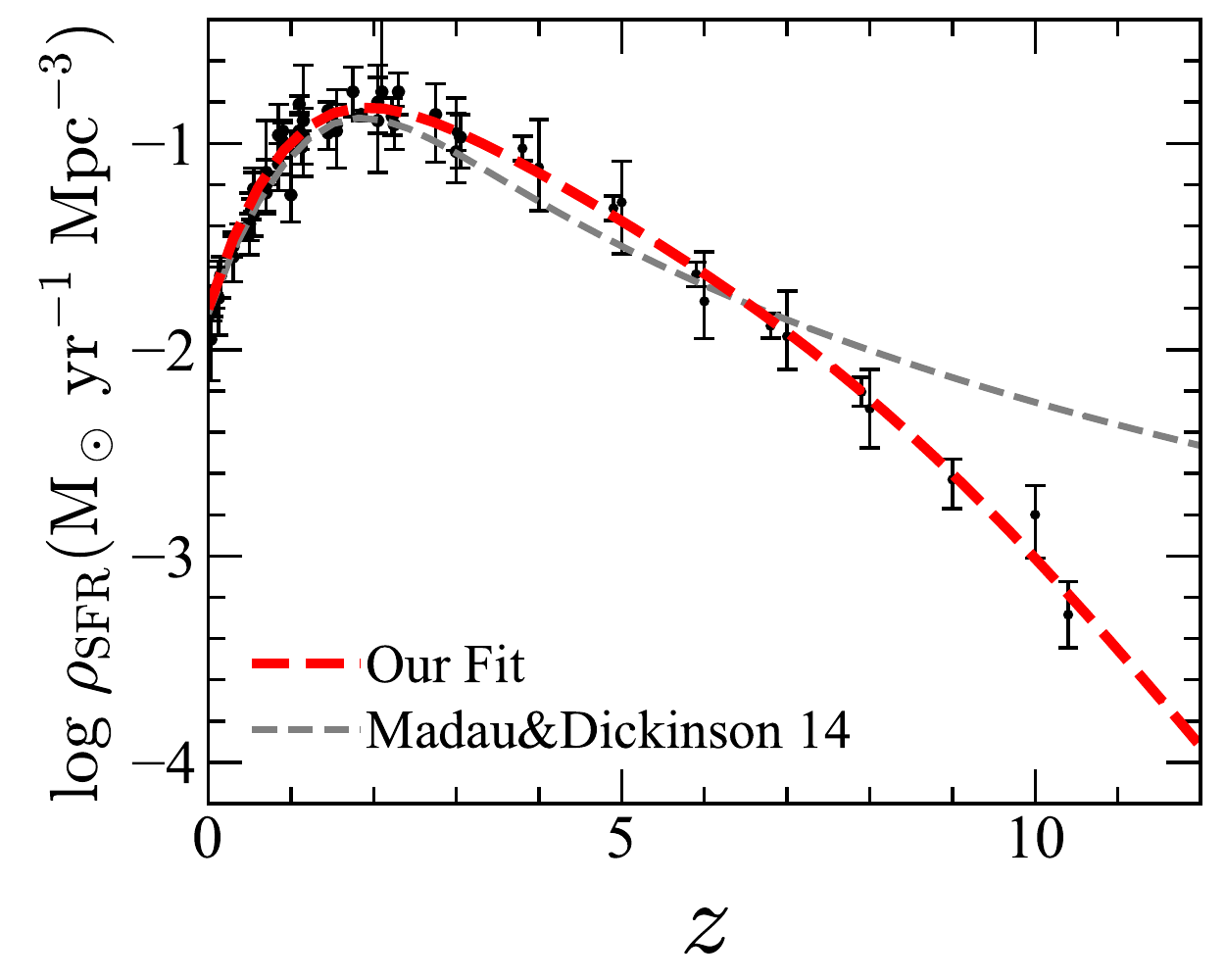}
\caption{
Fit to the observed cosmic SFR densities.
The red dashed curve represents our fit (Equation (\ref{eq_SFRD_fit})) to the observed cosmic SFR densities at $z\lesssim10$ and the calculated SFR densities at $z>10$ in this work.
The gray dashed curve shows the fit in \citet{2014ARA&A..52..415M}.
All results are converted to use the \citet{1955ApJ...121..161S} IMF (equation (\ref{eq_LUVSFR})).
}
\label{fig_cSFR_fit}
\end{figure}

Figure \ref{fig_cSFR_highz} compares our calculated SFR densities (the red curve) with predictions from models in the literature (\citealt[][]{2015ApJ...813...21M}, \citealt[][]{2016MNRAS.455.2101M}, \citealt[][]{2016MNRAS.460..417S}, \citealt[][]{2018ApJ...868...92T}, \citealt{2018ApJ...855..105O}, and \citealt[][]{2020MNRAS.499.5702B}).
We find that the cosmic SFR density based on the constant star formation efficiency rapidly decreases with increasing redshift as $\propto10^{-0.5(1+z)}$, similar to the predictions of other models.
More quantitatively, the SFR densities from observations at $z\lesssim10$ and our predictions at $z\gtrsim10$ are well-fitted with the following function:
\begin{align}
&\rho_\m{SFR}/[\m{M_\odot\ \m{yr^{-1}}\ \m{Mpc^{-3}}}]\notag\\
&\hspace{-0.5cm}=\frac{1}{61.7\times(1+z)^{-3.13} + 1.0\times10^{0.22(1+z)} + 2.4\times10^{0.50(1+z)-3.0}},
\label{eq_SFRD_fit}
\end{align}
as shown in Figure \ref{fig_cSFR_fit}.
This is contrast to the extrapolation of the fitting function in \citet{2014ARA&A..52..415M} that shows a smooth decline as $\propto(1+z)^{-2.9}$ at $z>10$ (the gray dashed curve in Figures \ref{fig_cSFR_highz} and \ref{fig_cSFR_fit}), and possible estimates of the SFR densities at $z>12$ based on $z\sim6$ passive galaxies in \citet{2020ApJ...889..137M}.
{\it James Webb Space Telescope (JWST)} will allow us to directly observe galaxies at $z\sim10-20$, and investigate whether the SFR density rapidly decreases as predicted in many models or not, providing insights into star formation efficiencies in $z\sim10-20$ galaxies.

\begin{figure}
\centering
\includegraphics[width=0.99\hsize, bb=13 7 358 286]{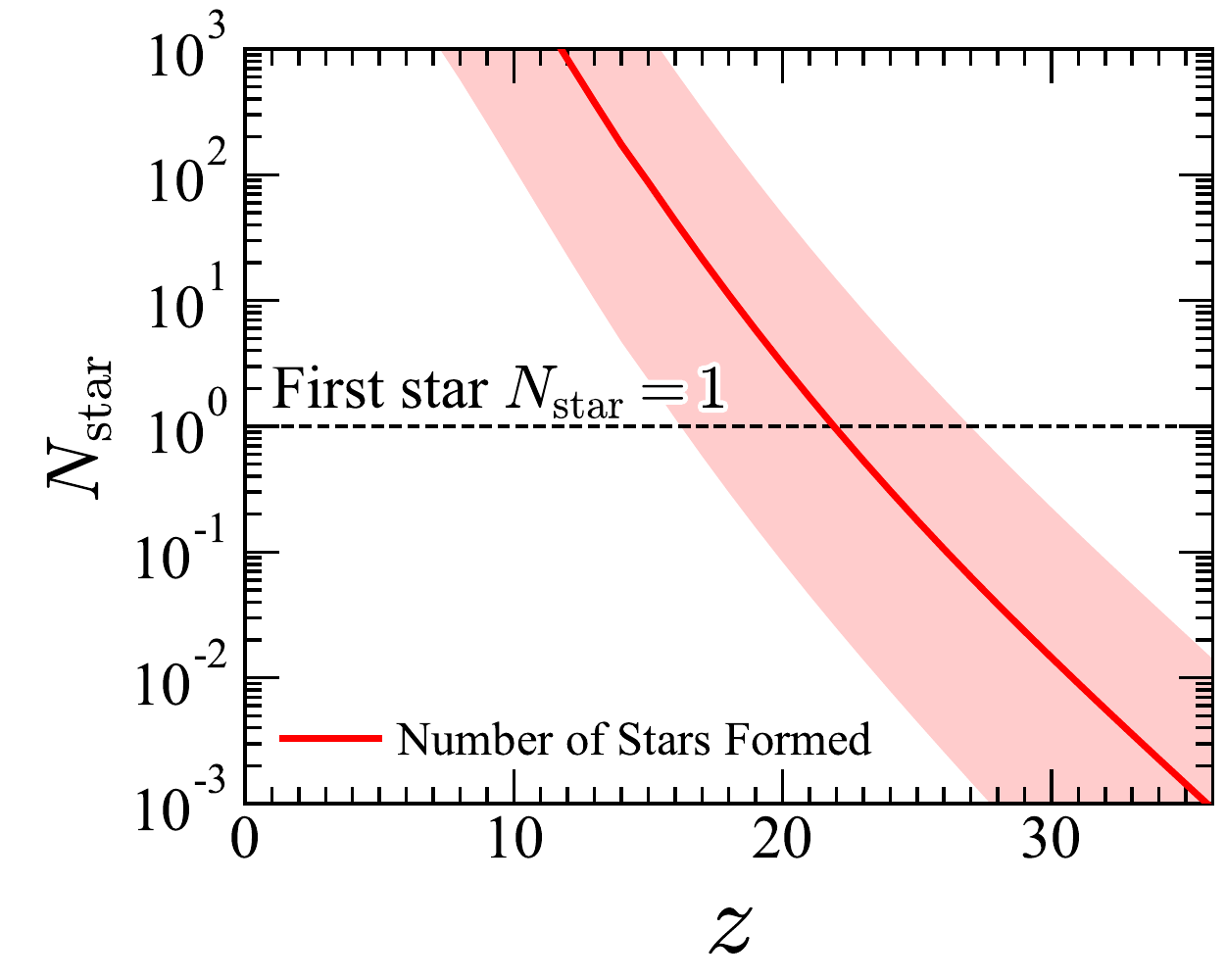}
\caption{
Cumulative number of formed stars as a function of the redshift.
The red line shows the cumulative number of formed stars calculated based on the SFR density (Equation (\ref{eq_firststar})).
We extrapolate the $SFR/\dot{M}_\m{h}-M_\m{h}$ relation both to the higher redshift and lower mass range, and the shaded region indicates possible uncertainties of the $SFR/\dot{M}_\m{h}$ ratio and the mass limit of the integration.
The cumulative number reaches 1 around $z\sim16-27$, implying the first star formation at this epoch.
}
\label{fig_z_firststar}
\end{figure}

Using this method to predict the SFR density at $z>10$, we can obtain a rough estimate of the epoch of the first star formation.
We calculate the cumulative number of formed stars as a function of redshift:
\begin{equation}
N_\m{star}(z)=V_\m{survey}\int^{t(z)}_{{t(z)}-t_\m{LF}}\!\!\!dt\rho_\m{SFR}/ M_\mathrm{FS},\label{eq_firststar}
\end{equation}
where $V_\m{survey}$ is the survey volume, $t_\m{LF}$ is the typical lifetime of the star, and $M_\mathrm{FS}$ is a typical mass of the first star.
Here we assume $t_\m{LF}=3\ \m{Myr}$ \citep{2002A&A...382...28S} and $M_\m{FS}=100\ M_\odot$ \citep[e.g.,][]{2015MNRAS.448..568H}.
We adopt $V_\m{survey}=(3\ h^{-1}\m{Mpc})^3$ that is the volume of the simulation box in \citet{2015MNRAS.448..568H}.
We calculate $\rho_\m{SFR}$ using Equation (\ref{eq_cSFR}) by extrapolating the $SFR/\dot{M}_\m{h}-M_\m{h}$ relation both to the higher redshift and lower mass range.
We integrate down to the halo mass of $10^5\ M_\odot$, comparable to halo masses of first stars in simulations \citep[e.g.,][]{2015MNRAS.448..568H}. 
Figure \ref{fig_z_firststar} shows the calculated cumulative number of formed stars.
The shaded region indicates possible uncertainties of the low-mass slope of the $SFR/\dot{M}_\m{h}-M_\m{h}$ relation and the mass limit of the integration that we adopt $10^4-10^6\ M_\odot$.
The number reaches 1 around $z\sim16-27$, implying the first star formation in this epoch.
This formation epoch agrees with theoretical simulations \citep{2014ApJ...781...60H,2015MNRAS.448..568H}, although this is a very rough estimate.
Especially, it is not clear whether the assumed $SFR/\dot{M}_\m{h}-M_\m{h}$ relation holds at $z>10$ or not,  because physics in the first star formation are expected to be different from star/galaxy formation at the epoch we currently observe due to the evolution of physical parameters such as metallicity.

\begin{figure*}
\centering
\begin{minipage}{0.34\hsize}
\begin{center}
\includegraphics[width=0.99\hsize, bb=12 3 358 286]{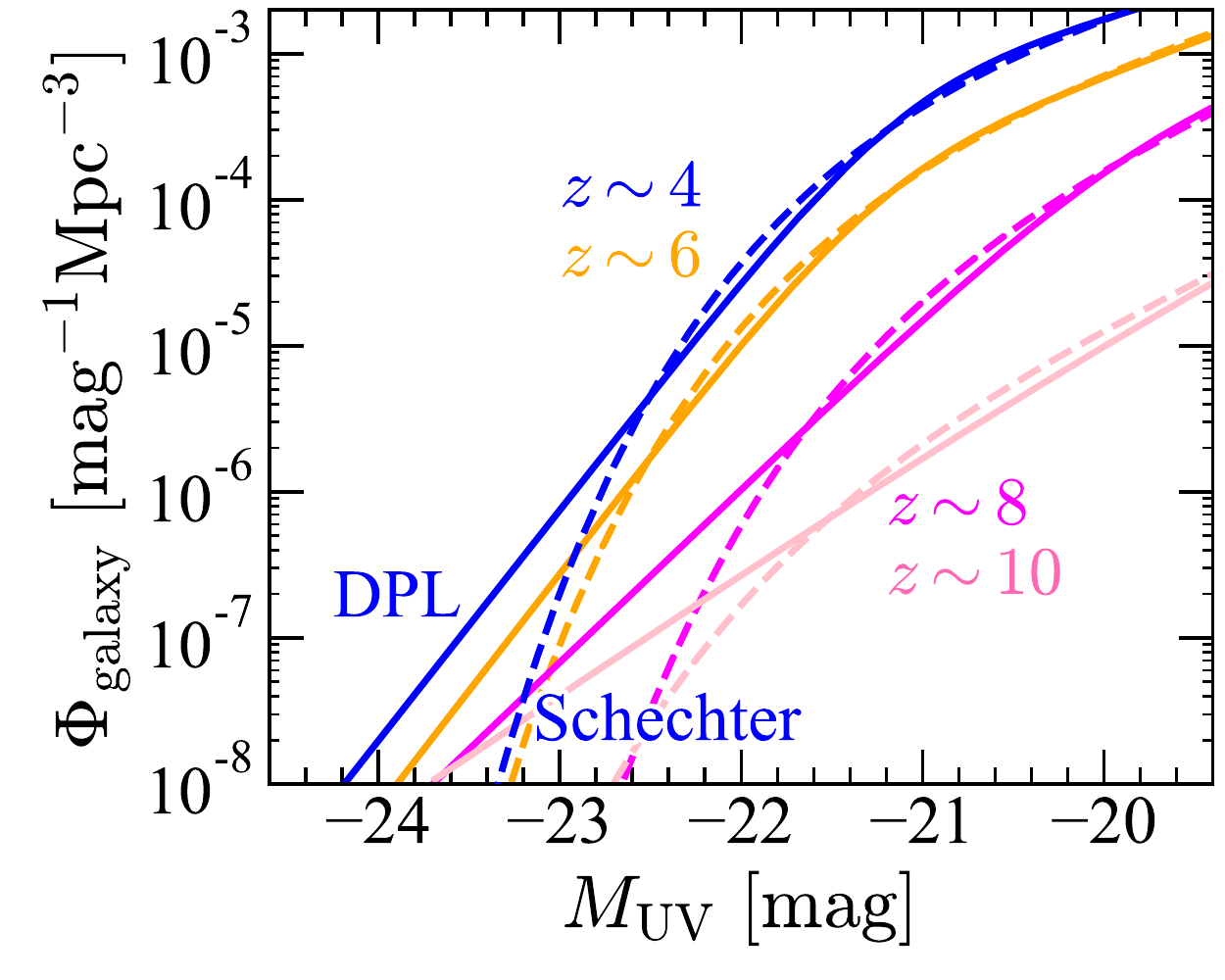}
\end{center}
\end{minipage}
\begin{minipage}{0.28\hsize}
\begin{center}
\includegraphics[width=0.99\hsize, bb=4 3 286 286]{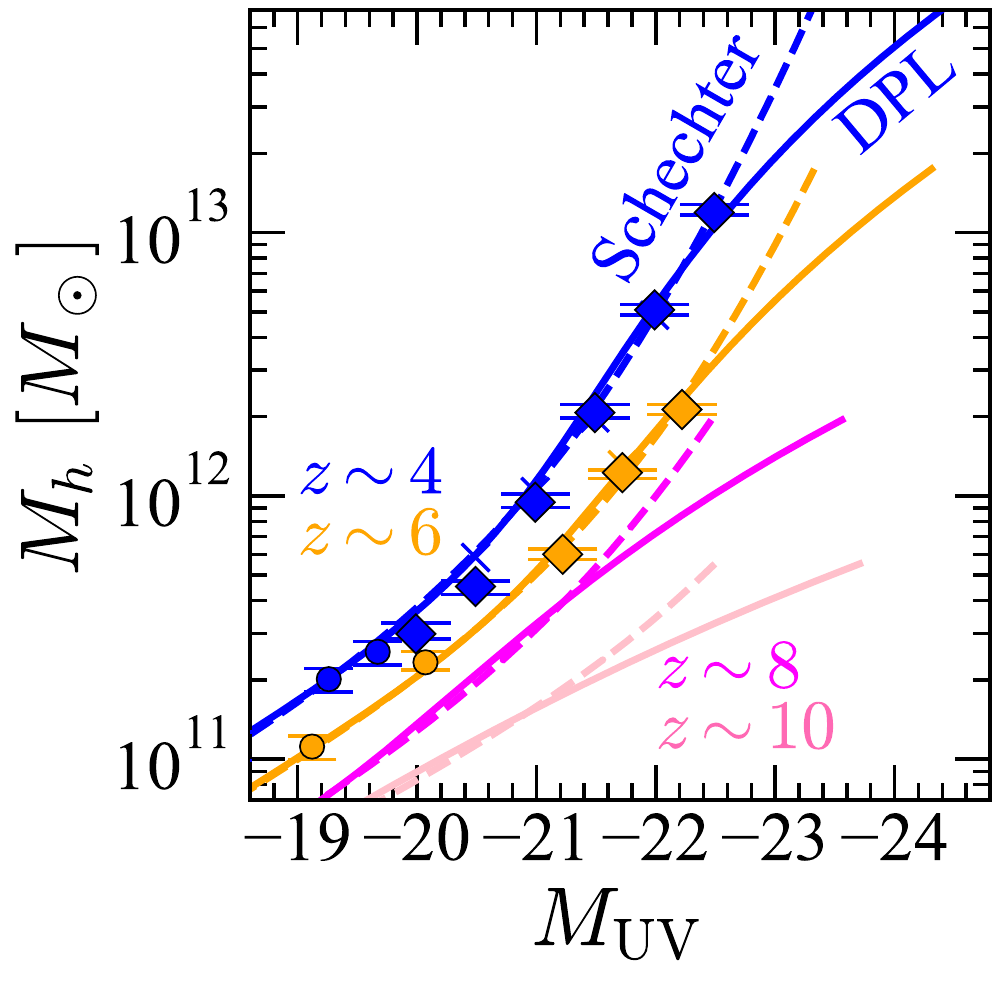}
\end{center}
\end{minipage}
\begin{minipage}{0.35\hsize}
\begin{center}
\includegraphics[width=0.99\hsize, bb=7 3 358 286]{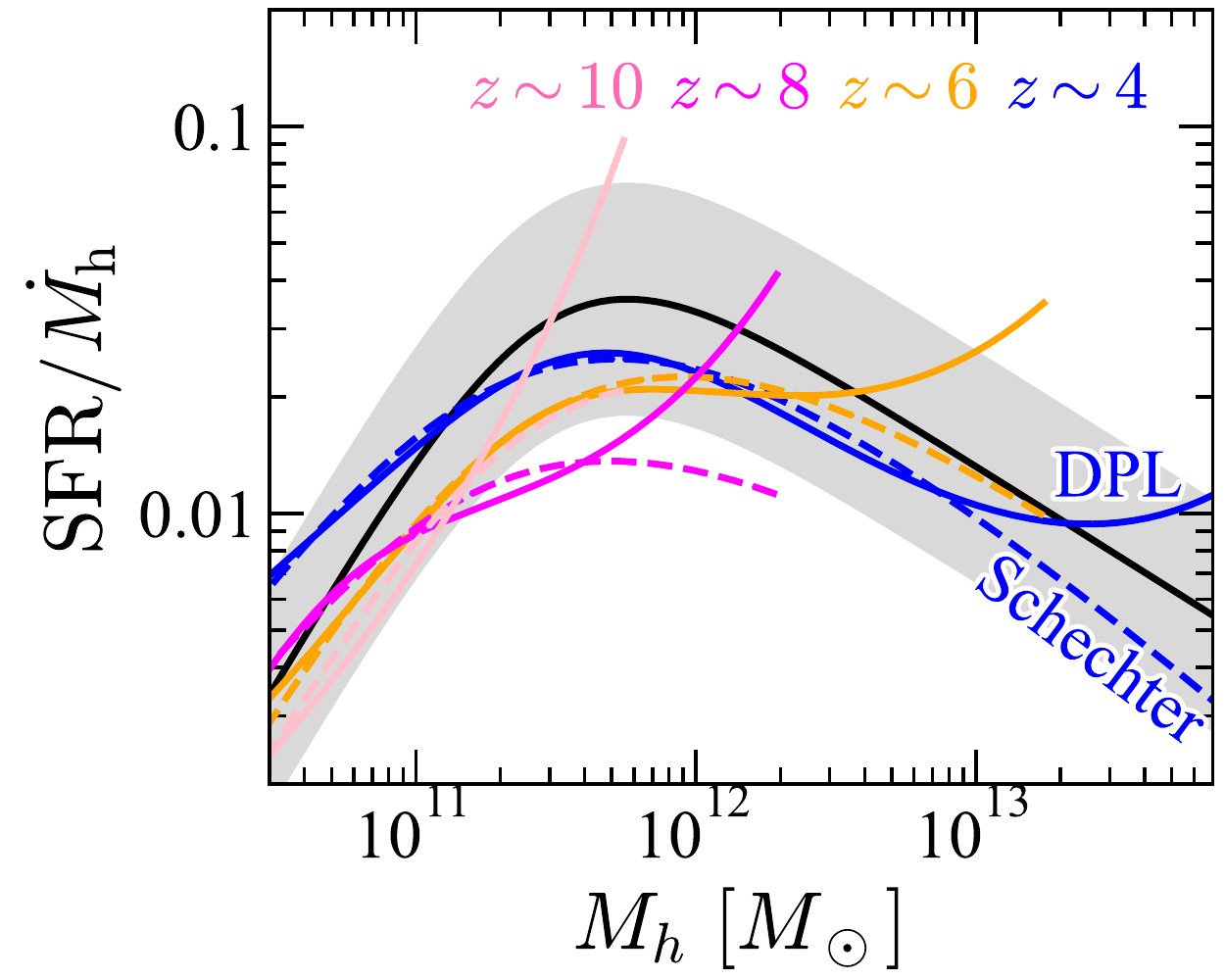}
\end{center}
\end{minipage}
\caption{Galaxy luminosity function (left), $M_\m{UV}-M_\m{h}$ relation (middle), and star formation efficiency ($SFR/\dot{M}_\m{h}-M_\m{h}$ relation, right) in cases of the DPL and Schechter luminosity functions.
The blue, orange, magenta, and pink solid (dashed) lines represent relations in the case of the DPL (Schechter) luminosity functions at $z\sim4$, $6$, $8$, and $10$, respectively.
To reproduce the observed bright end excess of the number density like the DPL function (the solid line in the left panel), the higher star formation efficiency (the solid line in the right panel) is needed at the massive end, compared to the case of the Schechter function (the dashed line).
The blue and orange symbols in the middle panel show the results of this study (diamonds), \citet[][crosses]{2018PASJ...70S..11H}, and \citet[][circles]{2016ApJ...821..123H}.
In the right panel, the black solid curve is the fitting formulae of Equation (\ref{eq_SFRdMdt_1}), and the gray shaded region represents the $2\sigma$ typical scatter ($0.3\ \m{dex}$).
}
\label{fig_PLSch}
\end{figure*}

\subsection{Origin of the Bright End Excess of the Galaxy Luminosity Function}\label{ss_dis_bright}

As presented in Section \ref{ss_galLF_fit}, the obtained galaxy UV luminosity functions cannot be explained by the Schechter functions at the bright end ($M_\m{UV}\lesssim-23$ mag), indicating the existence of the bright end excess of the number density beyond the Schechter function.
Since these luminosity functions are estimated based on the spectroscopic galaxy fractions, the bright end excess cannot be explained by apparent AGNs (e.g., quasars).
Here we discuss the following five possibilities for the origin of the bright end excess, (A) lensed galaxies, (B) mergers, (C) inefficient mass quenching, (D) low dust obscuration and (E) hidden AGN activity.

\begin{itemize}

\item[(A)] {\it Lensed galaxies.}
In Section \ref{ss_galLF_fit}, the lensed Schechter function can provide better fits than the Schechter functions.
This indicates that the bright end excess can be explained by gravitationally lensed galaxies that are apparently bright.
However, in our visual inspection of galaxies located at the bright end, we do not find a significant number of lensed galaxies that show elongated morphologies.
In addition, the $z\sim4$ luminosity function still shows the bright end excess at the $2.5-2.7\sigma$ levels beyond the lensed Schechter function, similar to the results of \citet{2018PASJ...70S..10O}.
As discussed in Section \ref{ss_galLF_fit}, the calculated lensed Schechter function is an upper limit, because some fraction of lensed galaxies might be too close to foreground lensing galaxies to be selected as dropouts in our samples. 
Thus the bright end excess does not seem to be easily explained by lensed galaxies, although high spatial resolution images are required to draw definitive conclusions.

\item[(B)] {\it Mergers.}
Some studies suggest that the major merger at high redshift is an important phase for formation of massive galaxies \cite[e.g.,][]{2020MNRAS.494.1366S}. 
Due to the limited spatial resolutions of the HSC images, some merging galaxies are not resolved and look like one bright galaxy.
If most of galaxies at the bright end turn out to be merging galaxies, the bright end excess can be explained by merging galaxies.
High spatial resolution images such as those obtained by {\it Hubble} are indeed useful to investigate this possibility \citep{2017MNRAS.466.3612B}, but only a small fraction of our sources are observed with {\it Hubble}.
Instead, \citet{2021arXiv210603728S} have made high spatial resolution images by using the super-resolution technique, and estimate a major merger fraction of bright galaxies in galaxy samples of  \citet{2018PASJ...70S..10O}.
They find that the major merger fraction is $10-70\%$ for bright galaxies ($-24\lesssim M_\m{UV}\lesssim-22$ mag) at $z\sim4-7$.
This can partly explain the number density excess of relatively bright galaxies ($M_\m{UV}\gtrsim-23.6$ mag), but is not sufficient to explain the excess of very bright galaxies ($M_\m{UV}\lesssim-23.6$ mag).
Thus, it seems that the bright end excess cannot be explained by mergers.

\item[(C)] {\it Inefficient mass quenching.}
It is thought that the exponential cutoff at the bright end of the Schechter function is caused by the mass quenching that suppresses star formation activity in massive halos (e.g., AGN feedback, virial shock heating). 
Thus the bright end excess beyond the Schechter function implies inefficient mass quenching in these high redshift bright galaxies as discussed in \citet{2018PASJ...70S..10O} and \citet{2020MNRAS.493.2059B}.
To investigate this possibility, we obtain the star formation efficiency, $SFR/\dot{M}_\m{h}$, as a function of the halo mass in cases of the Schechter and DPL luminosity functions.
Using the halo mass function and galaxy luminosity function, we obtain the $M_\m{UV}-M_\m{h}$ relation by the abundance matching technique:
\begin{eqnarray}
\int^\infty_{M_\m{h}} dM'_\m{h} \frac{dn}{dM_\m{h}}(M'_\m{h},z) (1+f_\m{sat})\notag\\
=\int^{M_\m{UV}}_{-\infty}dM'_\m{UV}\Phi_\m{galaxy}(M'_\m{UV},z).
\end{eqnarray}
Then we calculate the SFR and mass accretion rate in the same manner as Section \ref{ss_BCE}, and obtain the $SFR/\dot{M}_\m{h}$ ratio.

Figure \ref{fig_PLSch} presents the luminosity function, the $M_\m{UV}-M_\m{h}$ relation, and the star formation efficiency (the $SFR/\dot{M}_\m{h}-M_\m{h}$ relation) in cases of the DPL and Schechter functions at $z\sim4$, $6$, $8$, and $10$.
We find that estimated halo masses from our abundance matching agree well with those from the clustering analysis (the middle panel in Figure \ref{fig_PLSch}), as discussed in \citet{2016ApJ...821..123H}.
If we assume the steep decline of the star formation efficiency ($SFR/\dot{M}_\m{h}$) toward the massive end (the dashed line in the right panel), the calculated number density shows the exponential decline at the bright end similar to the Schechter function (the dashed line in the left panel), and cannot reproduce the excess of the observed number densities.
To reproduce the bright end excess of the number density like the DPL function (the solid line in the left panel), higher star formation efficiencies are needed at the massive end (the solid line in the right panel), compared to the case of the Schechter function.
For example, $\sim2$ times higher star formation efficiency is needed in halos of $M_\m{h}\simeq10^{13}\ (10^{12})\ M_\odot$ at $z\sim6$ (8).
The high star formation efficiency at the bright end can be made by the inefficient mass quenching.
Indeed the mass quenching is expected to be less efficient at higher redshift because of the shorter time scale of gas cooling and/or weaker AGN feedback due to the decreasing number of AGNs (i.e., quasar luminosity function) as discussed in Section \ref{ss_LFall_evol}.

\item[(D)] {\it Low dust obscuration.}
\citet{2020MNRAS.493.2059B} discuss the possibility that the intrinsic (without dust attenuation) UV luminosity function has a shallower decline at the bright end, and the dust obscuration controls the shape of the luminosity function.
In the calculations above (C), we assume the attenuation-$\beta_\m{UV}$ relation in \citet{1999ApJ...521...64M} and the $\beta_\m{UV}-M_\m{UV}$ relations in \citet{2014ApJ...793..115B}.
However, the attenuation curve of high redshift galaxies is not well-understood \citep[e.g.,][]{2019PASJ...71...71H,2020A&A...643A...4F,2020ApJ...896...93H,2020MNRAS.493.4294B}, and the $\beta_\m{UV}-M_\m{UV}$ relation is not well-constrained at this very bright magnitude range (i.e., $\sim-23$ mag).
\redc{In addition, some recent studies report dust-poor UV luminous star forming galaxies at $z>2$ \citep{2020MNRAS.499L.105M,2021MNRAS.507..524M}.}
Thus it is possible that the dust obscuration in these bright galaxies is lower than what we assumed, resulting the bright end excess beyond the Schechter function.

\item[(E)] {\it Hidden AGN activity.}
Although we subtract the number density of AGNs by the spectroscopic galaxy fractions, it is possible that there are still hidden AGNs in the galaxy luminosity function.
UV luminosities of such hidden AGNs could be boosted due to AGN activity \redc{\citep[e.g.,][]{2021ApJ...910L..11K}}, resulting in the bright end excess.
AGN activity of such sources can be probed only by deep spectroscopy covering several high ionization lines.
Indeed, some studies report possible AGN activity in bright ($M_\m{UV}\lesssim -22$ mag) galaxies at $z\gtrsim7$ \citep[e.g.,][]{2017ApJ...851...40L,2018MNRAS.479.1180M,2021NatAs...5..256J,2021MNRAS.502.6044E,2021arXiv210613807O}.
Since we do not know the fraction of such hidden AGNs in our sample due to the lack of deep spectroscopic data at the bright end, we cannot rule out the possibility that hidden AGNs make the bright end excess.

\end{itemize}

Based on these discussions above, we conclude that the bright end excess is possibly made by (C) inefficient mass quenching, (D) low dust obscuration and/or (E) hidden AGN activity, although  it is possible that the dominant effect in making the bright end excess changes with redshift.
We cannot distinguish these possibilities with the current datasets, and future large and deep observations are needed.
For example, {\it Euclid} and {\it Nancy Grace Roman Space Telescope} can identify a large number of high redshift galaxies located at the bright end, allowing us to investigate the clustering and the star formation efficiency of such bright galaxies.
ALMA follow-up observations for a statistical sample of bright galaxies will reveal the typical dust properties of these galaxies.
Deep spectroscopy for a large number of bright galaxies with Subaru/PFS will allow us to investigate hidden AGN activity in such bright galaxies.

\section{Summary}\label{ss_summary}
In this paper, we have identified 1,978,462 dropout candidates at $z\sim4-7$ from $\sim300$ deg$^2$ deep optical imaging data obtained in the HSC-SSP survey.
Among these dropout candidates, 1037 dropouts are spectroscopically identified in our follow-up observations and the literature.
\redccc{Typical contamination rates are $<20\%$ and $<40\%$ at $z\sim4-5$ and $z\sim6-7$, respectively}.
Combined with $z\sim2-3$ galaxy samples, we have a total of 4,100,221 sources at $z\sim2-7$, which is the largest sample of high redshift galaxies to date.
Using this sample, we have calculated the luminosity functions and angular correlation functions, and investigated statistical properties of these sources.

Our major findings are summarized below:
\begin{enumerate}

\item We have obtained rest-frame UV luminosity functions at $z\sim4-7$ (Figures \ref{fig_LFall_1} and \ref{fig_LFall_2}).
Combined with results based on the complementary ultra-deep {\it Hubble} data and wide-area SDSS data, we have probed the luminosity function in a very wide UV magnitude range of $-29\lesssim M_\m{UV}\lesssim -14$ mag, corresponding to the luminosity range of $0.002L^*_\m{UV}\lesssim L_\m{UV}\lesssim 2000L^*_\m{UV}$.

\item Spectroscopic galaxy fractions indicate that most of the sources are AGNs (galaxies) at $M_\m{UV}<-24$ ($M_\m{UV}>-22$) mag (Figures \ref{fig_LFall_1} and  \ref{fig_LFall_2}).
We have found that the luminosity function in this very wide magnitude range can be well-fitted by the DPL+DPL or DPL+Schechter functions (Figure \ref{fig_LFall_wfit}), indicating that the dropout luminosity function is a superposition of the AGN luminosity function (dominant at the bright end) and the galaxy luminosity function (dominant at the faint end).

\item We have estimated the galaxy luminosity functions by subtracting the AGN contributions using the spectroscopic galaxy fractions.
The obtained galaxy luminosity functions show the bright end excess of the number density beyond the Schechter function at $\gtrsim2\sigma$ levels (Figure \ref{fig_LFgal_wfit}), which is possibly made by the inefficient mass quenching, low dust obscuration, and/or hidden AGN activity (Section \ref{ss_dis_bright}).

\item We have derived angular correlation functions of galaxies at $z\sim2-6$ (Figures \ref{fig_ACF_lowz} and \ref{fig_ACF_highz}).
Combined with the HOD model analyses and previous clustering measurements for faint galaxies at $z\sim4-7$, we have obtained the relation between the dark matter halo mass and the UV magnitude over two orders of magnitude in the halo mass (Figure \ref{fig_MUV_Mh}).

\item We have calculated the ratio of the SFR to the dark matter accretion rate, $SFR/\dot{M}_\m{h}$, and identified an $SFR/\dot{M}_\m{h}-M_\m{h}$ relation which does not show strong redshift evolution beyond 0.3 dex at $z\sim2-7$ (Figure \ref{fig_SFR_dMdt}).
This weak evolution indicates that the star formation efficiency does not significantly change at high redshift, and star formation activities are regulated by the dark matter mass assembly, as suggested by our earlier work at $z\sim4-7$ \citep{2018PASJ...70S..11H}.
Meanwhile, the $SFR/\dot{M}_\m{h}$ ratio gradually increases with decreasing redshift from $z\sim5$ to $2$ within 0.3 dex.

\item We have found that the $SFR/\dot{M}_\m{h}-M_\m{h}$ relation can reproduce the redshift evolution of the cosmic SFR density and the UV luminosity function (Figures \ref{fig_cSFR} and \ref{fig_LF_SFRdMdt}).
These good agreements indicate that the evolution of the cosmic SFR densities is primarily driven by the steep increase of the halo number density from $z\sim10$ to $z\lesssim4$ due to the structure formation, and the decrease of the accretion rate due to the cosmic expansion with a steep drop from $z\sim2$ to $0$ (Figure \ref{fig_HMF_dMdt_panel}).

\end{enumerate}

We have further showed that the cosmic SFR density at $z>10$ decreases towards higher redshift more rapidly than the extrapolation of the fitting function in \citet{2014ARA&A..52..415M} if we assume the constant star formation efficiency (Figures \ref{fig_cSFR_highz} and \ref{fig_cSFR_fit}).
{\it JWST} observations allow us to directly investigate this rapid evolution by measuring the cosmic SFR densities at $z>10$, providing insights into star formation efficiency in the early universe.

\acknowledgments
We thank the anonymous referee for a careful reading and valuable comments that improved the clarity of the paper.
We thank Masao Hayashi, Yoshiyuki Inoue, Kentaro Nagamine, Rui Marques Coelho Chaves, Shun Saito, Masayuki Tanaka, Yoshihiko Yamada for useful comments and discussions, and Thibaud Moutard and Yechi Zhang for providing their data of UV luminosity functions.
This work is supported by the World Premier International Research Center Initiative (WPI Initiative), the Ministry of Education, Culture, Sports, Science and Technology, Japan, as well as KAKENHI Grant-in-Aid for Scientific Research (19J01222, 20H00180, 21H04467, and 21K13953) through the Japan Society for the Promotion of Science (JSPS).

The HSC collaboration includes the astronomical communities of Japan and Taiwan, and Princeton University.  The HSC instrumentation and software were developed by the National Astronomical Observatory of Japan (NAOJ), the Kavli Institute for the Physics and Mathematics of the Universe (Kavli IPMU), the University of Tokyo, the High Energy Accelerator Research Organization (KEK), the Academia Sinica Institute for Astronomy and Astrophysics in Taiwan (ASIAA), and Princeton University.  Funding was contributed by the FIRST program from the Japanese Cabinet Office, MEXT, the Japan Society for the Promotion of Science (JSPS), Japan Science and Technology Agency  (JST), the Toray Science  Foundation, NAOJ, Kavli IPMU, KEK, ASIAA, and Princeton University.

This paper makes use of software developed for the Large Synoptic Survey Telescope. We thank the LSST Project for making their code available as free software at  http://dm.lsst.org

This paper is based on data collected at the Subaru Telescope and retrieved from the HSC data archive system, which is operated by Subaru Telescope and Astronomy Data Center (ADC) at NAOJ. Data analysis was in part carried out with the cooperation of Center for Computational Astrophysics (CfCA), NAOJ.
This work was supported by the joint research program of the Institute for Cosmic Ray Research (ICRR), University of Tokyo.

Part of the data were obtained and processed as part of the CFHT Large Area U-band Deep Survey (CLAUDS), which is a collaboration between astronomers from Canada, France, and China described in \citet{2019MNRAS.489.5202S}.  CLAUDS is based on observations obtained with MegaPrime/ MegaCam, a joint project of CFHT and CEA/DAPNIA, at the CFHT which is operated by the National Research Council (NRC) of Canada, the Institut National des Science de l’Univers of the Centre National de la Recherche Scientifique (CNRS) of France, and the University of Hawaii. CLAUDS uses data obtained in part through the Telescope Access Program (TAP), which has been funded by the National Astronomical Observatories, Chinese Academy of Sciences, and the Special Fund for Astronomy from the Ministry of Finance of China. CLAUDS uses data products from TERAPIX and the Canadian Astronomy Data Centre (CADC) and was carried out using resources from Compute Canada and Canadian Advanced Network For Astrophysical Research (CANFAR).
CADC is operated by the National Research Council of Canada with the support of the Canadian Space Agency.

The Pan-STARRS1 Surveys (PS1) and the PS1 public science archive have been made possible through contributions by the Institute for Astronomy, the University of Hawaii, the Pan-STARRS Project Office, the Max Planck Society and its participating institutes, the Max Planck Institute for Astronomy, Heidelberg, and the Max Planck Institute for Extraterrestrial Physics, Garching, The Johns Hopkins University, Durham University, the University of Edinburgh, the Queen’s University Belfast, the Harvard-Smithsonian Center for Astrophysics, the Las Cumbres Observatory Global Telescope Network Incorporated, the National Central University of Taiwan, the Space Telescope Science Institute, the National Aeronautics and Space Administration under grant No. NNX08AR22G issued through the Planetary Science Division of the NASA Science Mission Directorate, the National Science Foundation grant No. AST-1238877, the University of Maryland, Eotvos Lorand University (ELTE), the Los Alamos National Laboratory, and the Gordon and Betty Moore Foundation.

The Cosmic Dawn Center is funded by the Danish National Research Foundation under grant No. 140. This project has received funding from the European Union's Horizon 2020 research and innovation program under the Marie Sklodowska-Curie grant agreement No. 847523 `INTERACTIONS'.

\software{galaxev \citep{2003MNRAS.344.1000B}, GalSim \citep[v1.4;][]{2015A&C....10..121R}, hscpipe \citep{2018PASJ...70S...5B}, MIZUKI \citep{2015ApJ...801...20T,2018PASJ...70S...9T}, SynPipe \citep{2018PASJ...70S...6H}}

\appendix

\section{Catalog of Spectroscopic Sources}
Table \ref{tab_speccat} summarizes 1037 spectroscopically confirmed galaxies and AGNs in our dropout samples at $z\sim4-7$.

\LongTables
\begin{deluxetable*}{ccccccccc}
\tablecaption{Spectroscopically Identified Galaxies and AGNs in Our Dropout Samples.}
\tablehead{\colhead{ID} & \colhead{R.A. (J2000)} & \colhead{Decl. (J2000)} & \colhead{$z_\m{spec}$} & \colhead{$m$} & \colhead{$M_\m{UV}$} & \colhead{Sample} & \colhead{Flag} & \colhead{Reference}  \\
\colhead{}& \colhead{($^\m{h}$\ \ $^\m{m}$\ \ $^\m{s}$)}& \colhead{($\arcdeg$\ \ $\arcmin$\ \ $\arcsec$)}& \colhead{}& \colhead{(mag)}& \colhead{(mag)}& \colhead{}& \colhead{}& \colhead{} \\
\colhead{(1)}& \colhead{(2)}& \colhead{(3)}& \colhead{(4)}& \colhead{(5)}& \colhead{(6)}& \colhead{(7)}& \colhead{(8)}& \colhead{(9)} }
\startdata
\multicolumn{9}{c}{Galaxies}\\
HSC\ J020649$-$020618 & 02:06:49.98 & $-$02:06:18.20 & 5.72 & 23.1 & $-23.6$ & 3 & 1 & Matsuoka et al. (2019b)\\
HSC\ J020834$-$021239 & 02:08:34.37 & $-$02:12:39.83 & 4.088 & 24.7 & $-21.4$ & 1 & 1 & This work\\
HSC\ J021041$-$055917 & 02:10:41.28 & $-$05:59:17.93 & 5.82 & 24.0 & $-22.6$ & 3 & 1 & Matsuoka et al. (2016)\\
HSC\ J021244$-$015824 & 02:12:44.76 & $-$01:58:24.62 & 6.0 & 22.9 & $-23.9$ & 3 & 1 & Matsuoka et al. (2018a)\\
HSC\ J021542$-$051008 & 02:15:42.43 & $-$05:10:08.71 & 4.851 & 25.5 & $-20.9$ & 2 & 1 & This work\\
HSC\ J021545$-$055529 & 02:15:45.22 & $-$05:55:29.14 & 5.744 & 23.6 & $-23.0$ & 3 & 1 & Harikane et al. (2020a)\\
HSC\ J021549$-$045741 & 02:15:49.28 & $-$04:57:41.81 & 3.449 & 24.8 & $-21.0$ & 1 & 1 & This work\\
\enddata
\tablecomments{(1) Object ID. (2) Right ascension. (3) Declination. (4) Spectroscopic redshift. (5) Apparent magnitude. (6) Absolute UV magnitude. (7) The dropout sample in which the source is selected: 1 = $g$-dropout, 2 = $r$-dropout, 3 = $i$-dropout, and 4 = $z$-dropout. (8) Galaxy/AGN flag (1 = galaxy; 2 = AGN). (9) Reference for the spectroscopic redshift.}
\label{tab_speccat}
\end{deluxetable*}

\bibliographystyle{apj}
\bibliography{apj-jour,reference}

\end{document}